\documentclass[11pt]{article}

\usepackage[top=2.5cm,bottom=2.5cm,left=1.4cm,right=2.0cm]{geometry}
\usepackage{cite}
\usepackage[T1]{fontenc}
\usepackage[american]{babel}
\usepackage{layout}
\usepackage{amsmath}
\usepackage{amssymb,amstext, amsthm, simplewick}
\usepackage{framed}
\usepackage[svgnames,dvipsnames,x11names]{xcolor}
\usepackage{amsfonts}
\usepackage{color} 
\usepackage{braket}
\usepackage{multicol} 
\usepackage{epsfig}
\usepackage{cancel}
\usepackage{caption}
\usepackage{epsf}
\usepackage{feynmf} 
\usepackage{frontespizio}
\usepackage{rotating}
\usepackage{subfigure}
\usepackage{pstricks}
\usepackage{type1ec}
\usepackage{lettrine}
\usepackage{bbold}
\usepackage{calligra}
\usepackage{tikz}
\usepackage{subfigure}
\usepackage{mathrsfs}
\usepackage{curve2e}
\usepackage{setspace}
\usepackage{indentfirst}
\usepackage{emptypage}
\usepackage{relsize}
\usepackage{mathrsfs}
\usepackage{stackengine}
\usepackage{calc}
\usepackage{hyperref}
\hypersetup{
	colorlinks,
	citecolor=green,
	filecolor=black,
	linkcolor=blue,
	urlcolor=black
}
\numberwithin{equation}{subsection} 
\makeatletter
\@addtoreset{equation}{section}
\makeatother

\def\eps{\varepsilon}

\def\dif{\text{d}}

\def\d4x{\dif^4 x}

\newcommand{\la}{\lambda}

\newcommand{\beq}{\begin{equation}}
\newcommand{\eeq}{\end{equation}}
\let\a=\alpha   \let\b=\beta      \let\d=\delta
         
\let\i=\iota      \let\l=\lambda  \let\m=\mu
\let\n=\nu                 
\let\s=\sigma        

\newcommand{\lag}{\langle}
\newcommand{\rag}{\rangle}
\newcommand{\E}{\boldsymbol{E}}

\newcommand{\B}{\boldsymbol{B}}

\newcommand{\bea}{\begin{eqnarray}}
\newcommand{\eea}{\end{eqnarray}}

\newcommand{\nn}{\nonumber}
\newcommand{\veps}{\varepsilon}
\newcommand{\be}{\begin{equation}}
\newcommand{\ee}{\end{equation}}
\newcommand{\beqa}{\begin{eqnarray}}
\newcommand{\eeqa}{\end{eqnarray}}

\newcommand{\secref}[1]{Section~\ref{#1}}		% for sections
\newcommand{\ul}{\underline}
\def\nbox#1#2{\vcenter{\hrule \hbox{\vrule height#2in
			\kern#1in \vrule} \hrule}}
\def\sq{\,\raise.5pt\hbox{$\nbox{.09}{.09}$}\,}
\def\sqb{\,\raise.5pt\hbox{$\overline{\nbox{.09}{.09}}$}\,}
\def\Box{\sq}
\usepackage{accents}

\begin{document}
\begin{center}
{\bf \Large Axion-like Interactions and CFT in Topological Matter,\\}
\vspace{0.3cm}
{\bf \Large Anomaly Sum Rules and the Faraday Effect \\}
\vspace{1cm}
{\bf $^{(1,2)}$Claudio Corian\`o, $^{(1) (3)}$Mario Cret\`i, $^{(1)}$Stefano Lionetti, \\ $^{(1)}$Dario Melle
and $^{(1)}$Riccardo Tommasi \\}

\vspace{1cm}
{\it  $^{(1)}$Dipartimento di Matematica e Fisica, Universit\`{a} del Salento \\
and INFN Sezione di Lecce, Via Arnesano 73100 Lecce, Italy\\
National Center for HPC, Big Data and Quantum Computing\\}
\vspace{0.5cm}
{\it  $^{(2)}$  Institute of Nanotechnology, \\ National Research Council (CNR-NANOTEC), Lecce 73100\\}

\vspace{0.5cm}
{\it $^{(3)}$Center for Biomolecular Nanotechnologies,\\ Istituto Italiano di Tecnologia, Via Barsanti 14,
73010 Arnesano, Lecce, Italy\\}
\vspace{0.5cm}

\begin{abstract}
We discuss fundamental aspects of chiral anomaly-driven interactions in conformal field theory (CFT) in four spacetime dimensions. They find application in very general contexts, from early universe plasma to topological condensed matter. We outline the key shared characteristics of these interactions, specifically addressing the case of chiral anomalies, both for vector currents and gravitons. 
In the case of topological materials, the gravitational chiral anomaly is generated by thermal gradients via the (Tolman-Ehrenfest) Luttinger relation. In the CFT framework, a nonlocal effective action, derived through perturbation theory, indicates that the interaction is mediated by an excitation in the form of an anomaly pole, which appears in the conformal limit of the vertex. To illustrate this, we demonstrate how conformal Ward identities (CWIs) in momentum space allow us to reconstruct the entire chiral anomaly interaction in its longitudinal and transverse sectors just by inclusion of a pole in the longitudinal sector. Both sectors are coupled in amplitudes with an intermediate chiral fermion or a bilinear Chern-Simons current with intermediate photons. In the presence of fermion mass corrections, the pole transforms into a cut, but the absorption amplitude in the axial-vector channel satisfies mass-independent sum rules related to the anomaly in any chiral interaction. The detection of an axion-like/quasiparticle in these materials may rely on a combined investigation of these sum rules, along with the measurement of the angle of rotation of the plane of polarization of incident light when subjected to a chiral perturbation. This phenomenon serves as an analogue of a similar one in ordinary axion physics, in the presence of an axion-like condensate, that we rederive using axion electrodynamics.  
\end{abstract}
\newpage
\tableofcontents

\end{center}
\newpage
\section{Introduction} 
In contrast to the traditional classification based on the presence of band gaps in conducting and insulating materials, a novel category of materials known as topological insulators (TIs) 
emerged at the turn of the century (see \cite{Chernodub:2021nff,Neupert:2018oma,Burkov:2015hba,Wehling:2014cla} for an overview). The analysis of Hamiltonians in 3D for time-reversal invariant electrons \cite{Kane2005,Fu2007,Essin2007} \cite{Moore2007}, showed that, similar to the integer quantum Hall effect, band-structure integrals can be used to classify insulators in both 2D and 3D as ordinary or  topological, based on their $\mathbb{Z}_2$ topological invariants \cite{Thouless1982,Kohmoto1985}. These invariants are remarkably robust, persisting even in the presence of disorder, which is a key feature of topological insulators (and superconductors). In \cite{Schnyder2008,Kitaev:2009mg} was presented a complete classification of topological insulators and superconductors in any dimension \cite{Altland:1997zz}. They can exhibit topological insulating phases with gapless surface states protected by topology.\\ 
Materials like HgTe\cite{Konig:2007stx} Bi$_x$Sb$_{1-x}$  \cite{Hsieh2008}, Bi$_2$Se$_3$, and Bi$_2$Te$_3$ \cite{Hsieh2009,Xia2009,Hsieh2009b,Chen2009} provide direct realization of these phenomena. In some cases they exhibit a quantized magnetoelectric response proportional to a parameter $(\theta$), due to electron orbital motion. The phase can be identified by the bulk polarization's response to an applied magnetic field \cite{Qi:2008ew}, which is described by axion electrodynamics \cite{Wilczek1987}.\\
TIs exhibit band gaps within their bulk structure while also featuring boundary states devoid of energy gaps. Unlike conventional insulating metals, transitioning between the phases of TIs and ordinary insulators requires transformations involving the opening and closing of band gaps, rather than adiabatic processes. \\The conventional method of characterizing different phases of matter, which relies on a local order parameter and the notion of spontaneous symmetry breaking, proves inadequate in distinguishing between TIs and other materials. \\
For instance, if both systems possess time-reversal invariance (TRI) symmetry, they cannot be discerned solely through this approach. However, alternative methods exist for distinguishing between them. \\
For example, the magnetoelectric response parameter $\theta$ can be quantized. This coefficient takes the value 0 (mod $2\pi$) for TIs, signifying a distinct response compared to conventional insulators where $\theta = 0$.\\
\subsection{Classification}
For insulators lacking time-reversal symmetry (TRI), two categories emerge: 1) axion-like and 2) magnetic. In axion-like insulators, an effective form of TRI arises when combined with a lattice translation, even with the explicit breaking of TRI. Also these materials possess a quantized topological response with $\theta = 0$ (mod $2\pi$). Conversely, magnetic insulators exhibit a broken TRI and a non-vanishing topological phase ($\theta \neq 0$).\\
 This non-zero phase is proportional to the material's magnetization, $M(\mathbf{x}, t)$.\\ If the magnetization acts as a local field, the response manifests as an axion-like interaction, expressed in terms of the electric and magnetic fields as 
 \beq
 \label{anom}
 \theta \tilde{F}F \approx \theta \,\mathbf{E} \cdot \mathbf{B}.
 \eeq
This interaction finds an analogy in chiral quantum field theories (QFT) affected by a global chiral anomaly. Indeed, in the local formulation of the anomaly effective action, \ref{anom} is the standard way in which an asymptotic pseudoscalar field, identified with $\theta$, couples to the anomaly $\tilde{F}F$.\\
We will adddress anomalies related to continue symmetries rather than to discrete ones. For a detailed discussion on the implications of a discrete anomaly on the quantum anomalous Hall effect, such as the parity anomaly, we refer to \cite{Bottcher:2019rrz} and references therein.\\
The study of the topological response in the context of anomaly interaction within ordinary QFT is the goal of this review.

\subsection{Harnessing chirality in topological insulators}
Central to unlocking the full potential of such materials is the generation of chiral currents, wherein electrons flow unidirectionally along the edges or boundaries of the material (see for instance \cite{PhysRevLett.114.256801}). Achieving this result relies upon breaking the TRI of these materials. \\
One possibility is the inclusion of magnetic atoms into the crystal lattice or the application of an external magnetic field, in this way TRI can be effectively broken within the TI. This facilitates the emergence of chiral edge states, where electrons navigate in a preferred direction along the boundaries of the material. This approach offers versatility in determing the chiral current behavior.\\
Another approach consist in the application of a mechanical strain to the TI material. The introduction of strain induces a non-zero Chern number, thereby fostering chiral edge currents. This method underscores the interplay between mechanical deformation and electronic properties, opening avenues for tunable chiral transport.\\
The combination of different TIs or their integration with other materials creates an ideal environment for investigating chiral phenomena. At the interface of these materials, complex band structures interact, leading to the emergence of chiral edge states. Manipulating heterostructures provides a pathway to engineer customized chiral transport pathways.\\
Circularly polarized light has the ability to impart angular momentum to electrons within the TI, disrupting Time-Reversal Invariance (TRI) and potentially generating a chiral photocurrent. This non-invasive method holds promise for dynamically modulating chiral behavior, paving the way for optically controlled electronics.\\
Tailoring specific geometries within the TI material, such as sharp corners or edges, enables the localization of chiral states. These topological corner states facilitate directed current flow, offering opportunities for miniaturized devices and robust electronic circuitry. The selection of a particular method depends on the intrinsic properties of the TI material and the desired characteristics of the chiral current.

 \subsection{Topological response, chiral and conformal anomalies}
If the topological response of TIs to chiral perturbations allows us to establish a link with topological aspects of QFT and anomalies
\cite{Qi:2010qag, Qi:2008ew,Sekine:2020ixs, Zyuzin:2012tv,Tutschku:2020rjq}, then the investigation of the interpolating chiral anomaly vertex plays an essential role.\\
In general, the analysis of these interactions is based on perturbative approaches. 
However, we are going to show that an independent analysis can be performed nonperturbatively, without resorting to a Lagrangian realization, using CFT methods. This 
opens the way to new anomalies, as shown in the case of the parity odd-trace anomaly \cite{Coriano:2023cvf}.\\
Previous perturbative analysis  have provided a in-depth characterization of the corresponding effective actions for chiral and conformal anomalies \cite{Giannotti:2008cv,Armillis:2010pa,Armillis:2009pq,Armillis:2009sm,Coriano:2020ees}, both of relevance for topological materials. 
It is well-known that anomalies \cite{Adler:1969gk,Bell:1969ts} (see \cite{Bonora:2023soh,Scrucca:2004jn} for overviews)  in ordinary gauge theories, are related with the presence of certain interactions, in a given gauge theory, that need to be canceled by the choice of appropriate charge assignments for the fermion spectrum. \\
The ordinary anomaly cancellation mechanism in the Standard Model of the elementary particles, indeed, bans interactions carrying gauge anomalies.  Anomalies don’t carry any scale and this defines an important link between this phenomenon and conformal symmetry, which we are going to explore 
and is the main motivation of our analysis. \\
As we are going to show, the presence of conformal symmetry and of a quasiparticle pole in the chiral correlator exhibiting the anomaly, are the only two ingredients that allow to completely characterize the anomaly behaviour in all those cases where an anomaly is present, in a certain field theory action. \\
\subsection{Scale independence of the interaction}
Chiral anomalies are scale independent, a feature that can be straightforwardly confirmed within perturbation theory. One can show quite straightforwardly that mass corrections, typically associated with the spontaneous symmetry breaking of gauge symmetry in a gauge theory, do not alter the anomaly. \\
We remind that the fermion mass, in a gauge theory such as the Standard Model, arises as a consequence of a order parameter generated by a local interaction, namely the vacuum expectation value (vev) of the Higgs field, after spontaneous symmetry breaking. The independence of the anomaly from such a vev, parallels the previous discussion wherein topological phase transitions remain distinct from the spontaneous breaking of local symmetries.\\
Thus, the presence of a quantized or continuous dimensionless constant or field in the topological response of a given material, interpreted as an axion-like interaction, serves as a monitor for essential topological features of such materials, which are not inherently associated with local operators.\\
Furthermore, as we delve into our discussion, as just mentioned, it becomes evident that the anomaly phenomenon is intimately linked to the exchange of a massless pole in the anomaly vertex. This picture emerges from a dispersion relation involving the spectral density of the anomaly form factor, present in the same vertex, in the conformal limit, a point we shall thoroughly explore in the forthcoming sections. \\
\subsection{Content of this work}
Our study begins with an examination of the nonlocal structure found in the chiral anomaly effective action, which is regarded as an induced action derived from an anomaly pole \cite{Giannotti:2008cv,2009PhLB..682..322A}. This formulation stems from a variational approach to the anomaly constraint, which is given by the magnetoelectric response in \eqref{anom}. \\
The anomaly constraint describes how a condensed matter system responds when subjected to variations induced by an external axial-vector field. Initially discussed within the context of kinetic mixing \cite{Giannotti:2008cv}, it was later reformulated using Stuckelberg axions \cite{Armillis:2011hj}, representing a broken symmetry characterized by a non-zero Ward identity in the axial-vector channel. Our analysis reveals the emergence of a ghost when the kinetic mixing of the local action is reformulated as separate degrees of freedom, a point discussed in \secref{mixing}. An alternative variational approach, utilizing a hydrodynamic framework with variations performed with respect to the external chiral current rather than the axial-vector source, has been explored in \cite{Mottola:2019nui}\cite{Mottola:2023emy}.\\
We delve into the structure of the anomaly vertex using two different representations in \secref{conf1}, followed by an analysis of the dispersion relation and sum rule satisfied by the extension of the interaction once we include massive fermions in \secref{spectral1}. Similar sum rules are demonstrated to hold for the gravitational chiral anomaly, where the correlator involves two stress-energy tensors alongside the chiral current. \secref{conf2} and \secref{conf3} are dedicated to outlining the relationship between CFT and both the chiral and gravitational anomalies. In \secref{conf4}, we briefly illustrate the relevance of gravitational correlators in the analysis of TI subjected to thermal stress, as predicted by the Luttinger relation between gravity and thermal gradients \cite{Rovelli:2010mv,Bermond:2022mjo}.\\
The last two sections focus on the local action of axion electrodynamics, summarizing the interaction between an asymptotic pseudoscalar axion-like field and the anomaly. We outline two key features of this action, one related to the rotation of the plane of polarization of light incident on a material, for which we provide a detailed derivation, addressing a gap in the literature. This effect was initially predicted in \cite{Harari:1992ea} within axion physics, in the presence of an axion condensate. In the case of a TI, this effect would similarly signify the presence of an effective axion condensate within a system. In the final section, before drawing our conclusions, we describe an intriguing aspect of the light-cone behavior of the electromagnetic propagator in the presence of a timelike condensed axion field.  The equations of axion electrodynamics exhibit oscillations across the entire light-cone, a phenomenon we elucidate before concluding our study.

\section{Anomalies and nonlocal versus local actions}
It's noteworthy that chiral interactions, as derived in perturbative QFT, are described by nonlocal actions.\\
Their local formulations have been extensively explored in prior studies, employing various approaches. One such approach, directly derived from perturbative analysis of the anomaly form factor, describes the interaction in terms of a kinetic mixing of two entangled degrees of freedom \cite{Giannotti:2008cv}. This aspect will be examined in detail in the upcoming section, where we show that the interaction can be reformulated in terms of two Stuckelberg axions, one of which manifests as a ghost. The effective action is quite similar to the one investigated in \cite{Qi:2012cs}. \\
The approach, however, is not unique. A modified variational principle that leads to a local action has been recently proposed in \cite{Mottola:2019nui}, thereby distinguishing between the 1PI description of the interaction and its local counterpart. In this local description, the longitudinal component of the chiral perturbation is assimilated into a local pseudoscalar field, $\varphi$, an axion.\\
 However, establishing this local description involves a rewriting the effective action through a field redefinition that depends on the external axial-vector background. This redefinition fundamentally alters the nonlocal interaction, reformulating it into a local one, which forms the foundation of the Lagrangian of axion electrodynamics, a topic to be explored further in the last sections of this work.\\
In this framework, axion electrodynamics can be regarded as describing the on-shell behavior of the perturbative (axial-vector/vector/vector) $AVV$ interaction, incorporating a pseudoscalar field $\varphi$ interpreted as an axion-like excitation. This local action finds widespread application in conventional axion physics and holds promise for experimental investigations in topological materials. In this context, the coupling of an axion to the anomaly is simply of the form $\varphi F\tilde{F}$, where F is the electromagnetic field strength, as common in axion physics. \\
The local action predicts the rotation of the plane of polarization of a light beam in the presence of a condensed axion field, particularly when the condensate exhibits a gradient. This aspect will be thoroughly examined in the last segment of the review.
\subsection{Chiral gravitational anomalies}
Interestingly, the charge assignments in the Standard Model of elementary particles also resolve another anomaly known as the gravitational anomaly. This anomaly involves a correlator with one axial-vector current and two gravitons, represented by the $ TTJ_5$ vertex, where $T$ is the stess energy tensor and $J_5$ is a chiral current. Within the Standard Model, this interaction is mediated by the hypercharge, and its cancellation can be understood in two ways.\\
One explanation involves considering the stress-energy tensor as a composite operator within the Standard Model. In this view, vertices where this operator is inserted alongside gauge interactions need to be carefully defined for consistency. \\ 
Alternatively, the resolution of this anomaly can be seen as evidence that the Standard Model and gravity are inherently consistent with each other.\\
It's important to emphasize subtle differences in the behavior of chiral and conformal anomaly correlators 
\cite{Coriano:2020ees}. The chiral anomaly is purely topological, whereas conformal symmetry is exact. In the case of conformal anomalies, the situation is more complex, because the corresponding anomaly is topological only in its Gauss-Bonnet part. This topological aspect emerges in the anomaly functional only after applying the Wess-Zumino consistency condition. The origin of the conformal anomaly is directly linked to the renormalization of the corresponding correlators in $d=3+1$, a step not required for any chiral anomaly.\\
The potential to replicate fundamental phenomena of quantum field theory (QFT) in a laboratory setting, which are typically observed indirectly in high-energy physics through particle accelerators or inferred via elaborate theoretical constructs, presents an exciting opportunity that we aim to explore in this work.\\
%To achieve this, it's necessary to establish a realization of the chirality constraints in an experimental setup. %This ensures that a specific material is subjected to an excitation capable of inducing a particular %anomalous response.\\

\subsection{The topological field theory perspective}
In our approach, we start from the continuum limit, where we discuss a QFT description of the phenomenon. 
We outline the potential for a comprehensive reconstruction of the vertex, operating on the sole assumption that it arises from the presence of just two fundamental ingredients: 1) the emergence of an anomaly pole in the longitudinal sector of the correlator  and 2) of conformal symmetry. \\
A third important ingredient of the interaction, as we move away from the conformal point, as already mentioned, is 3) the presence of a mass -independent sum rule, satisfied by the absorbitive part  of the transition  amplitude in the axial-vector channel. \\ 
The reconstruction of the interaction, which is performed with no reference to a Lagrangian realization and obtained by a solution of the conformal Ward identities (CWIs), is verified in free field theory at one loop by a direct perturbative computation of the interaction vertex. A similar analysis has been performed in the context of the gravitational chiral anomaly, where, again, the reconstruction of the interaction proceeds from the longitudinal sector. \\
The presence of sum rules for chiral currents will be illustrated in \secref{sumrules}, where we show that the pattern is shared by correlators such as the gravitational anomaly correlator $TTJ_5$ and $TTJ_{CS}$, where $J_{CS}$ is a Chern-Simons current \cite{Coriano:2023gxa} associated with a gauge field $V_\mu$ 
\beq
J_{CS}^{\lambda} =\epsilon^{\lambda \mu\nu\rho} V_{\mu}\partial_{\nu} V_{\rho}.
 \eeq 
Relying on a specific parameterization of the anomaly vertex, with the inclusion of an anomaly pole, whose residue is the anomaly, we propose that such an interpolating state should be viewed as essential in the response function of the material once it is subjected to an external chiral excitation, and should be properly searched in the experiments. We are going to elaborate on this interpretation first by resorting to the perturbative expansion of the vertex, and then illustrate the sum rules  satisfied by the chiral interactions, as soon as we move away from the conformal point with massive fermions. In other words, the axion pole is present in the conformal limit. \\ 
Indeed, the possible experimental verification of the sum rule could provide a direct check of the quasiparticle nature of the key dynamical feature of the anomaly vertex.\\

\section{QFT anomalies and the effective action}
Conformal and chiral anomalies \cite{Coriano:2020ees} \cite{Bertlmann:1996xk}
(see \cite{Chernodub:2021nff,Arouca:2022psl} for applications in condensed matter physics)
are generated from the partition function of a certain theory, $\mathcal{Z}(\chi_i)$, 
\beq
\label{one}
\mathcal{Z}(\chi_i) \equiv \int D\psi D\bar{\psi} e^{i\mathcal{S}_0(\psi, \bar{\psi}, \chi_i)}
\eeq
after integration over the matter sector, the Dirac fermion field $\psi$ in this case, whenever the functional integral is not invariant  under either Weyl or axial-vector gauge transformations - for conformal and chiral anomalies respectively -  or under both, if these are symmetries of the original action $\mathcal{S}_0$. \\
$\chi_i$ denote external background fields. \\
The role played by the anomaly in the  transport equations has been discussed in several works and in various contexts, from condensed matter physics to astrophysics \cite{Landsteiner:2013sja,Arjona:2019lxz,Kamada:2022nyt}, the different realizations indicating its universality. Notice that in transport equation the axion/anomaly pole is not present in the description, since the equations of motion are on-shell and incorporate the longitudinal axial-vector Ward identity from the start. In other words, they are based on a local action. In order to uncover the mechanism of the pole exchange, one should analyze the entire vertex, prior to any Ward identity, discussing its complete tensor structure, as we are going to show. \\
In the case of the chiral anomaly, the  underlying interaction is that of an ${AVV}$ (axial-vector/vector/vector) anomaly diagram, appearing at $O(e^2)$ in the expansion of \eqref{one}, when $\mathcal{S}_0$ is the fundamental action of axial QED 
\beq
\label{aqed}
\mathcal{S}_k=\int d^4 x \left(\bar{\psi}\gamma^\mu(\partial_\mu -ie A_\mu )\psi - m\bar{\psi}\psi\right),\qquad 
\mathcal{S}_0\equiv =\mathcal{S}_k - i\int d^4 x  \bar{\psi}\gamma_5\gamma^\mu \psi B_\mu \\\
\eeq
once we integrate out the Dirac fermion. $A_\mu$ is a vector gauge field, the photon, and $B_\mu$ is an axial-vector external source. If $B_\mu$ were a propagating quantum field, then the partition function generated by this theory would be clearly inconsistent, for breaking local gauge invariance. For this reason we will treat it exclusively as a non-propagating source. \\
Our discussion will be confined to axial QED, as given by \eqref{aqed} in a context relevant for topological materials, 
where, as already mentioned, the expectation value of the axial-vector current is generated by varying the effective action 
\beq
\mathcal{S}_{eff}=\log \mathcal{Z}(\chi_i)
\eeq
with respect to $B_\mu$. Three-point functions with external vector interactions can be obtained, similarly, by varying the same functional with respect to the external vector field $A_\mu$, one or multiple times. 
The $\chi_i$'s in \eqref{one} include all these external gauge sources, that take the form of axial-vector ($B_\mu$) or vector ($A_\mu$) fields. In a more general setup, one needs to also include the metric tensor $g_{\mu\nu}$, in the case of a conformal theory. In this second case, the action needs to be extended in a curved spacetime. This point will be briefly illustrated in \secref{metric}.
\\
In the following, we will be concerned with a correlator of the form $AVV$, where the analysis remains valid for any abelian vector current and for a single axial-vector one. The vector current couples to the photon, which can be on-shell, while the axial-vector one is coupled to a non-propagating axial-vector field, 
$B_{\mu}$, that takes the role of an external off-shell source. In the measurement of the optical activity of the material induced by the anomaly, which can be described within axion electrodynamics and is the topic of the last section, the photons are obviously required to be on-shell, since these are experimentally detected.\\
As already mentioned, the effective action underlining the fermion dynamics is characterised by a global $U(1)_A $ and by a gauge $U(1)_V$ symmetries, with the $U(1)_A$ broken by the anomaly. 
The breaking of such symmetry, at quantum level, is expressed in terms of an anomaly functional which contains the topological term \eqref{anom}. Under a variation of the external sources, the effective action
changes in the form 
\beq
\label{tvar}
\delta \mathcal{S}_{eff}=\int d^4 x \frac{\delta \mathcal{S}_{eff}}{\delta \theta(x)}\delta\theta(x),
\eeq
with the finite part of the variation
\beq
\label{varvar}
\frac{\delta\mathcal{S}_{eff}}{\delta \theta(x)}=\mathcal{A}(\chi_i)
\eeq
given by the anomaly $\mathcal{A}(\chi_i)$. 
The anomaly is result is due to the anomalous variation of $\mathcal{S}_{eff}$ under a gauge transformation of the external gauge field 
$B_{\mu}$, with gauge parameter $\xi(x)$,  $\delta B_\mu=\partial_\mu \theta(x)$
\beq
d\mu \rightarrow d\mu\,  \exp\left(-\frac i {8\pi^2} \int dx\,
  \theta (x) F(x)_{\mu\nu}\tilde F(x)^{\mu\nu} \right)
\eeq
where we have indicated with $d\mu$ the variation of the fermion integration measure in the partition function, in the background of the electromagnetic field. 
This may involve the Euler-Poincar\`e density  $E_d$ plus other terms in the case of a conformal anomaly, whose explicit expressions depend on the spacetime dimensions  or the divergence of a topological Pontryagin current $K^\mu$, with  $\partial \cdot K=F\tilde{F}\sim \vec{E}\cdot \vec{B}$, for the chiral anomaly. \\
Notice that \eqref{tvar} is the usual way in which the topological response is discussed in all the literature on 
the chiral anomaly in topological materials (see for example \cite{Coriano:2024nhv}). This is equivalent to a description of the longitudinal Ward identity and does not describe the entire vertex as it appears in the interaction in the original current correlator. While this reduced description is at the basis of the local action, it erases the pole from the off-shell effective action. In particular, the link between conformal symmetry, the anomaly pole and the sum rule is absent. \\
The presence or the breaking of these symmetries manifests with an infinite set of either exact or anomalous Ward identities among the quantum correlation functions of the theory \cite{Coriano:2020ees}. \\
As mentioned in the Introduction, the analysis of such interactions is fundamental for the consistency of the gauge theories of particle physics, and anomaly cancelation is crucial for the identification of physics beyond the Standard Model, once the gauge interactions are extended with wider symmetries.  \\
%\subsection{The generation of interpolating currents} 
A common character of the topological terms, in a perturbative expansion of the partition function, is that they are not directly required by the regularization procedure of the corresponding Feynman diagrams, but appear as a result of external conditions associated with the Ward identities, imposed on the diagrammatic expansion.\\
 Both the Wess-Zumino consistency condition - for the conformal anomaly - or the condition of conserved vector currents (CVC) - for the chiral anomaly - can be formulated as external Ward identities, necessary for the consistency of the quantum theory. 
 %There is a direct relation between such conditions and the presence of an entangled state, in the form of an anomaly pole, in the anomaly %action. \\ 
 The process is exhausted at one-loop in both cases: at trilinear level in the external fields for the abelian chiral anomaly, at fourth order in the nonabelian case, while it goes to all orders in the case of the conformal anomaly. The anomaly functional, however, even in this second case, can also be inferred by the analysis of the first lower-point functions with external gravitons.  \\
In the case of the conformal anomaly, this perturbative procedure does not identify the entire contribution to the anomaly functional, once the matter sector is integrated out - in our case this corresponds just to fermions - since other, non topological terms appear, such as the square of the Weyl tensor in $d=4$, which are necessary for the renormalization of the quantum action, once all the loop corrections are taken into account.\\
Obviously, in topological materials  such interactions are emergent, and are realized artificially by soliciting the sample either with chemical potentials or with thermal gradients that interpolate with 
specific currents or tensor operators, such as the the chiral current $J^\mu_{5}$ or the stress energy tensor $T^{\mu\nu}$. For instance, thermal gradients can be treated as an artificial metric, using Luttinger's relation \cite{Luttinger:1964zz}\cite{Rovelli:2010mv,Bermond:2022mjo}. \\
In general, such studies, from the field theory perspective, are performed in the conformal limit, when all the masses of the matter system are neglected. 
This is also the limit in which the effective action  manifests key features of the anomalous behaviour. 
\subsection{Correlators from the partition function} 
$\mathcal{S}_{eff}$ can be expanded perturbatively in the two coupling constants $e$ and $g_B$. Here we will consider the contribution 
only of the $AVV$ diagram, which is the only one responsible for the anomalous interaction of the electromagnetic field $A_{\mu}$ with the axial vector source $B_{\lambda}$.\\
In the chiral anomaly case, the only violation of the axial-vector current Ward identities takes place at trilinear level and it is summarised  by the two simple diagrams in Fig. \ref{VAA1}. The anomalous variation of $\mathcal{S}_{eff}$ is essentially defined by this contribution and all the relevant physical implications should be extracted at this level $(O(e^2))$, quadratic in the electric charge $e$. \\
\begin{figure}[t]
{\centering \resizebox*{08cm}{!}{\rotatebox{0}
{\includegraphics{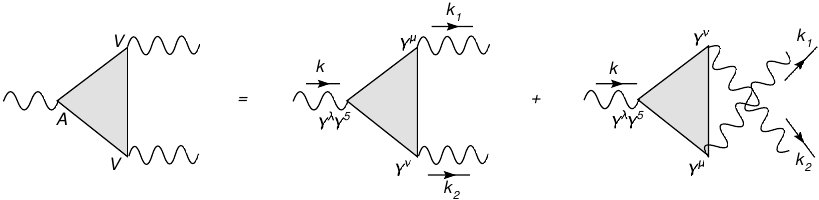}}}\par}
\caption{The ${\bf AVV}$ diagrams (direct and exchanged) in the expansion of the anomaly action $\mathcal{S}_{eff}$ }
\label{VAA1}
\end{figure}
%with a fundamental action which is a function of the background fields $B$ and $A$, where $B$ is an axial vector and $A$ is, for our purposes, the electromagnetic gauge potential. 
In the limit of vanishing fermion mass $m\rightarrow 0$, the classical Lagrangian has a
$U(1)$ global symmetry under $\psi \rightarrow e^{i\alpha\gamma^5}\psi$, in 
addition to $U(1)$ local gauge invariance, and $J_5^{\mu}$ is the Noether current
corresponding to this chiral symmetry. 
The two currents, vector and axial-vector 
\bea
J^{\mu}(x) = \bar\psi (x) \gamma^{\mu} \psi (x)\,\\
J_5^{\mu}(x) = \bar\psi (x) \gamma^{\mu} \gamma^5 \psi (x)\,
\eea\label{currents1}
satisfy the conditions
\beq
\partial_{\mu} J^{\mu} = 0\,,
\label{vecons}
\eeq
while the classical conservation equation for the axial-vector current is explicitly broken by the mass term
\beq
\partial_{\mu} J_5^{\mu} = 2 i m\, \bar\psi\gamma^5 \psi.
\label{classic}
\eeq
One of the two symmetries cannot be 
maintained at the quantum level. 
If we denote by $\lag J_5^{\mu}(z) \rag_{{A}}$ 
the quantum average of $J_5^{\mu}$, with 
\beq
\lag J_5^{\mu}(z) \rag_{_A} =\frac{\delta \mathcal{S}_{eff}}{\delta B_{\mu}}\vert_{B=0}
\eeq 
then the quantum version of \eqref{classic} becomes

\beq
\partial_\mu J_5^\mu = 2m {\bar{\psi} \gamma_5 \psi}  - 2 \mathcal {A} \, ,
\eeq
which in the fermion zero mass limit reduces to 
\beq
\partial_{\mu}\lag J_5^{\mu}\rag_{_A}\Big\vert_{m=0} = \frac{e^2}{16\pi^2} \
\epsilon^{\mu\nu\rho\sigma}F_{\mu\nu}F_{\rho\sigma} = \frac{e^2}{2\pi^2}\,{\bf \E \cdot \B}.
\label{axanom}
\eeq
The second variation of $\lag J_5^{\mu}(z) \rag_{_A}$ projects over the $\bf{AVV}$ correlation function 
\beqa
\Gamma^{\mu \alpha\beta}(z,x,y) &\equiv&  -i\frac{\delta^2 \lag J_5^{\mu}(z) \rag_{_A}}
{\delta A_{\alpha}(x) \delta A_{\beta}(y)}\Bigg\vert_{A=0}\nonumber \\
&=& -i(ie)^2  \lag { T} J_5^{\mu}(z) J^{\alpha}(x) J^{\beta}(y)\rag\big\vert_{A=0}
\label{axvarpos}
\eeqa
which is affected by an anomaly.
\bea
&&\Gamma^{\mu \alpha\beta}(p,q) \equiv - i \int d^4x\int d^4y\, e^{ip\cdot x+ i q\cdot y}\  
\frac{\delta^2 \lag J_5^{\mu}(0) \rag_{_A}}{\delta A_{\alpha}(x) \delta A_{\beta}(y)} 
\nonumber\Bigg\vert_{A=0}\\
&&= ie^2 \int d^4 x \int d^4 y\, e^{i p\cdot x + i q \cdot y}\ \lag { T}
J_5^{\mu}(0) J^{\alpha}(x) J^{\beta}(y)\rag\big\vert_{A=0}\,.
\label{GJJJ}
\eea
Then we have the expansion 
\beqa
\lag J_5^{\mu}(z) \rag_{_A} &=& \frac{i}{2} \int \frac{d^4 p}{(2\pi)^4} \int \frac{d^4 q}{(2\pi)^4}
\int d^4x \int d^4 y \, e^{-i p\cdot (x-z)}\,\times \nonumber \\
 && e^{- i q \cdot (y-z)}\, \Gamma^{\mu \alpha\beta}(p,q)
\,A_{\alpha}(x) A_{\beta}(y) + \dots
\eeqa
up to second order in the gauge field background $A_{\mu}$. 
Focusing our discussion on the chiral case, we expect that the anomaly effective action, computed from the anomaly functional, should be expressed in terms of the complete components of the external axial-vector source $B_{\lambda}$, rather than just its longitudinal part. This point can be illustrated as follows. \\
The quantum average of the axial-vector current can be extracted from the expression

\beq
\frac{\delta {\cal S}_{eff}} {\delta {B}_{\mu}}= \lag J_5^{\mu}\rag_{_A} \,.
\label{dec}
\eeq
If we decompose the axial vector 
${ B}_{\mu}$ into its transverse and longitudinal parts,
\beq
{B}_{\mu} = {B}_{\mu}^{\perp} + \partial_{\mu} {\varphi}
\eeq
with $\partial^{\mu} { B}_{\mu}^{\perp} =0$ and $\varphi$ a pseudoscalar that will take the role of an axion. The axial-vector interaction in 
$\mathcal{S}_0$ can then be re-expressed in the form 
\beq
\int d^4 x J_5^\mu B_{\mu}= -\int d^4 x \partial_\mu J_5^\mu \varphi +\int d^4 x J_5^\mu B^\perp_\mu 
\eeq
using the functional chain rule in the differentiation with repsect to the pseudoscalar component $\phi$ we obtain
\beq
\partial_{\mu} \lag J_5^{\mu}\rag_{_A} = -\frac{\delta {\cal S}} {\delta {\varphi}}\,
\label{var}
\eeq
whose variational solution can be immediately chosen of the form 
\beq
{\cal S}_{eff} = -\frac{e^2}{16\pi^2} \,\int d^4x \, \epsilon^{\mu\nu\rho\sigma}F_{\mu\nu}F_{\rho\sigma}\,{\varphi}\,,
\label{var}
\eeq
and describes the coupling of an axion ($\varphi$) to the anomaly.  Notice that $\varphi$ behaves like a Nambu-Goldstone (NG) mode under gauge variations of the external source since, for a generic gauge function $\theta(x)$, we have  
\beq
B_\mu\to B'_\mu= B_\mu +\partial_\mu\theta
\eeq
since it undergoes a shift, while $B^\perp_\mu$ is invariant

\beq
\label{gd}
\varphi\to \varphi + \theta  \qquad B^\perp_\mu \to B^\perp_\mu.
\eeq
Notice also that the solution \eqref{var} is excluding contributions which are proportional to the transverse component of the source, $B_\mu^\perp$. Such extra corrections correspond to homogenous terms which do not contribute to \eqref{var}, for being transverse, but are part of the 
expression of the current $\langle J_5^\mu\rangle $.\\
Even within the limitations of \eqref{var}, we need to relate $\varphi$ to the source $B_\lambda$ and use the constraint
 $\partial^{\lambda} { B}_{\lambda} = \Box{ \varphi}$,
to derive the nonlocal form of the action $\eqref{var}$
\beq
{\cal S}_{eff} = -\frac{e^2}{16\pi^2} \,\int d^4 x\int d^4y \, [\epsilon^{\mu\nu\rho\sigma}F_{\mu\nu}F_{\rho\sigma}]_x
\Box^{-1}_{xy} \,[\partial^{\lambda} { B}_{\lambda}]_y,
\label{nonl}
\eeq
rather than to a local one. Notice that the coupling of a NG mode to topological densities, such as $F\wedge F$ or the Euler-Poincar\`e density $E_d$, is  a common feature of the method of derivation of the effective action, with significant differences respect to those formulation that do not follow this procedure.\\
%There is no doubt, however, that the only consistent derivation of the 1PI (perturbative) effective action is %the one based directly on the computation of the $AVV$ diagram, performing the integration on the fermion %sector, rather than relying only on a variational solution as in 
%\eqref{var}. If we consider the perturbative expansion of $\mathcal{Z}$ or $\mathcal{S}_{eff}$, then the only %consistent approach that we need to follow is the diagrammatic one. This allows to identify the structure of %$\langle J_5^\mu\rangle $ directly from a computation of simple Feynman diagrams. \\
Being the chiral anomaly contribution present just at trilinear level, in the ${AVV}$ and ${AAA}$ sectors, the details can be worked out within a simple perturbative picture.    

%Another important point concerns the inclusion of a kinetic term for $\varphi$, that one would expect to %come from the only gauge invariant kinetic contribution allowed in the partition function \eqref{one}, via a %suitable extension of \eqref{aqed} in the form 
%\beq
%\label{aqed1}
%\mathcal{S}_0\to  -\frac{1}{4}B_{\mu\nu}B^{\mu\nu} + \bar{\psi}\gamma^\mu(\partial_\mu -ie A_\mu - i %\gamma_5 B_\mu)\psi - m\bar{\psi}\psi,
%\eeq
% with $B_{\mu\nu}$ denoting the field strength of the field $B_{\mu}$. Even for a non-propagating $B_{\mu}%$ field, the partition function should be gauge invariant.  It is clear, however, that this is not possible, since %the kinetic term for $B_\mu$ for \eqref{aqed1} is only expressed in terms of $B^\perp_\mu$, without any 
 %contribution from $\varphi$, which is, instead, according to \eqref{gd}, gauge dependent. 

\begin{figure}[t]
\centering
\subfigure[]{\includegraphics[scale=0.8]{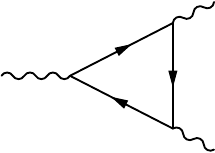}}  \hspace{2cm}
\subfigure[]{\includegraphics[scale=0.8]{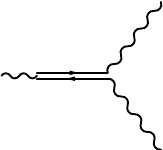}} \hspace{2cm}
\subfigure[]{\includegraphics[scale=0.8]{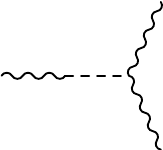}}
\caption{The fermion loop (a); the collinear region in the loop integration (b); the effective pseudoscalar exchange as an effective axion (c). }
\label{ddiag}
\end{figure}

\subsection{Kinetic mixing, Stuckelberg axions and a ghost} 
\label{mixing}
A first proposal for rewriting the nonlocal action in local form was presented in \cite{Giannotti:2008cv} in terms of two auxiliary fields, kinetically mixed 
\beq
{\cal S}_{eff}[\eta, \chi; A, {B}] =  \int d^4x\, 
\left\{ (\partial^{\mu}\eta)\,(\partial_{\mu}\chi) - \chi\, \partial^{\mu}{ B}_{\mu} 
+ \frac{e^2}{8\pi^2}\,\eta\,F_{\mu\nu}\tilde F^{\mu\nu} 
 \right\}
\label{b1}
\eeq
On can immediately show that the equations of motion of the two pseudoscalars $\eta$ and $\chi$ take the form

\beqa
&&\Box\,\eta = -\partial^{\lambda}{ B}_{\lambda}\,,\\
&&\Box\, \chi = \frac{e^2}{8\pi^2}\, F_{\mu\nu}\tilde F^{\mu\nu} 
= \frac{e^2}{16\pi^2}\,\epsilon^{\mu\nu\rho\sigma} F_{\mu\nu}F_{\rho\sigma}\,.
\label{chiaxeom}
\eeqa
At the same time, after an integration by parts and removing a total divergence, one can show that \eqref{b1} 
reproduces the nonlocal action \eqref{nonl}. \\
In an ordinary quantum field theory 
pseudoscalar states at $d=4$ should be characterized by a canonical mass dimension equal to one. Notice, however that in the 
local Lagrangian introduced by Giannotti and Mottola $\chi$ \cite{Giannotti:2008cv} has mass dimension two and $\eta$ has mass dimension zero. By an appropriate scaling, we introduce two fields $\bar{\chi}$ and $\bar{\eta}$ defined as 
\beq
\chi=\frac{1}{\sqrt{2}}\bar{\chi} M  \qquad \eta=\frac{1}{\sqrt 2}\frac{ \bar{\eta}}{M},
\eeq
 canonically normalised, with the original action \eqref{nonl} that  takes the form 
  
 \beqa
{\cal S}_{eff}[\bar\eta, \bar\chi; A, { B}] &=&  \int d^4x\, 
\left\{ (\partial^{\mu}\bar\eta)\,(\partial_{\mu}\bar\chi) + M\partial^{\mu}\bar\chi\, {B}_{\mu}\right. \nonumber \\
&& \left. +\alpha\frac{\bar\eta}{M}\,F_{\mu\nu}\tilde F^{\mu\nu} 
 \right\}
\label{lag1}
\eeqa
with $\alpha\equiv {e^2}/{8\pi^2}$. 
 In the context of axion physics, the coupling of $\bar{\eta} $ to the topological density $F\tilde{F}$ is naturally suppressed by $M$, which defines the axion decay constant, as in ordinary axion models. Now introduce the two linear combinations 
 \beq
 \bar{\eta}=a(x)-b(x)   \qquad \bar{\chi}=b(x) + a(x)
 \eeq
and we rewrite \eqref{lag1} in the form 
\beqa
{\cal S}_{eff}[a,b; A, { B}] &=&\int d^4 x \left( \frac{1}{2}(\partial_\mu a  -
  \bar{M} B_\mu)^2  -  \frac{1}{2}(\partial_\mu b - \bar{M} B_\mu)^2 +
\frac{a - b}{\bar M}F\tilde{F}\right)\nn \\
\eeqa
where $\bar{M}=M/\sqrt{2}$.
This action has close resemblance with the action introduced in \cite{Qi:2012cs}, except for the absence of the $b$ ghost, that there is formulated as an ordinary kinetic term, as well as and for the inclusion of a periodic potential, that here is absent. The two degrees of freedom, in this formulation, are entangled.\\
Notice that the action above takes the ordinary St\"uckelberg form, in terms of two 
St\"uckelberg axions $a(x)$ and $b(x)$. The kinetic terms are invariant under the gauge transformations 
\beq
a\to a - \bar{M}\theta(x),  \qquad b\to b + \bar{M}\theta, \qquad B_\mu\to B_\mu + \partial_\mu\theta
\eeq
 the second of them being ghost-like and given by $b$. In ordinary St\"uckelberg models \cite{Coriano:2007fw}, the symmetry 
 with a single axion is sufficient in order to provide a gauge invariant mass to a $U(1)$ field, in this case identified with $B_\mu$. 
% It is clear, in our case, that the emergence of a ghost in the action, after a rewriting of the interaction and %fields in tn terms of their 
% canonically dimensioned ones $(\bar{\chi}, \bar{\eta})$. 
  %Entanglement in coordinate space in field theory has been discussed recently in several papers %%\cite{MohammadiMozaffar:2012dsy} 
 %\cite{Taylor:2015kda}.
 For a different approach to the solution of the variational problem concerning the local and the nonlocal actions in this context we refer to \cite{Mottola:2019nui}. 
 
\section{The chiral anomaly from perturbation theory and covariance}
\label{conf1}
In this section, we delineate the basic parameterization of the chiral correlator affected by a chiral anomaly at perturbative level (the vector/vector/axial-vector or $VVA$ diagram).  The diagrammatic expression of the interaction is described at perturbative level in Fig. \eqref{ddiag} a). We have indicated with $p_3$ the momentum of the axial-vector current  $J_5$ and with $p_1$ and $p_2$  those of the vector currents.\\
\begin{figure}[t]
{\centering \resizebox*{10.8cm}{!}{\rotatebox{0}
{\includegraphics{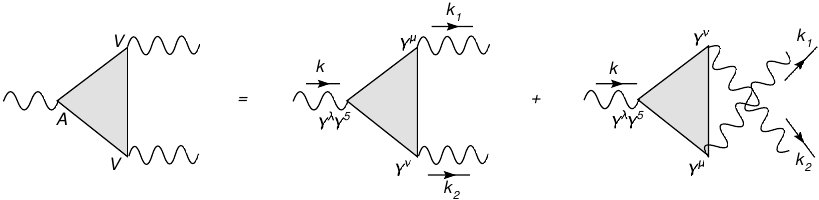}}}\par}
\caption{The ${ AVV}$ diagrams (direct and exchanged) in the expansion of the anomaly action $\mathcal{S}_{eff}$ }
\label{VAA1}
\end{figure}The original parameterization of the $VVA$ vertex was presented in \cite{Rosenberg:1962pp}. 
These identities establish connections between two commonly used representations of this vertex. The first, introduced long ago by Rosenberg \cite{Rosenberg:1962pp}, is expressed in terms of 6 tensor structures and form factors. The second representation \cite{Knecht:2003xy}, more recent and particularly valuable from a physical standpoint, enables the attribution of the anomaly to the exchange of a pole in the longitudinal channel \cite{Dolgov:1971ri} \cite{Giannotti:2008cv,Armillis:2009pq}.\\
In this second parameterization of the vertex, the decomposition identifies longitudinal and transverse components:
\begin{equation}
\langle J^{\mu_1}(p_1)J^{\mu_2}(p_2)J^{\mu_3}_5(p_3) \rangle=\frac{1}{8\pi^2}\left(
W_L^{\mu_1\mu_2\mu_3}- W_T^{\mu_1\mu_2\mu_3}\right)
\label{sec}
\end{equation}
where $W_T$ represents the transverse part, while the longitudinal tensor structure is given by \eqref{refe}.
Here, $w_L$ represents the anomaly form factor, which exhibits a ${1}/{p_3^2}$ pole in the massless (chiral or conformal) case.\\
Lorentz symmetry and parity fix the correlation function in the following form:
\begin{align}
\langle J^{\mu_1}(p_1)J^{\mu_2}(p_2)J^{\mu_3}_5(p_3) \rangle &= B_1 (p_1, p_2) \varepsilon^{p_1\mu_1\mu_2\mu_3} + B _2 (p_1, p_2)\varepsilon^{p_2\mu_1\mu_2\mu_3} \nonumber \\
&\quad + B_3 (p_1, p_2) \varepsilon^{p_1p_2\mu_1\mu_3}{p_1}^{\mu_2} \nonumber \\
&\quad +  B_4 (p_1, p_2) \varepsilon^{p_1p_2\mu_1\mu_3}p_2^{\mu_2} \nonumber \\
&\quad + B_5 (p_1, p_2)\varepsilon^{p_1p_2\mu_2\mu_3}p_1^{\mu_1} \nonumber \\
&\quad + B_6 (p_1, p_2) \varepsilon^{p_1p_2\mu_2\mu_3}p_2^{\mu_1},
\label{Ross}
\end{align}
where $B_1$ and $B_2$ are divergent by power counting and we used the notation $\varepsilon^{p_1\mu_1\mu_2\mu_3}\equiv\varepsilon^{\alpha\mu_1\mu_2\mu_3}p_{1\alpha}$. The four invariant amplitudes $B_i$ for $i\geq3$ are given by explicit parametric integrals \cite{Rosenberg:1962pp}:
\begin{align}
B_3(p_1, p_2) &= - B_6 (p_2, p_1) =   16 \pi^2 I_{11}(p_1, p_2), \nonumber \\
B_4(p_1,p_2) &= - B_5 (p_2, p_1) =- 16 \pi^2 \left[ I_{20}(p_1,p_2) - I_{10}(p_1,p_2) \right], 
\end{align}
that respect the Bose symmetry of the two vector lines, with $I_{st}$ defined by
\begin{equation}
I_{st}(p_1,p_2) = \int_0^1 dw \int_0^{1-w} dz w^s z^t \left[ z(1-z) p_1^2 + w(1-w) p_2^2 + 2 w z (p_1\cdot p_2) - m^2 \right]^{-1}.
\end{equation}
By power counting, one immediately notices that both $B_1$ and $B_2$ are ill-defined, but they can be rendered finite by imposing the conservation Ward identities on the two vector lines, giving
\begin{align}
B_1 (p_1,p_2) &= p_1 \cdot p_2 \, B_3 (p_1,p_2) + p_2^2 \, B_4 (p_1,p_2),
\label{WI1} \\
B_2 (p_1,p_2) &= p_1^2 \, B_5 (p_1,p_2) + p_1 \cdot p_2 \, B_6 (p_1,p_2).
\label{WI2}
\end{align}
One of the main characteristics of a chiral anomaly interaction is that it can be rendered finite by imposing suitable Ward identities on the corresponding correlators. 
These relations allow us to re-express the formally divergent amplitudes in terms of the convergent ones without the need of introducing counterterms. This is the main reason why the diagrammatic contribution described above is conformal. The chiral case is different from other correlators, for instance those characterised by the insertion of an energy momentum tensor, where an anomaly, the conformal anomaly, is directly associated with their renormalization.  While the description of the interaction, at this level, is purely perturbative, we are going to see how the vertex can be inferred from the solution of the CWIs in momentum space, with no reference to perturbation theory or to free field theory realizations.  \\
Using the conservation Ward identities for the vector currents, we obtain the convergent expansion \cite{Armillis:2009sm}
\begin{align}
&\langle J^{\mu_1}J^{\mu_2}J^{\mu_3}_5 \rangle = B_3 (p_1 \cdot p_2 \varepsilon^{p_1\mu_1\mu_2\mu_3} + p_1^{\mu_2} \varepsilon^{p_1p_2\mu_1 \mu_3}) \nonumber \\
&\quad + B_4 (p_2 \cdot p_2 \varepsilon^{p_1\mu_1\mu_2\mu_3} + p_2^{\mu_2} \varepsilon^{p_1p_2\mu_1 \mu_3} ) \nonumber \\
&\quad + B_5 (p_1 \cdot p_1 \varepsilon^{p_2\mu_1\mu_2\mu_3}+ p_1^{\mu_1} \varepsilon^{p_1p_2\mu_2\mu_3}) \nonumber \\
&\quad + B_6 (p_1 \cdot p_2 \varepsilon^{p_2\mu_1\mu_2\mu_3} + p_2^{\mu_1} \varepsilon^{p_1p_2\mu_2\mu_3} ) \notag\\
& \equiv B_3 \, \eta_3^{\mu_1\mu_2\mu_3}(p_1,p_2) + B_4 \, \eta_4^{\mu_1\mu_2\mu_3}(p_1,p_2) \nonumber \\
&\quad + B_5 \, \eta_5^{\mu_1\mu_2\mu_3}(p_1,p_2) + B_6 \, \eta_6^{\mu_1\mu_2\mu_3}(p_1,p_2),
\label{reduced}
\end{align}
\begin{table}[t]
\begin{center}
\begin{tabular}{|c|c|}
\hline
$\eta_1$ & $ p_1^{\mu_3} \, \varepsilon^{p_1p_2\mu_1\mu_2} $  \\ \hline
$\eta_2$ & $ p_2^{\mu_3} \, \varepsilon^{p_1p_2\mu_1\mu_2}$ \\ \hline\hline
$\eta_3$         &           $p_1 \cdot p_2 \varepsilon^{p_1\mu_1\mu_2\mu_3} + p_1^{\mu_2} \varepsilon^{p_1p_2\mu_1 \mu_3}$ \\ \hline
$\eta_4$ & $p_2 \cdot p_2 \varepsilon^{p_1\mu_1\mu_2\mu_3} + p_2^{\mu_2} \varepsilon^{p_1p_2\mu_1 \mu_3} $ \\ \hline
$\eta_5$ & $ p_1 \cdot p_1 \varepsilon^{p_2\mu_1\mu_2\mu_3}+ p_1^{\mu_1} \varepsilon^{p_1p_2\mu_2\mu_3} $  \\ \hline
$\eta_6$ & $p_1 \cdot p_2 \varepsilon^{p_2\mu_1\mu_2\mu_3} + p_2^{\mu_1} \varepsilon^{p_1p_2\mu_2\mu_3} $  \\ \hline
\end{tabular}
\caption{Tensor structures of odd parity in the expansion of the $VVA$ with conserved vector currents. \label{table2}}
\end{center}
\end{table}In the last step of Eq. \eqref{reduced}, we  have introduced four tensor structures that are mapped into one another under the Bose symmetry of the two vector lines. We can identify six of them, as indicated in Table \ref{table2}, but two of them
\begin{align}
\eta_1^{\mu_1\mu_2\mu_3}(p_1,p_2) &=  p_1 ^{\mu_3} \, \varepsilon^{p_1p_2\mu_1\mu_2},  \\
\eta_2^{\mu_1\mu_2\mu_3}(p_1,p_2) &=  p_2 ^{\mu_3} \, \varepsilon^{p_1p_2\mu_1 \mu_2},
\end{align}
are related by the Schouten relations to the other four, $\eta_3,\ldots \eta_6$. Indeed, we have
\begin{align}
\eta_1 ^{\mu_1\mu_2\mu_3}(p_1,p_2) &= \eta_3 ^{\mu_1\mu_2\mu_3}(p_1,p_2) - \eta_5 ^{\mu_1\mu_2\mu_3}(p_1,p_2),    \\
\eta_2 ^{\mu_1\mu_2\mu_3}(p_1,p_2) &= \eta_4 ^{\mu_1\mu_2\mu_3}(p_1,p_2) - \eta_6 ^{\mu_1\mu_2\mu_3}(p_1,p_2).
\end{align}
The remaining tensor structures are inter-related by the Bose symmetry:
\begin{align}
\eta^{\mu_1\mu_2\mu_3}_3(p_1,p_2) &=-\eta^{\mu_2\mu_1\mu_3}_6(p_2,p_1),  \notag \\
\eta_4^{\mu_1\mu_2\mu_3}(p_1,p_2) &=-\eta_5^{\mu_2\mu_1\mu_3}(p_2,p_1). 
\end{align}
The correct counting of the independent form factors/tensor structures can be done only after we split each of them into their symmetric and antisymmetric components:
\begin{align}
\eta^{\mu_1\mu_2\mu_3}_i &= \eta_i^{S \, \, \mu_1\mu_2\mu_3} +\eta_i^{A \, \, \mu_1\mu_2\mu_3}, \notag \\
\eta_i^{S/A \, \, \mu_1\mu_2\mu_3} &\equiv \frac{1}{2}\left( \eta^{\mu_1\mu_2\mu_3}_i(p_1,p_2) \pm \eta^{ \mu_2\mu_1\mu_3}_i(p_2,p_1)\right) \equiv \eta_i^{\pm \, \, \mu_1\mu_2\mu_3},
\end{align}
with $i\geq 3$, giving:
\begin{align}
\eta_3^+(p_1,p_2)&=-\eta_6^+(p_1,p_2),  \notag \\
\eta_3^-(p_1,p_2)&=\eta_6^-(p_1,p_2), \notag \\
\eta_4^+(p_1,p_2)&=-\eta_5^+(p_1,p_2), \notag \\
\eta_4^-(p_1,p_2)&=\eta_5^-(p_1,p_2). 
\end{align}
We can then re-express the correlator as:
\begin{equation}
\langle VVA \rangle = B_3^+\eta_3^+ + B_3^-\eta_3^- +B_4^+\eta_4^+ + B_4^-\eta_4^-
\end{equation}
in terms of four tensor structures of definite symmetry times four independent form factors. 
\subsection{Finite density extensions}
More recently, the structure of the correlator at finite density has also been investigated \cite{Coriano:2024nhv}. The analysis involves also the 4-momentum of the heat bath $\eta_\mu$ that renders the analysis of the parameterization far more involved compared to the expression known in the vacuum case. \\
As we move to finite density, the original 60 tensor structures, by the imposition of the Schouten identities are reduced to 28. At a third stage, after imposing the Bose symmetry, this number is reduced to 16. Finally, imposing the vector WIs, the final expression can be given in terms of 10 tensor structures. The same steps can be repeated by assuming some special conditions on the momenta of the two vector lines with respect to the thermal bath. These take the form 
\begin{equation}
p_1\cdot\eta=p_2\cdot\eta=p\cdot\eta,\end{equation}\begin{equation}
p_1^2=p_2^2=p^2
\end{equation}   
that allows to perform further simplification of the tensor structure. One can show that even in the presence of chemical potentials, the structure of the anomaly vertex preserves its topological nature and the anomaly pole is protecetd by such corrections.  
These extensions offer the possibility at experimental level to investigate the response functions of topological materials in more general experimental situations, where the fermion chemical potential $\mu$ can be fine-tuned in order to characterize the response function of such materials in more realistic environments.  Details of this analysis can be found in \cite{Coriano:2024nhv}.

\subsection{Chern-Simons terms}
Another important feature of the $VVA$ vertex is the possibility of redefining the partial WIs in each channel by the inclusion of 
Chern-Simons (CS) forms. The identification of such extra interactions, which allow to "move around" the anomaly from one vertex to the other, can be more easily understood by considering directly the Lagrangian formulation of this gauge-dependent interaction. 
Introducing external gauge fields $B_\lambda$ and $A_\mu$, the first an axial-vector and the second a vector, the effective action for a chiral anomaly interaction can be modifed by CS terms of the form 
\beq
V_{CS}\equiv  i \int dx A^{\lambda}(x) B^{\nu}(x) F^{A}_{\rho \sigma}(x) \varepsilon^{\lambda \nu \rho \sigma}
\eeq
that in momentum space generates the vertex
\beq
\varepsilon^{\lambda \mu \nu \alpha} \,( p^{\alpha}_1 -p_2^\alpha),
\eeq
identifying the CS vertex.  In perturbation theory, the identification of such extra contributions in the form factor decomposition of the $VVA$ diagram  proceeds rather easily if, in momentum space, one performs an arbitrary shift in the loop integral. 
For example, if we proceed with a specific momentum parameterization of the loop  we obtain 
\beqa
p_{1\mu} \mathcal{W}^{\la\mu\nu}(p_1,p_2) = a_1 \epsilon^{\lambda\nu\alpha\beta} 
p_1^\alpha p_2^\beta \nonumber \\
p_{2\nu}\mathcal{W}^{\la\mu\nu}(p_1,p_2) = a_2 \epsilon^{\lambda\mu\alpha\beta} 
p_2^\alpha p_1^\beta \nonumber \\
p_{3\,{\la}} \mathcal{W}^{\la\mu\nu}(p_1,p_2) = a_3 \epsilon^{\mu\nu\alpha\beta} 
p_1^\alpha p_2^\beta, \nonumber \\
\eeqa
where
\beq
 a_1=-\frac{i}{8 \pi^2} \qquad a_2=-\frac{i}{8 \pi^2} \qquad a_3=-\frac{i}{4 \pi^2}.
\label{basic}
\eeq
Notice that $a_1=a_2 $, as expected from the Bose symmetry of the two vector lines. It is also well known that the total anomaly 
$a_1+a_2 + a_3 \equiv  a_n$ is regularization scheme independent. 
We recall that a shift of the momentum in the integrand $(p\rightarrow p + a)$ where $a$ is the most general momentum written in terms of the two independent external momenta of the triangle diagram $(a=\alpha (p_1 + p_2) + \beta(p_1 - p_2))$ induces on  $\Delta$ changes that 
appear only through a dependence on one of the two parameters characterizing $a$, that is 
\beq
\mathcal{W}^{\la\mu\nu}(\beta,p_1,p_2)=\mathcal{W}^{\la\mu\nu}(p_1,p_2) - \frac{i}{4 \pi^2}\beta \epsilon^{\lambda\mu\nu\sigma}\left( p_{1\sigma} - 
p_{2\sigma}\right).
\eeq
We have introduced the notation $\mathcal{W}^{\la\mu\nu} (\beta,p_1,p_2)$ to denote the shifted 3-point function, while 
$\mathcal{W}^{\la\mu\nu}(p_1,p_2)$ denotes the original one, with a 
vanishing shift. Noticd that under this momentum shift the differenc of the two form factors $B_1$ and $B_2$ 
\beqa
p_{1\mu}\mathcal{W}^{\lambda\mu\nu}(\beta',p_1,p_2)&=& (a_1 -\frac{i \beta'}{4 \pi^2})
\varepsilon^{\lambda\nu\alpha\beta}p_1^\alpha p_2^\beta,\nonumber\\
p_{2\nu}\mathcal{W}^{\lambda\mu\nu}(\beta',p_1,p_2)&=&(a_2-\frac{i \beta'}{4 \pi^2})
\varepsilon^{\lambda\mu\alpha\beta}p_2^\alpha p_1^\beta,\nonumber\\
k_\lambda\mathcal{W}^{\lambda\mu\nu}(\beta',p_1,p_2)&=&(a_3+\frac{i \beta'}{2 \pi^2})
\varepsilon^{\mu\nu\alpha\beta}p_1^\alpha p_2^\beta.
\label{bbshift}
\eeqa
All these manipulations can be performed in four spacetime dimensions. The inclusion of external Ward identities is what saves us from dealing directly with such CS contributions.  In the case of a TI, the parameterization of the underlying $VVA$ vertex, obviously, should 
respect the QED hgauge symmetry, with the conservation of the vector currents. For this obvious reason, one does not have to deal with such CS terms in an experimental setting.  

\subsection{Duality symmetry and the CS current}
Here we pause for a moment to remark that the CS current plays a role in specific correlators, for example in the $TTJ_5$, where a chiral current is coupled to two stress energy tensors, a vertex which is responsible for the gravitational anomaly. We will comment on this interaction in a follow-up section. If the chiral current is of the CS form, then, from the perturbative viewpoint, this vertex is pictured as a triangle diagram with a spin-1 (photon) field running in the loop. This is clearly possible from the perturbative viewpoint if the Maxwell action is coupled to an external gravitational metric.  \\
We recall that the Maxwell equations in the absence of charges and currents satisfy the duality symmetry 
($E\to B$ and $B\to -E$). The symmetry can be viewed as a special case of a continuous symmetry

\beq
\delta F^{\mu\nu}=\beta \tilde{F}^{\mu\nu}
\eeq
where $\delta\beta$ is an infinitesimal $SO(2)$ rotation and $\tilde{F}^{\mu\nu}= \epsilon^{\mu\nu\rho\sigma}F_{\rho\sigma}/2$. Its finite form 

\begin{equation}
\begin{pmatrix}
 E\\
 B \\
\end{pmatrix}=
\begin{pmatrix}
 \cos\beta&\sin\beta\\
 -\sin\beta&\cos\beta \\
\end{pmatrix}\begin{pmatrix}
E\\B\\
\end{pmatrix}
\end{equation}
is indeed a symmetry of the equations of motion, but not of the Maxwell action.  
The action
\beq
\mathcal{S}=\int d^4 x F^{\mu\nu} F_{ \mu\nu}^{}
\eeq
 is invariant under an infinitesmal transformation modulo a total derivative. For $\beta=\pi/2$, the discrete case, then the action flips sign since $(F^2\to - \tilde{F}^2)$, while its infinitesimal variation takes the form 
\beq
\label{sym}
\delta_\beta {S}_{0}=-\beta \int d^4 x \partial_\mu \left(\tilde{F}^{\mu\nu} \,A_\nu\right).
\eeq
Using the dual Bianchi identity
\beq
\partial_\nu F^{\mu\nu}=0 \leftrightarrow \epsilon^{\mu\nu\rho\sigma}\partial_\nu \tilde{F}_{\rho\sigma}=0,
\eeq
we can introduce the dual gauge field $\tilde{A}^\mu$ 
\beq
\tilde{F}^{\mu\nu}=\partial^\mu \tilde{A}^\nu - \partial^\nu \tilde{A}^\mu,
\eeq
which is related to the original $A_\mu$ one by 
\beq
\partial^\mu \tilde{A}^\nu - \partial^\nu \tilde{A}^\mu=\epsilon^{\mu\nu\rho\sigma}\partial_{\rho}A_\sigma.
\eeq
The current corresponding to the infinitesimal symmetry \eqref{sym} can be expressed in the form
\beq
J^\mu=\tilde{F}^{\mu\nu}A_\nu - F^{\mu\nu}\tilde{A}_\nu, 
\eeq
whose conserved charge is gauge invariant 
\beq
Q_5=\int d^3 x \left( A\cdot \nabla  \times A -  \tilde{A}\cdot\nabla \times \tilde{A}\right)
\label{kk}
\eeq
after an integration by parts. The two terms on the equation above count the linking number of magnetic and electric lines respectively. In fluid mechanics, helicity is the volume integral of the scalar product of the velocity field with its curl given by
\beq
\mathcal{H}_{fluid}=\int d^3 x \, \vec{v}\cdot \nabla \times \vec{v}
\eeq
and one recognizes in \eqref{kk} the expression 
\beq
\label{qq5}
Q_5=\int d^3 x \left( B\cdot A  - E\cdot \tilde{A}\right)
\eeq
with $B=\nabla\times A$ and $E=-\nabla\times \tilde{A}$, that coincides with the optical helicity of the electromagnetic field \cite{delRio:2020cmv}. \\

\subsection{L/T decomposition} 

An alternative parameterization of the $VVA$ correlator brings us closer to the main point of our discussion, i.e. on its most fundamental characterization, which involves an anomaly pole. A pole shows in the parameterization only after the use of the Schouten identities. Indeed, the topological behaviour of the chiral interaction can surely be attributed to these relations. One can insert poles as well as remove them by using these identities. This is the reason why it takes some effort in order to show the physical relevance of such terms.\\
This takes us to consider a parameterization in which the vertex is separated into its longitudinal and transverse sectors. The longitudinal sector naturally contains a pole, as we are going to see, while the transverse part is homogeneous with respect to the axial-vector WI. The meaning of this separation will become clear once we let a massive fermion run in the loop, and introduce a dispersion relations in the parameterization of the anomaly form factor.  This point will be addressed later in this review. The longitudinal transverse (L/T) decomposition takes the following form \cite{Knecht:2003xy}
\beq
\langle J^{\mu_1}(p_1)J^{\mu_2}(p_2)J^{\mu_3}_5(p_3) \rangle=\frac{1}{8\pi^2} \left({W^{L}}^{\mu_1\mu_2\mu_3} - {  \mathcal \, W^{T}}^{\mu_1\mu_2\mu_3} \right)
\eeq
where the longitudinal component is specified as 

\begin{equation}
\label{refe}
W_L^{\mu_1\mu_2\mu_3}=w_L \, {p_3^{\mu_3}}\epsilon^{\mu_1\mu_2 \rho\sigma}{p_{1\rho}p_{2\sigma}}\equiv w_L\,
{p_3^{\mu_3}}\epsilon^{\mu_1\mu_2  p_1 p_2}
\end{equation}
while the transverse component is given by
\begin{align}
	{  \mathcal \, W^{T}}^{\mu_1\mu_2\mu_3}(p_1,p_2,p_3^2) &=
	w_T^{(+)}\left(p_1^2, p_2^2,p_3^2 \right)\,t^{(+)\,\mu_1\mu_2\mu_3}(p_1,p_2)
	+\,w_T^{(-)}\left(p_1^2,p_2^2,p_3^2\right)\,t^{(-)\,\mu_1\mu_2\mu_3}(p_1,p_2) \nonumber \\
	& +\,\, {\widetilde{w}}_T^{(-)}\left(p_1^2, p_2^2,p_3^2 \right)\,{\widetilde{t}}^{(-)\,\mu_1\mu_2\mu_3}(p_1,p_2).
\end{align}
This decomposition automatically account for all the symmetries of the correlator
with the transverse tensors given by
\begin{equation}\label{tensors}
	\begin{aligned}
		t^{(+) \, \mu_1\mu_2\mu_3}(p_1,p_2) &=
		p_{1}^{\mu_2}\, \veps^{ \mu_1\mu_3 p_1p_2}  -
		p_{2}^{\mu_1}\,\veps^{\mu_2\mu_3 p_1 p_2}  - (p_{1} \cdot p_2)\,\veps^{\mu_1\mu_2\mu_3(p_1 - p_2)}\\ &\hspace{4cm}
		+  \frac{p_1^2 + p_2^2 - p_3^2}{p_3^2}\,  (p_1+p_2)^{\mu_3} \, 
		\veps^{\mu_1\mu_2 p_1 p_2}
		\nonumber  , \\
		t^{(-)\,\mu_1\mu_2\mu_3}(p_1,p_2) &= \left[ (p_1 - p_2)^{\mu_3} - \frac{p_1^2 - p_2^2}{p_3^2}\, (p_{1}+p_2)^{ \mu_3} \right] \,\veps^{\mu_1\mu_2 p_1 p_2}
		\nonumber\\
		{\widetilde{t}}^{(-)\, \mu_1\mu_2\mu_3}(p_1,p_2) &= p_{1}^{\mu_2}\,\veps^{ \mu_1\mu_3 p_1p_2} +
		p_{2}^{\mu_1}\,\veps^{\mu_2\mu_3 p_1 p_2} 
		- (p_{1}\cdot p_2)\,\veps^{ \mu_1 \mu_2 \mu_3 (p_1+p_2)}.
	\end{aligned}
\end{equation}
The map between the representation in \eqref{reduced} and the current one is given by the relations 
\begin{align}
B_3 (p_1, p_2) &= \frac{1}{8 \pi^2} \left[ w_L - \tilde{w}_T^{(-)}
-\frac{p_1^2+p_2^2}{p_3^2}     w_T^{(+)}
- 2 \,  \frac{p_1 \cdot p_2 + p_2^2 }{p_3^2 }w_T^{(-)}  \right],  \\
B_4 (p_1, p_2) &= \frac{1}{8 \pi^2} \left[  w_L
+ 2 \, \frac{p_1 \cdot p_2}{p_3^2}       w_T^{(+)}
+ 2 \, \frac{p_1 \cdot p_2 + p_1^2}{p_3^2 }w_T^{(-)}  \right], 
\end{align}
and viceversa
\begin{equation}
\label{ppo}
w_L (p_1^2, \, p_2^2,p_3^2) = \frac{8 \pi^2}{p_3^2} \left[B_1 - B_2 \right]
\end{equation}
or, after the imposition of the Ward identities in Eqs.(\ref{WI1},\ref{WI2}),
\begin{align}
w_L ( p_1^2, \, p_2^2,p_3^2) &= \frac{8 \pi^2}{p_3^2}
\left[ (B_3-B_6) p_1 \cdot p_2 + B_4 \, p_2^2 - B_5 \, p_1^2 \right],
\label{wL}\\
w_T^{(+)} (p_1^2, \, p_2^2,p_3^2)  &= - 4 \pi^2 \left(B_3 - B_4 + B_5 - B_6 \right),
\label{wTp}\\
w_T^{(-)} ( p_1^2, \, p_2^2,p_3^2)  &=  4 \pi^2 \left(B_4+B_5 \right),
\label{wTm}\\
\tilde{w}_T^{(-)} ( p_1^2, \, p_2^2,p_3^2)  &= - 4\pi^2 \left( B_3 + B_4 + B_5 + B_6 \right),
\label{wTt}
\end{align}
where $B_i\equiv B_i(p_1,p_2)$. As already mentioned, \eqref{ppo} is a special relation, since it shows that the pole is not affected by Chern-Simons forms, telling us of the physical character of this part of the interaction. \\
Also in this case, the counting of the form factor is four, one for the longitudinal pole part and three for the transverse part. Notice that all of them are either symmetric or antisymmetric by construction 
\beqa
w_L(p_1^2,p_2^2,p_3^2)&=& w_L(p_2^2,p_1^2,p_3^2) \nn\\
w_T^{(+)}(p_1^2,p_2^2,p_3^2)&=& w_T^{(+)}(p_2^2,p_1^2,p_3^2) \nn\\
w_T^{(-)}(p_1^2,p_2^2,p_3^2)&=& - w_T^{(-)}(p_2^2,p_1^2,p_3^2) \nn \\
\tilde{w}_T^{(-)}(p_1^2,p_2^2,p_3^2)&=&- \tilde{w}_T^{(-)}(p_2^2,p_1^2,p_3^2). 
\eeqa
There is an important point to note: Eq. \eqref{ppo} is not affected by CS forms. This implies that even if we do not impose the vector WI, which allows us to re-express the divergent form factors $B_1$ and $B_2$ in terms of the converging ones, the pole is not affected by the parameterization of the loop momentum due to its relation to the difference of the two form factors. 
This is the first indication of the significance of the $1/p_3^2$ term, known as the anomaly pole, in the anomaly vertex. This has some remarkable implications regarding the structure of the anomaly effective action and the origin of the axion-like interaction generated in the response function of topological materials. \\
Such an effective action has been developed in the form of an expansion in terms of dimensionless composite operators, as we will explain. These operators encode the absence of scales in the anomaly and highlight the dominance of the phenomenon in light-cone physics. The expansion is governed by the insertion of interactions of nonlocal operators of the form $R\Box^{-1}$ for the conformal anomaly and $\partial\cdot B \Box^{-1}$ for the chiral anomaly. It is reproduced both in perturbation theory and non-perturbatively using conformal field theory (CFT) methods, through the solution of the CWIs. Here, $R$ represents the Ricci scalar and $B$ is an external axial-vector source, driving the response of a system to external chiral or conformal interactions.
Before diving into a discussion of the anomaly effective action for chiral interactions, we will briefly discuss CFT in momentum space, which offers an independent perspective capable of reproducing all the highlighted results.

\section{The Dispersive sum rule and the spectral density flow}
\label{spectral1}
As highlighted extensively in the literature on the chiral anomaly \cite{Dolgov:1971ri}, a notable perturbative manifestation of this phenomenon is the emergence of a massless pole, referred to as an anomaly pole, within the spectrum of the $\langle AVV \rangle$ diagram. This pole manifests solely in a specific kinematic scenario—namely, at zero fermion mass and with on-shell vector lines—within perturbation theory. The intermediate state exchanged in the effective action, depicted in Fig. (\ref{ddiag}) and mediated by the $\langle AVV \rangle$ diagram, is characterized by two massless fermions moving on the light-cone. Within the confines of the perturbative framework, it is highly suggestive to interpret the occurrence of this intermediate configuration as indicative of the potential exchange of a bound state in the quantum effective action.
\\
In a phenomenological context, the broader significance of this kinematic mechanism stems from the emergence of a certain kinematical duality accompanying any perturbative anomaly. In this specific case, it is commonly referred to as $Q^2$-duality, which establishes a connection between the resonance and asymptotic regions of a particular correlator in a nontrivial manner \cite{Bertlmann:1981by}. This property typically finds justification in the existence of a sum rule for the spectral density $\rho(s, m^2)$ of the correlator, often taking the form
\begin{equation}
\label{sr}
\frac{1}{\pi} \int_0^{\infty}\rho(s, m^2) ds = f,
\end{equation}
where the constant $f$ remains independent of any mass or other parameter characterizing the thresholds or strengths of the resonant states potentially present in the integration region ($s > 0$). It is important to emphasize that sum rules formulated for the analysis of resonance structures, such as their strengths and masses as observed in the context of QCD, the theory of the strong interactions, necessitate a parameterization of the resonant behavior of $\rho(s,m)$ at low $s$ values through a phenomenological approach, coupled with the inclusion of the asymptotic behavior of the correlator, amenable to perturbation theory for larger $s$. This significant interplay between the infrared (IR) and ultraviolet (UV) regions aptly warrants the term "duality" to describe the implications of a given sum rule.\\
It has been noted for some time that a specific aspect of the chiral anomaly lies in the existence of a sum rule for the $\langle AVV \rangle$ diagram \cite{Horejsi:1985qu}, later extended to a similar investigation of the trace of the energy-momentum tensor ($\textrm{tr}T$) for the $\langle \textrm{tr}TVV \rangle$ in QED (with $V$ representing a vector current), particularly at zero momentum transfers \cite{Horejsi:1997yn,Horejsi:1994aj}. This scrutiny has provided substantial evidence that the sum rule, in conjunction with the initial identification of the anomaly pole from the perturbative spectral density \cite{Dolgov:1971ri}, constitutes two significant and interconnected aspects of the anomaly phenomenon. It is worth recalling that the exploration of these correlators has a lengthy history \cite{Freedman:1974ze, Adler:1976zt,Chanowitz:1972da}.\\
More recently, comprehensive perturbative analyses of the $\langle TVV \rangle$ correlator (or the graviton-gauge-gauge vertex), conducted off-shell and at nonzero momentum transfer, have revealed that the general characteristics observed in the $\langle AVV \rangle$ and $\langle \textrm{tr}TVV \rangle$ cases remain preserved \cite{Giannotti:2008cv,Armillis:2009pq, Armillis:2010qk, Armillis:2009im}.\\
A distinct feature of the spectral density associated with the chiral and conformal anomalies is the introduction of a pole in the spectrum under specific kinematic limits, representing a degeneracy of the two-particle cut when any secondary scale (such as the fermion mass) tends to zero. The phenomenon of the "cut turning into a pole" is characteristic of finite (non-superconvergent) sum rules. It is associated with a spectral density normalized by the sum rule akin to an ordinary weighted distribution, with its support positioned at the edge of the allowed phase space ($s=0$) as the conformal deformation approaches zero. This enables the isolation of a unique interpolating state among all possible exchanges permitted in the continuum, particularly for $s> 4 m^2$, as the theory transitions towards its conformal/superconformal point.
\\
As previously discussed, the presence of a sum rule for the form factor responsible for a particular anomaly indicates a UV/IR connection manifested by the corresponding spectral density. However, this connection is not exclusively tied to the anomaly phenomenon. Indeed, non-anomalous form factors, in certain cases, exhibit similar behaviors. Nonetheless, the breaking of symmetry should ideally result in the emergence of a massless state in the effective theory's spectrum, lending a distinctive significance to the saturation of the spectral density with a single resonance in an anomaly form factor.
\\
Expanding upon Eq. (\ref{sr}), it becomes apparent that the anomaly's effect generally correlates with the behavior of the spectral density across all values of $s$, albeit in certain kinematical limits, particularly around the light cone ($s\sim 0$), dominating the sum rule and constituting a resonant contribution. The combination of the scaling behavior of the corresponding form factor $F(Q^2)$ (or equivalently, its density $\rho$) with the requirement of integrability of the spectral density essentially fixes $f$ as a constant and ensures that the sum rule (\ref{sr}) is saturated by a single massless resonance. Conversely, a superconvergent sum rule obtained for $f=0$ would not exhibit this behavior. Importantly, the absence of subtractions in the dispersion relations underscores the significance of the sum rule, independent of any ultraviolet cutoff.
\\
It is straightforward to demonstrate that Eq. (\ref{sr}) imposes a constraint on the asymptotic behavior of the associated form factor. The proof involves observing the dispersion relation for a form factor in the spacelike region ($Q^2=-k^2> 0$), which, upon expansion and utilization of Eq. (\ref{sr}), induces an asymptotic behavior on $F(Q^2,m^2)$ as $Q^2\to \infty$. This behavior, $F\sim f/Q^2$ at large $Q^2$, with $f$ independent of $m$, highlights the dominance of the pole in $F$ as $Q^2 \rightarrow \infty$. The UV/IR conspiracy of the anomaly, as discussed in \cite{Armillis:2009im, Armillis:2009sm,Armillis:2009pq}, is evidenced by the reappearance of the pole contribution at very large values of the invariant $Q^2$, even for a nonzero mass $m$. Notably, the spectral density exhibits support around the $s=0$ region $(\rho(s)\sim \delta(s))$, akin to the massless $(m=0)$ case. This subtle point underscores the decoupling of the anomaly pole for a nonzero mass, termed as "decoupling," referring to the non-resonant behavior of $\rho$.\\
Thus, the presence of a $1/Q^2$ term in anomaly form factors reflects the entire flow, converging to a localized massless state ($\rho(s)\sim\delta(s)$) as $m\to 0$, while the presence of a non-vanishing sum rule validates the asymptotic constraint. Importantly, although the independence of the asymptotic value $f$ on $m$ for conformal deformations driven by a single mass parameter is a straightforward consequence of the scaling behavior of $F(Q^2,m^2)$, it holds generally even for completely off-shell kinematics.
\\
In summary,  the two fundamental features of anomalous behavior of a certain form factor responsible for chiral or conformal anomalies are: 1) the existence of a spectral flow that transforms a dispersive cut into a pole as $m$ approaches zero, and 2) the presence of a sum rule relating the asymptotic behavior of the anomaly form factor to the strength of the pole resonance.

\subsection{The case of the $AVV$ }
We illustrate this point by an analysis of the anomaly form factor in the $AVV$ interaction for a massive fermion in the loop and on-shell photons. A direct computation gives
\beq
\label{RChiralOSMassive}
\Gamma_{}^{\mu\alpha\beta}(p,q) =  i \frac{g^2 \,}{12 \pi^2} \, \phi_1(k^2,m^2) \, \frac{k^\mu}{k^2}\eps[p, q, \alpha ,\beta]  \,, \\
\eeq
with
\beq
\Phi_1(k^2,m^2) = - 1 - 2\, m^2 \, \mathcal C_0(k^2,m^2) \,, \nn \\
\eeq
and with the anomaly form factor given by 
\beq
 \chi(k^2, m^2)\equiv \Phi_1(k^2,m^2)/k^2. 
 \eeq

We start by introducing the spectral density $\rho(k^2)$, which is the discontinuity of $\mathcal C_0$ along the cut $(k^2>4 m^2)$, as 
\beq
\rho(k^2,m^2)=\frac{1}{2 i} \textrm{Disc}\, \mathcal C_0(k^2, m^2) \,,
\label{spect}
\eeq
with the usual $i\epsilon$ prescription ($\epsilon >0$)
\beq
\label{discdef}
 \textrm{Disc} \, \mathcal C_0(k^2, m^2)\equiv \mathcal C_0(k^2 +i \epsilon,m^2) - \mathcal C_0(k^2 -i \epsilon,m^2).
\eeq
\begin{figure}[t]
\centering
\subfigure{\includegraphics[scale=0.8]{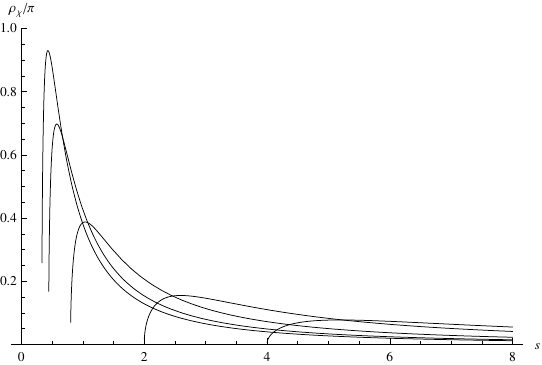}} \hspace{.5cm}
\caption{Representatives of the family of spectral densities $\frac{{{\rho}_\chi}^{(n)}}{\pi}(s)$ plotted versus $ s $ in units of $m^2$. The family "flows"  towards the $s=0$ region becoming a $\delta(s)$ function as $m^2$ goes to zero.  The are under the curve is preserved by the flow and equals the anomaly. A similar result holds been shown holds also for the gravitational chiral anomaly.}
\label{seq}
\end{figure}
To determine the discontinuity above the two-particle cut we can proceed in two different ways. We can use the unitarity cutting rules and therefore compute the integral
\bea
\label{disco}
\textrm{Disc} \, \mathcal C_0(k^2, m^2)&=& \frac{1}{i \pi^2} 
 \int d^4 l \frac{2 \pi i \delta_+(l^2 - m^2) 2 \pi i \delta_+((l - k)^2- m^2)}{(l-p)^2 - m^2 + i \epsilon} \nn\\
 &=&  \frac{2 \pi}{i k^2}\log\left(\frac{1 + \sqrt{\tau(k^2 ,m^2)}}{1-\sqrt{\tau(k^2 ,m^2)}}\right)\theta(k^2 - 4 m^2) \,,
\label{cut}
 \eea
where $\tau(k^2,m^2) = \sqrt{1 -4m^2/k^2}$.
 The integral has been computed by sitting in the rest frame of the off-shell line of momentum $k$.
Alternatively, we can exploit directly the analytic continuation of the explicit expression of the $\mathcal C_0(k^2,m^2)$ integral in the various regions. This is given by
\bea  
\label{ti}
 \mathcal C_0(k^2\pm i\epsilon,m^2) = \left\{ 
\begin{array}{ll} 
\frac{1}{2 k^2}\log^2 \frac{\sqrt{\tau(k^2,m^2)}+1}{\sqrt{\tau(k^2,m^2)}-1} & \mbox{for} \quad k^2 < 0 \,, \\
- \frac{2}{k^2} \arctan^2{\frac{1}{\sqrt{-\tau(k^2,m^2)}}} & \mbox{for} \quad 0 < k^2 < 4 m^2 \,, \\
\frac{1}{2 k^2} \left( \log \frac{1 + \sqrt{\tau(k^2,m^2)}}{1 - \sqrt{\tau(k^2,m^2)} } \mp i \, \pi\right)^2 & \mbox{for} \quad k^2 > 4 m^2 \,.  
\end{array} 
\right.
\eea
From the two branches encountered with the $\pm i \epsilon$ prescriptions, the discontinuity is then present only for $k^2> 4m^2$, as expected from unitarity arguments, and the result for the discontinuity, obtained using the definition in Eq. (\ref{discdef}), 
clearly agrees with Eq. (\ref{disco}), computed instead by the cutting rules. \\
The dispersive representation of $\mathcal C_0(k^2,m^2)$ in this case is written as
\beq
\mathcal C_0(k^2,m^2)=\frac{1}{\pi} \int_{4 m^2}^{\infty} d s \frac{\rho(s, m^2)}{s - k^2 }.
\eeq
We can reconstruct the scalar integral 
$\mathcal C_0(k^2,m^2)$ from its dispersive part. \\
Having determined the spectral function of the scalar integral $\mathcal C_0(k^2,m^2)$, we extract the spectral density associated with the anomaly form factor in Eq. (\ref{RChiralOSMassive}), which is given by
and which can be computed as
\bea
\textrm{Disc}\, \chi(k^2, m^2)= \chi(k^2+i\epsilon,m^2)-\chi(k^2-i\epsilon, m^2) = - \textrm{Disc}\left( \frac{1}{k^2} \right) - 2 m^2 \textrm{Disc}\left(\frac{\mathcal C_0(k^2,m^2)}{k^2}\right).
\label{cancel1}
\eea
Using the principal value prescription 
\beq
\frac{1}{x\pm i\epsilon}=P\left(\frac{1}{x} \right) \mp i\pi \delta(x),
\eeq
we obtain
\bea
&& \textrm{Disc} \left( \frac{1}{k^2} \right) = - 2 i\pi\delta(k^2) \nn\\
&& \textrm{Disc}\left( \frac{\mathcal C_0(k^2,m^2)}{k^2}\right) = P\left(\frac{1}{k^2}\right)\textrm{Disc}\,\mathcal C_0(k^2,m^2) - i\pi \delta(k^2) A(0) \,,
\eea 
where we have defined  
\beq
\label{As}
A(k^2)\equiv C_0(k^2+i\epsilon,m^2)+C_0(k^2-i\epsilon,m^2),
\eeq
and
\bea
A(0) &=& \lim_{k^2\to 0} A(k^2) =  -\frac{1}{m^2}. 
\eea
This gives, together with the discontinuity of $\mathcal C_0(k^2,m^2)$ which we have computed previously in Eq. (\ref{cut}),
\beq
\label{discC0}
\textrm{Disc}\left( \frac{\mathcal C_0(k^2,m^2)}{k^2}\right)=-2 i\frac {\pi}{(k^2)^2} \log \frac{1 + \sqrt{\tau(k^2,m^2)}}{1 - \sqrt{\tau(k^2,m^2)}}\theta(k^2-4 m^2)
+i\frac{\pi}{m^2}\delta(k^2).
\eeq
The discontinuity of the anomalous form factor $\chi(k^2,m^2)$ is then given by
\beq
\textrm{Disc} \, \chi(k^2,m^2)=4 i \pi \frac{m^2}{ (k^2)^2}\log \frac{1 + \sqrt{\tau(k^2,m^2)}}{1 - \sqrt{\tau(k^2,m^2)} }\theta(k^2- 4 m^2).
 \eeq 
The total discontinuity of $\chi(k^2,m^2)$, as seen from the result above, is characterized just by a single cut for $k^2> 4 m^2$, since the $\delta(k^2)$ (massless resonance) contributions cancel between the first and the second term of Eq. (\ref{cancel1}). This result proves the {\em decoupling } of the anomaly pole at $k^2=0$ in the massive case due to the disappearance of the resonant state. \\
The function describing the anomaly form factor, $\chi(k^2,m^2)$, then admits a dispersive representation over a single branch cut
\beq
\chi(k^2,m^2)=\frac{1}{\pi}\int_{4 m^2}^{\infty}\frac{{\rho}_\chi(s,m^2)}{s- k^2 }ds
\eeq
corresponding to the ordinary threshold at $k^2=4 m^2$,
with 
\bea
\label{spectralrho}
{\rho}_\chi(s,m^2) = \frac{1}{2 i} \textrm{Disc} \, \chi(s,m^2)  
=\frac{2 \pi m^2}{s^2}\log\left(\frac{1 + \sqrt{\tau(s,m^2)}}{1-\sqrt{\tau(s,m^2)}}\right)\theta(s- 4 m^2). 
\eea
As we have anticipated above, a crucial feature of these spectral densities is the existence of a sum rule. In this case it is given by
\beq
\label{superc}
\frac{1}{\pi}\int_{4 m^2}^{\infty}ds{{\rho}_\chi(s,m^2)}=1.
\eeq
At this point, to show the convergence of the family of spectral densities to a resonant behaviour, it is convenient to extract a  
discrete sequence of functions, parameterized by an integer $n$ and then let $n$ go to infinity
\bea
\rho^{(n)}_\chi(s) &\equiv& \rho_\chi(s,m_n^2)  \qquad \mbox{with} \quad m_n^2 =\frac{4 m^2}{n}.
\eea
One can show that this sequence $\{\rho^{(n)}_\chi\}$ then converges to a Dirac delta function 
\bea
\lim_{m\to 0} \rho_\chi(s,m^2) = \lim_{m\to 0} \frac{2 \pi m^2}{s^2}\log\left(\frac{1 + \sqrt{\tau(s,m^2)}}{1-\sqrt{\tau(s,m^2)}}\right)\theta(s- 4 m^2) = \pi \delta(s)
\eea
in a distributional sense. We have shown in Fig. (\ref{seq}), on the left, the sequel of spectral densities which characterize the flow as we turn the mass parameter to zero. The area under each curve is fixed by the sum rule and is a characteristic of the entire flow. Clearly, the $\rho^{(n)}$ are normalized distributions for each given value of $m$. They describe, for each invariant mass value $s$, the absolute weight of the intermediate state - of that specific invariant mass - to a given anomaly form factor. 

\subsection{Other chiral anomaly sum rules} 
\label{sumrules}
Here we are going to prove the existence of other chiral sum rules for similar correlators. In each example, one needs to move away from the conformal point, where the correlation function is fixed by the pole and the CWIs, by giving a mass $m$ to the particle in the loop.  
Chiral gravitational anomalies for spin 1 fields (photons) have been discussed long ago by a perturbative analysis of the $\langle TTJ_{CS}\rangle$ \cite{Dolgov:1987yp}. \\
The presence of an anomaly pole in this correlator can indeed be extracted from \cite{Dolgov:1987yp}, in agreement with our result, based on the CFT reconstruction of the corresponding vertex. One can show that similar anomaly sum rules are present in all the vertices as we move away from the conformal limit, reproducing the phenomenon that we have described for the $AVV$.
Indeed, for on-shell gravitons ($g$) and photons ($\gamma$), the authors obtain, with the inclusion of mass effects in the $\langle AVV\rangle $, $\langle TTJ_{f}\rangle $ and $\langle TT J_{CS}\rangle$ the following expressions for the matrix elements
\beq
\langle 0|J^\mu_f |\gamma \gamma\rangle = f_1(q^2)\frac{q^\mu}{q^2} F_{\kappa\la}\tilde{F}^{\kappa\lambda} (q)
\eeq
\beq
\langle 0|J^\mu_f | g g\rangle = f_2(q^2)\frac{q^\mu}{q^2} R_{\kappa\la\rho\sigma}\tilde{R}^{\kappa\lambda\rho\sigma}(q) 
\label{oone}
\eeq
\beq
\langle 0|J^\mu_{CS} | g g\rangle = f_3(q^2)\frac{q^\mu}{q^2} R_{\kappa\la\rho\sigma}\tilde{R}^{\kappa\lambda\rho\sigma}(q),
\label{oone2}
\eeq
where $q$ is the momentum of the chiral current. $R_{\kappa\la\rho\sigma}$ is the Riemann tensor. 
The terms $R\tilde{R}(q)$ denotes the Fourier transform to momentum space of $R\tilde{R}$, differentiated twice with respect to the gravitational metric and contracted with physical polarizations of the external gravitational field. 
The anomaly poles are extracted by including a mass $m$ in the propagators of the loop corrections, in the form of either a fermion mass for the $\langle AVV\rangle$ and the $\langle TTJ_f\rangle $, or working with a 
Proca spin-1 in the case of $\langle TTJ_{CS} \rangle$, and then taking the limit for $m\to 0$.
A dispersive analysis gives for the corresponding spectral densities \cite{Dolgov:1987yp}
\beqa
\Delta_{AVV}(q^2,m)\equiv \textrm{Im} f_1(q^2)&=& \frac{ d_{AVV}}{ q^2}(1- v^2) \log\frac{1 +v }{1-v}\nn\\
\Delta_{TT{J_f}}(q^2,m)\equiv \textrm{Im} f_2(q^2)&=&\frac{d_{TT{J_f}}}{ q^2}(1- v^2)^2 \log\frac{1 +v }{1-v}\nn\\
\Delta_{TTJ_{CS}}(q^2,m)\equiv \textrm{Im} f_3(q^2)&=&\frac{d_{TTJ_{CS}}}{ q^2}v^2(1- v^2)^2 \log\frac{1 +v }{1-v},\nn\\
\eeqa
with $v=\sqrt{1-4 m^2/q^2}$ and $d_{AVV}=-1/2\, \alpha_{em}$, $d_{TT{J_f}}=1/(192 \pi)$ and $d_{TT J_{CS}}=1/(96 \pi)$ being the corresponding anomaly coefficients in the normalization of the currents of \cite{Dolgov:1987yp}, with $\alpha_{em}$ the electromagnetic coupling. \\
Notice the different functional forms of $\Delta_{TT{J_f}}(q^2,m)$ and $\Delta_{TTJ_{CS}}(q^2,m)$ away from the conformal limit, when the mass $m$ is nonzero.
One can easily check that in the massless limit the branch cut present in the previous spectral densities at $q^2= 4m^2$ turns into a pole 
\beq
\lim_{m\to 0}\Delta(q^2,m)\to \delta(q^2) 
\eeq
in all the three cases. Beside, one can easily show that the same spectral densities satisfy three sum rules 
\cite{Coriano:2023gxa}
\beqa 
\int_{4 m^2}^{\infty} ds\,\Delta_{AVV}(s,m)&=& 2 \, d_{AVV}\\
\int_{4 m^2}^{\infty}ds\,\Delta_{TTJ_f}(s,m)&=&\frac{2}{3}\,  d_{TT{J_f}}\\
\int_{4 m^2}^{\infty}ds\,\Delta_{TTJ_{CS}}(s,m)&=& \frac{14}{45} \, d_{TTJ_{CS}},
\eeqa
indicating that for any deformation $m$ from the conformal limit, the integral under $\Delta(s,m)$ is mass independent. Therefore, the numerical value of the area equals the value of the anomaly coeeficient in each case.\\
One can verify, by taking the on-shell photon/graviton limit, that the transverse sector of $\langle TTJ_5\rangle $, vanishes. Then it is clear that, in general, the structure of the anomaly action responsible for the generation of the gravitational chiral anomaly can be expressed in the form
\beq
\mathcal{S}_{anom}\sim \int d^4 x \, d^4 y\, \partial_\lambda A^\lambda\, \frac{1}{\Box}(x,y) R\tilde{R}(y) +\ldots
\eeq 
where the ellipses stand for the transverse sector, and $A_\lambda$ is a spin-1 external source. For on-shell gravitons, as remarked above, this action summarizes the effect of the entire chiral gravitational anomaly vertex, being exactly given by the exchange of a single anomaly pole.

\section{CFT in momentum space}
\label{conf2}
In this section we turn to a nonperturbative analysis of the chiral correlators using the fundamental costraints of CFT, here formulated in momentum space. \\
The formulation in coordinate space of the theory allows to reconstruct the correlators quite efficiently, but does not provide any hint regarding the origin of the anomaly, that, as we have seen in the previous sections, 
should be attributed to the correlated exchange of a fermion/antifermion pair, as viewed from the axial-vector channel. \\
Anomalies arise from the regions in coordinate space where all the points of a certain correlator coalesce, and, as shown long ago in \cite{Osborn:1993cr}, they need to be included by hand in a certain chiral or conformal correlator, as non-homogeneous contributions to the ordinary solutions of the CWIs, in the form of product of Dirac delta functions. For this reason, there is no much information, from the kinematical point of view, about the nature of the  correlated exchange which is responsible for the phenomenon on the light-cone. \\
Momentum space methods are very important in order to overcome this limitation and have been formulated 
independently of the previous approach. An overview of these methods can be found in \cite{Coriano:2020ees}. \\
In $d=4$, it has been investigated in \cite{Coriano:2013jba, Bzowski:2013sza} and \cite{Coriano:2018bbe, Coriano:2018bsy, Bzowski:2015pba, Bzowski:2018fql}, and more recent work can be found in \cite{Bzowski:2019kwd, Maglio:2019grh}. The conformal constraints can be reduced to a set of differential equations with regular solutions whose exponents are fixed by the Fuchsian points of singular differential equations \cite{Coriano:2018bbe,Coriano:2018bsy}. \\
The extension of the methods to tensor correlators have been originally formulated for conformal anomalies, corresponding to parity even sectors \cite{Bzowski:2015pba}, while their extension to the non-conserved parity-odd ones has been discussed in several works \cite{Marotta:2022jrp}. The extension of the method to chiral correlators, with the direct inclusion of the anomaly constraint, has been presented in \cite{Coriano:2023hts} for the ordinary chiral anomaly and in \cite{Coriano:2023gxa} for the gravitational chiral anomaly. More general parity-odd anomalies \cite{Coriano:2023cvf}, whose existence, from the perturbative picture, has been debated recently, due to opposite conclusions \cite{Armillis:2010pa,Bonora:2015nqa,Bastianelli:2019zrq,Ferrero:2023unz,Larue:2023qxw},  has also been discussed within CFT in \cite{Coriano:2023cvf}, nonperturbatively. 
 \\
The hypergeometric structure of the CWIs has been identified independently in \cite{Coriano:2013jba} and \cite{Bzowski:2013sza}, as mentioned earlier, in the case of 3-point functions. The identification of generalized hypergeometric solutions of the CWIs for 4-point functions, which share a structure typical of 3-point functions, and of the homogeneous solutions of Lauricella type, has been discussed in \cite{Maglio:2019grh}.
\\
The CWIs are composed of special conformal and dilatation WIs, besides the ordinary (canonical) WIs corresponding to Lorentz and translational symmetries, which we specify below. We recall that, in $d=4$, conformal symmetry is realized by the action of $15$ generators, $10$ of them corresponding to the Poincar\'e subgroup, $4$ to the special conformal transformations, and $1$ to the dilatation operator.
\\
In infinitesimal form, they are given by
\begin{equation}
x'_\mu(x)=x_\mu+a_\mu+\omega_{\mu\nu}x^\nu+\sigma x_\mu+b_\mu x^2-2b\cdot x\,x_\mu, \label{transf}
\end{equation}
and they can be expressed as a local rotation
\begin{equation}
\frac{\partial x '^{\mu}}{\partial x^\nu}=\Omega(x) R^\mu_{\nu}(x),
\end{equation}
where $\mu=1,2\ldots d$, and $\Omega(x)$ and $R^\mu_\nu(x)$ are, respectively, finite position-dependent rescalings and rotations with
\begin{equation}
\Omega(x)=1-\l(x),\quad \l(x)=\s-2b\cdot x,\label{Om}
\end{equation}
and $b_\mu$ is a constant $d$-vector.\\
The transformation in \eqref{transf} is composed of the parameters $a_\mu$ for the translations, $\omega_{\mu\nu}=-\omega_{\nu\mu}$ for boosts and rotations, $\sigma$ for the dilatations, and $b_\mu$ for the special conformal transformations. The first three terms in \eqref{transf} define the Poincar\'e subgroup, obtained for $\Omega(x)=1$, which leaves the infinitesimal length invariant. For a general $d$, the counting of the parameters of the transformation is straightforward. We have $d(d-1)/2$ ordinary rotations associated with a $SO(d)$ symmetry in $\mathbb{R}^d$ - with parameters $\omega_{\mu\nu}$ - $d$ translations ($P_\mu$) with parameters $a_\mu$, $d$ special conformal transformations $K^\mu$ (with parameters $b_\mu$), and one dilatation $D$ whose corresponding parameter is $\sigma$, for a total of $(d+1)(d+2)/2$ parameters. This is exactly the number of parameters appearing in general $SO(2,d)$ transformation. Indeed, one can embed the actions of the conformal group of $d$ dimensions into a larger $\mathbb{R}^{d+2}$ space, where the action of the generators is linear on the coordinates $x^M$ ($M=1,2,\ldots,d+2$) of such space, using a projective representation. This is the basis of the so-called embedding formalism. For more details, we refer to \cite{Simmons-Duffin:2016gjk}.
\\
By including the inversion $(\mathcal{I})$
\begin{equation}
x_\mu\to x'_\mu=\frac{x_\mu}{x^2},\qquad \Omega(x)=x^2,
\end{equation}
we can enlarge the conformal group to $O(2,d)$. Special conformal transformations can be realized by considering a translation preceded and followed by an inversion.
\\
We will focus our discussion mostly on scalar primary operators of a quantum CFT, acting on a certain Hilbert space, which under a conformal transformation will transform as
\begin{equation}
O_i(\mathbf{x})\to O'(\mathbf{x}')=\lambda^{-\Delta_i} O(\mathbf{x}),
\end{equation}
with specific scaling dimensions $\Delta_i$. We start this excursus on the implication of such symmetry on the quantum correlation functions of a CFT, by considering the simple case of a correlator of $n$ primary scalar fields $O_i(\mathbf{x}_i)$, each of scaling dimension $\Delta_i$
\begin{equation}
\label{defop}
\Phi(\mathbf{x}_1,\mathbf{x}_2,\ldots,\mathbf{x}_n)=\braket{O_1(\mathbf{x}_1)O_2(\mathbf{x}_2)\ldots O_n(\mathbf{x}_n)}.
\end{equation}
In all the equations, covariant variables will be shown in boldface. Here we are going to summarize some basic results, whiole few additional details can be found in the appendix.
\\
3- and 4-point functions (besides 2-point functions) in any CFT are significantly constrained in their general structures due to such CWIs. For scalar correlators, the special CWIs are given by first-order differential equations
\begin{equation}
\label{SCWI0}
K^\kappa(x_i) \Phi(\mathbf{x}_1,\mathbf{x}_2,\ldots,\mathbf{x}_n) =0,
\end{equation}
with
\begin{equation}
\label{transf1}
K^\kappa(x_i) \equiv \sum_{j=1}^{n} \left(2 \Delta_j x_j^\kappa- x_j^2\frac{\partial}{\partial x_j^\kappa}+ 2 x_j^\kappa x_j^\alpha \frac{\partial}
{\partial x_j^\alpha} \right)
\end{equation}
being the expression of the special conformal generator in coordinate space.
\\
The corresponding dilatation WI on the same $n$-point function $\Phi$ is given by
\begin{equation}
\label{scale12}
D(x_i)
\Phi(\mathbf{x}_1,\ldots \mathbf{x}_n)=0,
\end{equation}
with
\begin{equation}
\label{scale11}
D(x_i)\equiv\sum_{i=1}^n\left( x_i^\alpha \frac{\partial}{\partial x_i^\alpha} +\Delta_i\right)
\end{equation}
for scale covariant correlators. In the case of scale invariance, the dilatation WI takes the form
\begin{equation}
D_0(x_i)
\Phi(\mathbf{x}_1,\ldots \mathbf{x}_n)=0,
\end{equation}
with $D_0(x_i)$ given by
\begin{equation}
D_0(x_i)\equiv\sum_{i=1}^n\left( x_i^\alpha \frac{\partial}{\partial x_i^\alpha}\right).
\end{equation}
Such CWIs are sufficient to completely determine the expression of a scalar 3-point function of primary operators $\mathcal{O}_i$ of scaling dimensions $\Delta_i$ $(i=1,2,3)$ in the form
\begin{equation}
\label{corr}
\langle \mathcal{O}_1(\mathbf{x}_1)\mathcal{O}_2(\mathbf{x}_2)\mathcal{O}_3(\mathbf{x}_3)\rangle =\frac{C_{123}}{ x_{12}^{\Delta_t - 2 \Delta_3}  x_{23}^{\Delta_t - 2 \Delta_1}x_{13}^{\Delta_t - 2 \Delta_2} },\qquad \Delta_t\equiv \sum_{i=1}^3 \Delta_i,
\end{equation}
where $x_{ij}=|\mathbf{x}_i - \mathbf{x}_j|$ and  $C_{123}$ is a constant that specifies the CFT. For 4-point functions, the same constraints are weaker, and the structure of a scalar correlator is identified modulo an arbitrary function of the two cross ratios
\begin{equation}
\label{uv}
u(x_i)=\frac{x_{12}^2 x_{34}^2}{x_{13}^2 x_{24}^2} \qquad v(x_i)=\frac{x_{23}^2 x_{41}^2}{x_{13}^2 x_{24}^2}.
\end{equation}
The general solution, allowed by the symmetry, can be written in the form
\begin{equation}
\label{general}
\langle \mathcal{O}_1(\mathbf{x}_1)\mathcal{O}_2(\mathbf{x}_2)\mathcal{O}_3(\mathbf{x}_3)\mathcal{O}_4(\mathbf{x}_4)\rangle= h(u(x_i),v(x_i))\, \frac{1}{\left(x_{12}^2\right)^\frac{\Delta_1 + \Delta_2}{2}\left(x_{3 4}^2\right)^\frac{\Delta_3 + \Delta_4}{2}},
\end{equation}
where $h(u(x_i),v(x_i))$ remains unspecified.\\
For the analysis of $n$-point functions, it is sometimes convenient to introduce more general notations. For instance, one may define
\begin{equation}
\label{conv}
\begin{aligned}
 &\Phi(\underline{\mathbf{x}})\equiv \langle \mathcal{O}_1(\mathbf{x}_1)\mathcal{O}_2(\mathbf{x}_2)\ldots \mathcal{O}_n(\mathbf{x}_n)\rangle, && e^{i \underline{\mathbf{p x}}}\equiv e^{i(\mathbf{p}_1\mathbf{x}_1 + \mathbf{p}_2 \mathbf{x}_2 + \ldots \mathbf{p}_n \mathbf{x}_n)}, \\
 &\ul{d p}\equiv dp_1 dp_2 \ldots d p_n, && 
\Phi(\ul{\mathbf{p}})\equiv \langle O_1(\mathbf{p}_1)O_2(\mathbf{p}_2)\ldots O_n(\mathbf{p}_n)\rangle ,\qquad 
\end{aligned}
\end{equation}
where each of the integrations $dp_i\equiv d^d p_i$ is performed on the $d$-dimensional components of the momenta  $\mathbf{p}_i=(p_i^1,p_i^2\ldots p_i^d)$. 
\\
It will also be useful to introduce the total momentum $\mathbf{P}=\sum_{j=1}^{n} \mathbf{p}_j$ characterizing a given correlator, which vanishes because of the translational symmetry of the correlator in $\mathbb{R}^d$. The momentum constraint in momentum space is enforced via a delta function $\delta(P)$. For instance, translational invariance of $\Phi(\ul{\mathbf{x}})$ gives
\begin{equation}
\label{ft1}
\Phi(\ul{\mathbf{x}})=\int \ul{dp}\ \delta(\mathbf{P}) \ e^{i\ul{p x}} \ \Phi(\ul{\mathbf{p}}).
\end{equation}
In general, for an $n$-point function $\Phi(\underline{\mathbf{x}})$, the condition of translational invariance generates an expression of the same correlator in momentum space of the form \eqref{ft1}. We can remove one of the momenta, conventionally selecting the last one, $\mathbf{p}_n$, which is replaced by its "on shell" version $\bar{\mathbf{p}}_n=-(\mathbf{p}_1+\mathbf{p}_2 +\ldots +\mathbf{p}_{n-1})$.\\
In general, for an $n$-point function $\Phi(\underline{\mathbf{x}})$, the condition of translational invariance is given by
\begin{equation}
\langle \mathcal{O}_1(\mathbf{x}_1)\mathcal{O}_2(\mathbf{x}_2),\ldots, \mathcal{O}_n(\mathbf{x}_n)\rangle = \langle\mathcal{O}_1(\mathbf{x}_1+\mathbf{a} )\mathcal{O}_2(\mathbf{x}_2+\mathbf{a})\ldots \mathcal{O}_n(\mathbf{x}_n+\mathbf{a})\rangle.
\end{equation}
This generates an expression of the same correlator in momentum space, given by Eq. \eqref{ft1}. By convention, we can remove one of the momenta, typically the last one, $\mathbf{p}_n$, which is replaced by its "on shell" version 
$\bar{\mathbf{p}}_n=-(\mathbf{p}_1+\mathbf{p}_2 +\ldots +\mathbf{p}_{n-1})$:
\begin{equation}
\Phi(\underline{\mathbf{x}})=\int dp_1 dp_2... dp_{n-1} e^{i(\mathbf{p}_1 \mathbf{x}_1 + \mathbf{p}_2 \mathbf{x}_2 +...\mathbf{p}_{n-1} \mathbf{x}_{n-1} + \bar{\mathbf{p}}_n \mathbf{x}_n)} \Phi(\mathbf{p}_1,\ldots \mathbf{p}_{n-1},\bar{\mathbf{p}}_n),
\end{equation}
where
\begin{equation}
\Phi(\mathbf{p}_1,\ldots \mathbf{p}_{n-1},\bar{\mathbf{p}}_n)=\langle O_1(\mathbf{p}_1)\ldots O_n(\bar{\mathbf{p}}_n)\rangle
\end{equation}
is the Fourier transform of the original correlator \eqref{defop}. Further discussions on the derivations of expressions in momentum space for dilatation and special conformal transformations can be found in \cite{Coriano:2018bbe}.
\\
The special conformal generator in momentum space is expressed as:
\begin{equation}
K^\kappa(p_i)\equiv\sum_{j=1}^{n-1}\left(2(\Delta_j- d)\frac{\partial}{\partial p_j^\kappa}+p_j^\kappa \frac{\partial^2}{\partial p_j^\alpha\partial p_j^\alpha} -2 p_j^\alpha\frac{\partial^2}{\partial p_j^\kappa \partial p_j^\alpha}\right).
\end{equation}
This corresponds to \eqref{transf1}, and thus the special CWIs are given by the equation
\begin{equation}
K^\kappa(p_i)\Phi(\mathbf{p}_1,\ldots \mathbf{p}_{n-1},\bar{\mathbf{p}}_n)=0. \label{SCWI}
\end{equation}
If the primary operator $\mathcal{O}_i$ transforms under scaling as
\begin{equation}
\mathcal{O}_i(\lambda\ \mathbf{x}_i)=\lambda^{-\Delta_i}\mathcal{O}_i(\mathbf{x}_i),
\end{equation}
then in momentum space, the same scaling takes the form
\begin{equation}
\Phi(\lambda\,\mathbf{p}_1,\ldots, \lambda\,\bar{\mathbf{p}}_n)=\lambda^{-\Delta'}\Phi(\mathbf{p}_1,\ldots ,\bar{\mathbf{p}}_n),
\end{equation}
where
\begin{equation}
\Delta'\equiv \left(-\sum_{i=1}^n \Delta_i +(n-1) d\right)=-\Delta_t +(n-1) d.
\end{equation}
In momentum space, the conditions of scale covariance and invariance respectively take the forms:
\begin{equation}
D(p_i)\,\Phi(\mathbf{p}_1,\ldots ,\bar{\mathbf{p}}_n)=0,
\end{equation}
where
\begin{equation}
D(p_i)\equiv\sum_{i=1}^{n-1}  p_i^\alpha \frac{\partial}{\partial p_i^\alpha} + \Delta',
\end{equation}
and
\begin{equation}
D_0(p_i)\,\Phi(\mathbf{p}_1,\ldots ,\bar{\mathbf{p}}_n)=0,
\end{equation}
where
\begin{equation}
D_0(p_i)\equiv\sum_{i=1}^{n-1} p_i^\alpha \frac{\partial}{\partial p_i^\alpha}.
\end{equation}
In the case of tensor correlators, the structure of the special CWIs involves the Lorentz generators $\Sigma^{\mu\nu}$ and takes the form
\begin{equation}
\begin{split}
& \sum_{r=1}^{n-1} \left( p_{r \, \mu} \, \frac{\partial^2}{\partial p_{r}^{\nu} \partial p_{r \, \nu}}  - 2 \, p_{r \, \nu} \, 
\frac{\partial^2}{ \partial p_{r}^{\mu} \partial p_{r \, \nu} }    + 2 (\Delta_r - d) \frac{\partial}{\partial p_{r}^{\mu}}  + 2 
(\Sigma_{\mu\nu}^{(r)})^{i_r}_{j_r} \frac{\partial}{\partial p_{r \, \nu}} \right) \\ 
& \hspace{7cm} \, \times \langle \mathcal O^{i_1}_1(\mathbf{p}_1) \ldots  \mathcal O^{j_r}_r(\mathbf{p}_r) \ldots \mathcal O^{i_n}_n(\mathbf{p}_n) \rangle = 0,
\end{split}
\end{equation}
where the indices $i_1,\ldots i_n$ and $j_1\ldots j_n$ run on the representation of the Lorentz group to which the operators belong. Note that the sum over the index $r$ selects in each term a specific momentum $p_r$, but the last momentum $p_n$ is not included, since the summation runs from 1 to $n-1$. Therefore, the differentiation with respect to the last momentum $p_n$, which has been chosen as the dependent one, is performed implicitly. Meanwhile, the action of the rotation (Lorentz) generators $\Sigma_{\mu\nu}^{(r)}$ of $SO(d)$ is performed on each of the primary operators $O_1, O_2, \ldots$, except the last one, $O_n$, which is treated like a singlet under such rotational symmetry \cite{Coriano:2018bbe}.

\subsection{2-point functions}
The simplest application of such equations are for 2-point functions \cite{Coriano:2013jba}
$G^{ij}(\mathbf{p}) \equiv \langle \mathcal O_1^i(\mathbf{p}) \mathcal O_2^j(-\mathbf{p}) \rangle$ of two primary fields, each of spin-1, here defined as $i$ and $j$. In this case, if we consider the correlator of two primary fields each of spin-1, the equations take the form
\begin{equation}
\label{ConformalEqMomTwoPoint}
\begin{split}
& \left( - p_{\mu} \, \frac{\partial}{\partial p_{\mu}}  + \Delta_1 + \Delta_2 - d \right) G^{ij}(\mathbf{p}) = 0 \,, \\
& \left(  p_{\mu} \, \frac{\partial^2}{\partial p^{\nu} \partial p_{\nu}}  - 2 \, p_{\nu} \, \frac{\partial^2}{ \partial p^{\mu} 
	\partial p_{\nu} }    + 2 (\Delta_1 - d) \frac{\partial}{\partial p^{\mu}} \right) G^{ij}(\mathbf{p}) + 2 (\Sigma_{\mu\nu})^{i}_{k} \frac{\partial}{\partial 
	p_{\nu}}   G^{kj}(\mathbf{p})  = 0 \,,
\end{split}
\end{equation}
For the 2-point function $G_S(\mathbf{p})$ of two scalar quasi primary fields, the invariance under the Poincar\'e group implies that the function $G_S$ depends on the scalar invariant $p^2$ and then $G_S(\mathbf{p})=G_S(p^2)$. Furthermore, the invariance
under scale transformations implies that $G_S(p^2)$ is a homogeneous function of degree 
$\alpha = \frac{1}{2}(\Delta_1 + \Delta_2 - d)$. It is easy to show that one of the two equations in (\ref{ConformalEqMomTwoPoint}) can be satisfied only if $\Delta_1 = \Delta_2$. 
Therefore conformal symmetry fixes the structure of the scalar two-point function up to an arbitrary overall constant $C$ as
\begin{equation}
\label{TwoPointScalar}
G_S(p^2) = \langle \mathcal O_1(\mathbf{p}) \mathcal O_2(-\mathbf{p}) \rangle = \delta_{\Delta_1 \Delta_2}  \, C\, (p^2)^{\Delta_1 - d/2} \, .
\end{equation}
If we redefine
\begin{equation}
C=c_{S 12} \,  \frac{\pi^{d/2}}{4^{\Delta_1 - d/2}} \frac{\Gamma(d/2 - \Delta_1)}{\Gamma(\Delta_1)} 
\end{equation}
in terms of the new integration constant $c_{S 12}$, the two-point function reads as
\begin{equation}
\label{TwoPointScalar2}
G_S(p^2) =  \delta_{\Delta_1 \Delta_2}  \, c_{S 12} \,  \frac{\pi^{d/2}}{4^{\Delta_1 - d/2}} \frac{\Gamma(d/2 - \Delta_1)}{\Gamma(\Delta_1)} 
(p^2)^{\Delta_1 - d/2} \,,
\end{equation}
and after a Fourier transform in coordinate space takes the familiar form
\begin{equation}
\langle \mathcal O_1(\mathbf{x}_1) \mathcal O_2(\mathbf{x}_2) \rangle \equiv \mathcal{F.T.}\left[ G_S(p^2) \right] =  \delta_{\Delta_1 \Delta_2} \,  c_{S 12} 
\frac{1}{x_{12}^{2\Delta_1}} \,,
\end{equation}
where $x_{12} = |\mathbf{x}_1 - \mathbf{x}_2|$. 

\subsection{The hypergeometric structure from 3-point functions \texorpdfstring{$F_4$}{F4}} 

In the case of a scalar correlator of 3-point functions, all the conformal WI's can 
be re-expressed in scalar form by taking as independent momenta the magnitude $ {p}_i=|\mathbf{p}_i|=\sqrt{\mathbf{p}_i^2}$. In fact, Lorentz invariance on the correlation function implies that 
\begin{equation}
\Phi(\mathbf{p}_1,\mathbf{p}_2,\bar{\mathbf{p}}_3)=\Phi(p_1,p_2,p_3),\notag
\end{equation}
i.e., it is a function which depends on the magnitude of the momenta $p_i$, $i=1,2,3$. 
In this case $\mathbf{p}_3$ is taken as the dependent momentum ($\bar{\mathbf{p}}_3=-\mathbf{p}_1-\mathbf{p}_2$) by momentum conservation, with $p_3=|\mathbf{p}_1+\mathbf{p}_2|$. The original equations, in the covariant version, take the form 
\begin{equation}
K^\kappa(p_i)\Phi(\mathbf{p}_1,\mathbf{p}_2,\bar{\mathbf{p}}_3)\equiv\sum_{j=1}^{2}\left(2(\Delta_j- d)\frac{\partial}{\partial p_j^\kappa}+p_j^\kappa \frac{\partial^2}{\partial p_j^\alpha\partial p_j^\alpha} -2 p_j^\alpha\frac{\partial^2}{\partial p_j^\kappa \partial p_j^\alpha}\right)\Phi(\mathbf{p}_1,\mathbf{p}_2,\bar{\mathbf{p}}_3)=0,\label{Scw}
\end{equation}
for the special conformal WI and

\begin{equation}
D(p_i)\Phi(\mathbf{p}_1,\mathbf{p}_2,\bar{\mathbf{p}}_3)\equiv\left(\sum_{i=1}^{2}  p_i^\alpha \frac{\partial}{\partial p_i^\alpha} + \Delta'\right)
\Phi(\mathbf{p}_1,\mathbf{p}_2,\bar{\mathbf{p}}_3)
\end{equation}
for the dilatation WI. In this case $K^\kappa(p_i)$ doesn't involve the spin part $\Sigma$, as illustrated in the general expression \eqref{GenFormSCWI}, because of the scalar nature of this particular correlation function. For this reason, the action of $K^\kappa$ is purely scalar  
$K^\kappa(p_i)\equiv K_{scalar}^\kappa(p_i)$
Using the chain rule
\begin{equation}
\label{chainr}
\frac{\partial \Phi}{\partial p_i^\mu}=\frac{p_i^\mu}{  p_i}\frac{\partial\Phi}{\partial  p_i} 
-\frac{\bar{p}_3^\mu}{  p_3}\frac{\partial\Phi}{\partial   p_3}\qquad i=1,2,
\end{equation}
and the properties of the scalar products
\begin{align}
\mathbf{p}_1\cdot \mathbf{p}_2&=\frac{1}{2}\left[p_3^2-p_1^2-p_2^2\right]\notag\\
\mathbf{p}_i\cdot \mathbf{p}_3&=\frac{1}{2}\left[p_j^2-p_3^2-p_i^2\right], \quad i\ne j,\  i,j=1,2\ ,\notag
\end{align}
one can re-express the differential operator for the dilatation WI as
\begin{equation}
p_1^\alpha \frac{\partial \Phi}{{p_1}^\alpha}+ p_2^\alpha \frac{\partial \Phi}{{p_2}^\alpha}=
{p}_1\frac{ \partial \Phi}{\partial   p_1} +   p_2\frac{ \partial \Phi}{\partial   p_2} +   p_3\frac{ \partial \Phi}{\partial   p_3},
\end{equation}
giving the equation 
\begin{equation}
\label{scale1}
\left(\sum_{i=1}^3\Delta_i -2 d - \sum_{i=1}^3    p_i \frac{ \partial}{\partial   p_i}\right)\Phi(p_1,p_2,p_3)=0.
\end{equation}
One can show that the special conformal transformations, summarised in \eqref{Scw},  
take the form
\begin{equation}
\sum_{i=1}^3 p_i^\kappa \left(\,{K}_i\, \Phi(p_1,p_2,p_3)\right)=0,
\label{kappa2}
\end{equation}
having introduced the operators
\begin{equation}
{ K}_i\equiv \frac{\partial^2}{\partial    p_i \partial    p_i} 
+\frac{d + 1 - 2 \Delta_i}{   p_i}\frac{\partial}{\partial   p_i}.
\end{equation}
It is easy to show that Eq. (\ref{kappa2}) can be split into the two independent equations
\begin{equation}
\frac{\partial^2\Phi}{\partial   p_i\partial   p_i}+
\frac{1}{  p_i}\frac{\partial\Phi}{\partial  p_i}(d+1-2 \Delta_1)-
\frac{\partial^2\Phi}{\partial   p_3\partial   p_3} -
\frac{1}{  p_3}\frac{\partial\Phi}{\partial  p_3}(d +1 -2 \Delta_3)=0\qquad i=1,2,
\label{3k1}
\end{equation}
having used the momentum conservation equation $p_3^\kappa=-p_1^\kappa-p_2^\kappa$. \\
By defining 
\begin{equation}
\label{kij}
K_{ij}\equiv {K}_i-{K}_j,
\end{equation}
Eqs. (\ref{3k1}) take the form 
\begin{equation}
\label{3k2}
K_{13}\,\Phi(p_1,p_2,p_3)=0 \qquad \textrm{and} \qquad K_{23}\,\Phi(p_1,p_2,p_3)=0,
\end{equation}
which are equivalent to a hypergeometric system of equations, with solutions given by linear combinations of Appell's functions $F_4$.

\section{The chiral anomaly interaction derived nonperturbatively}
\label{conf3}
In the case of the $AVV$, the invariance of the correlator with respect to the dilatations is encoded in the following equation
\begin{equation}
\begin{aligned}
\left(\sum_{i=1}^3\Delta_i-2d-\sum_{i=1}^2\,p_i^\mu\frac{\partial}{\partial p_i^\mu}\right)\braket{J^{\mu_1}(p_1)J^{\mu_2}(p_2) J^{\mu_3}_5 (p_3)}=0.\label{Diltt}
\end{aligned}
\end{equation}
while the special CWIs are given by
\begin{equation}
\begin{aligned}\label{eq:scwiAVV}
0=&\sum_{j=1}^2\left[-2\frac{\partial}{\partial p_{j\kappa}}-2p_j^\alpha\frac{\partial^2}{\partial p_j^\alpha\,\partial p_{j\kappa}}+p_j^\kappa\frac{\partial^2}{\partial p_j^\alpha\,\partial p_{j\alpha}}\right]\braket{J^{\mu_1}(p_1)J^{\mu_2}(p_2)J_5^{\mu_3}(p_3)}\\
&+2\left(\delta^{\mu_1\kappa}\,\frac{\partial}{\partial p_1^{\alpha_1}}-\delta^\kappa_{\alpha_1}\,\frac{\partial}{\partial p_{1\mu_1}}\right)\braket{J^{\alpha_1}(p_1)J^{\mu_2}(p_2)J_5^{\mu_3}(p_3)}\\
&+2\left(\delta^{\mu_2\kappa}\,\frac{\partial}{\partial p_2^{\alpha_2}}-\delta^\kappa_{\alpha_2}\,\frac{\partial}{\partial p_{2\mu_2}}\right)\braket{J^{\mu_1}(p_1)J^{\alpha_2}(p_2)J_5^{\mu_3}(p_3)}\equiv \mathcal{K}^\kappa\braket{J^{\mu_1}(p_1)J^{\mu_2}(p_2)J_5^{\mu_3}(p_3)}.
\end{aligned}
\end{equation}
The analysis of the conformal constraints for $\braket{JJJ_5}$, as already mentioned, is performed by applying the L/T decomposition to the 
correlator. 
 We focus our analysis on the $d=4$ case, where the conformal dimensions of the conserved currents $J^\mu$ are $\Delta=3$ and the 
 tensorial structures of the correlator will involve the antisymmetric tensor in four dimensions $\epsilon^{\mu\nu\alpha\beta}$.  The procedure to obtain the general structure of the correlator starts from the conservation Ward identities
\begin{align}\label{eq:CWIJJJ}
\nabla_\mu \braket{J^\mu}=0, \qquad\qquad \nabla_\mu \braket{J^\mu_5}=a_1 \, \varepsilon^{\mu\nu\rho\sigma}F_{\mu\nu}F_{\rho\sigma}
\end{align}
of the expectation value of the non anomalous $J^{\mu}$ and anomalous $J_5^{\mu}$ currents. The vector currents are coupled to the 
vector source $A_{\mu}$ and the axial-vector current to the source $B_\mu$.
Applying multiple functional derivatives to \eqref{eq:CWIJJJ} with respect to the source $A_{\mu}$, after a Fourier transform, we find the conservation Ward identities related to the entire correlator which are given by
\begin{equation}\label{eq:jjjconsward3p}
\begin{aligned}
&p_{i\mu_i}\,\braket{J^{\mu_1}(p_1)J^{\mu_2}(p_2)J_5^{\mu_3}(p_3)}=0,\quad \quad i=1,2\\[1.2ex]
&p_{3\mu_3}\,\braket{J^{\mu_1}(p_1)J^{\mu_2}(p_2)J_5^{\mu_3}(p_3)}=-8 \, a_1 \, i \, \varepsilon^{p_1p_2\mu_1\mu_2}
\end{aligned}
\end{equation}
From this relations we construct the general form of the correlator, splitting the operators into 
a transverse and a longitudinal part as
\begin{equation}
\begin{aligned}
\label{ex}
&J^{\mu}(p)=j^\mu(p)+j_{loc}^\mu(p),\\
&j^{\mu}=\pi^{\mu}_{\alpha}(p)\,J^{\alpha}(p),\quad \pi^{\mu}_\alpha(p)\equiv \delta^{\mu}_\alpha-\frac{p_\alpha\,p^\mu}{p^2},\\
&j_{loc}^{\mu}(p)=\frac{p^\mu}{p^2}\,p\cdot J(p)
\end{aligned}
\end{equation}
and
\begin{equation}
\begin{aligned}
\label{ex}
&J_5^{\mu}(p)=j_5^\mu(p)+j_{5 loc}^\mu(p),\\
&j_5^{\mu}=\pi^{\mu}_{\alpha}(p)\,J_5^{\alpha}(p),\\
&j_{ 5 loc}^{\mu}(p)=\frac{p^\mu}{p^2}\,p\cdot J_5(p)
\end{aligned}
\end{equation}
Due to \eqref{eq:jjjconsward3p}, the correlator is purely transverse in the vector currents. We then have the following decomposition
\begin{equation}
	\braket{ J^{\mu_1 }(p_1) J^{\mu_2 } (p_2)J_5^{\mu_3}(p_3)}=\braket{ j^{\mu_1 }(p_1) j^{\mu_2 }(p_2) j_5^{\mu_3}(p_3)}+\braket{J^{\mu_1 }(p_1) J^{\mu_2 }(p_2)\, j_{5 \text { loc }}^{\mu_3}(p_3)}\label{decomp}
\end{equation}
where the first term is completely transverse with respect to the momenta $p_{i\mu_i}$, $i=1,2,3$ and the second term is the longitudinal part that is proper of the anomaly contribution. Using the anomaly constraint on $j_{5 loc}$ we obtain
\begin{align}
\braket{J^{\mu_1 }(p_1) J^{\mu_2 }(p_2)\, j_{5 \text { loc }}^{\mu_3}(p_3)}=\frac{p_3^{\mu_3}}{p_3^2}\,p_{3\,\alpha_3}\,\braket{J^{\mu_1}(p_1)J^{\mu_2}(p_2)J_5^{\alpha_3}(p_3)}=-\frac{8\,a\,i}{p_3^2}\varepsilon^{p_1p_2\mu_1\mu_2}\,p_3^{\mu_3}
\end{align}
The general structure of the transverse part can instead be parametrized in the following way 
\begin{align}
	\braket{j^{\mu_1}(p_1)j^{\mu_2}(p_2) j^{\mu_3}_5 (p_3)} &=\pi^{\mu_1}_{\alpha_1}(p_1)
	\pi^{\mu_2}_{\alpha_2} (p_2) \pi^{\mu_3}_{\alpha_3}
	\left(p_3\right)\Big[ 	A_1(p_1,p_2,p_3)\,\varepsilon^{p_1p_2\alpha_1\alpha_2}p_1^{\alpha_3} \notag\\
	& \qquad+ 
	A_2(p_1,p_2,p_3)\, \varepsilon^{p_1 \alpha_1\alpha_2\alpha_3}  -
	A_2(p_2,p_1,p_3)\, \varepsilon^{p_2\alpha_1\alpha_2\alpha_3}  
	\Big]\label{decompFin}
\end{align}
where $	A_1(p_1,p_2,p_3)=-A_1(p_2,p_1,p_3)$. The form factors $A_1$ and $A_2$ can be completely fixed by imposing the conformal invariance on the correlator encoded in the eqs$.$ \eqref{Diltt} and \eqref{eq:scwiAVV}. The solution is found in terms of special $3K$ integrals, which are parametric integrals of three Bessel functions, that can be mapped into ordinary perturbative Feynman integrals.
Explicitly we have
\begin{equation}\label{bb1}
	\left\langle j^{\mu_1}\left(p_1\right) j^{\mu_2}\left(p_2\right) j_5^{\mu_3}\left(p_3\right)\right\rangle=8 i a_1 \pi_{\alpha_1}^{\mu_1}\left(p_1\right) \pi_{\alpha_2}^{\mu_2}\left(p_2\right) \pi_{\alpha_3}^{\mu_3}\left(p_3\right)\left[p_2^2 I_{3\{1,0,1\}} \varepsilon^{p_1 \alpha_1 \alpha_2 \alpha_3}-p_1^2 I_{3\{0,1,1\}} \varepsilon^{p_2 \alpha_1 \alpha_2 \alpha_3}\right]
\end{equation}
Notice how both the longitudinal and transverse sectors are proprotional to the same factor $a_1$, which is the residue at the anomaly pole in the longitudinal sector.   \\
One can show that the reconstruction by the pole plus the CWIs, is completely equivalent to the perturbative expression where $a_1$ is the ordinary value of the anomaly for a given type of fermion running in the loop.
\\
The construction of the entire correlator proceeds from the anomaly pole, which plays a fundamental role in any anomaly diagram. This serves as a pivot in the procedure, facilitating the straightforward resolution of the longitudinal anomalous Ward Identity (WI).\\
The tensorial expansion of a chiral vertex is not unique, owing to the presence of Schouten relations among its tensor components. For example, an anomaly pole in the virtuality of the axial-vector current (denoted as $1/p_3^2$ in our notation) can be introduced or removed from a given tensorial decomposition simply by utilizing these relations.\\
For gauge anomalies, the cancellation of the anomaly poles is entirely tied to the particle content of the theory and delineates the conditions for eliminating such massless interactions. The total residue at the pole thus identifies the total anomaly of a specific fermion multiplet.\\
A similar behavior is observed for conformal correlators with stress-energy tensors, where the residue at the pole aligns with the $\beta$-function of the Lagrangian field theory. This value is determined by the number of massless degrees of freedom included in the corresponding anomaly vertex, at the scale where the perturbative prescription holds \cite{Coriano:2014gja}.\\
Our parametrization can also be mapped to the Rosenberg one, using the following eqs$.$
\begin{equation}
	\begin{aligned}
		&A_1=B_3-B_6\\
		&A_2=p_2^2(B_6  + B_4).
	\end{aligned}
\end{equation}
Further details of this analysis can be found in \cite{Coriano:2023hts}. Therefore, CFT and a pole in the longitudnal axial-vector WI are sufficient to determine the entire vertex, without any reference to a Lagrangian realization of the interaction. A topological material exhibiting an anomaly, is therefore necessarily subjected to the same contraints. 

\section{The gravitational chiral anomaly and Luttinger's relation} 
\label{conf4}
Certain quantum anomalies, such as conformal and mixed axial-gravitational anomalies, may manifest themselves in curved spacetimes due to their involvement with the energy-momentum tensor and, consequently, the metric tensor. In condensed matter systems, these gravitational anomalies can be investigated in an off-equilibrium regime using the Luttinger theory of thermal transport coefficients \cite{Luttinger:1964zz,Stone:2012ud}, which has been applied, for instance, in studies of the thermal effects of the axial-gravitational anomaly.
\\
The fundamental concept is that the influence of a temperature gradient ${\boldsymbol{\nabla}} T$, which impels a system out of equilibrium, can be counteracted, at linear order, by a non-uniform gravitational potential $\Phi$:
\begin{equation}
\frac{1}{T} {\boldsymbol{\nabla}} T = - \frac{1}{c^2} {\boldsymbol{\nabla}}  \Phi\,,
\label{eq:Luttinger}
\end{equation}
where $c$ represents the speed of light. In the regime of a weak gravitational field (in the Newtonian limit), the gravitational potential $\Phi$ is defined as:
\begin{equation}
g_{00} = 1 + \frac{2 \Phi}{c^2}\,,
\label{eq:g00}
\end{equation}
which is linked to the $g_{00}$ component of the metric, while other components of the metric tensor remain unaltered. This observation is closely associated with the Tolman--Ehrenfest effect \cite{Tolman:1930ona}, which asserts that in a stationary gravitational field, the local temperature of a system at thermal equilibrium varies spatially. The temperature is spatially dependent according to the formula:
\begin{equation}
T(x) = \frac{T_0}{\sqrt{g_{00}(x)}},
\label{eq:TE}
\end{equation}
where $T_0$ denotes a reference temperature at a chosen point where $g_{00}=1$. The Luttinger formula \eqref{eq:Luttinger} can be deduced from simple thermodynamic considerations (see, for example, Ref.~\cite{Rovelli:2010mv}).

\subsection{Conservation and trace Ward identities}
The reconstruction of the correlator is performed using the anomaly pole present in the decomposition of this correlator in its anomaly sector as a pivot. In the case of $\langle JJJ_5\rangle$ or $\langle J_5 J_5 J_5\rangle$, we satisfy the anomaly constraint with one or three anomaly poles respectively, and all the remaining sectors are fixed by this choice. The general character of this reconstruction procedure, based on the inclusion of the anomaly constraint, will also be proven in the case of $\langle TT J_5\rangle$. In all these three cases, our analysis shows that the anomaly phenomenon, at least in the parity-odd case, is entirely associated with the presence of an anomaly pole. Once the coefficient in front of the anomaly is determined, then the entire correlator is fixed, if we impose the conformal symmetry. In $\langle TTJ_5\rangle$, if we include both a gauge field and a general metric background in the background, the anomalous WI that we will use for the definition of the correlator in CFT is given by the equation:
\begin{equation} \label{eq:anomaliachirale}
    \nabla_\mu\langle J_5^\mu\rangle =
    a_1\, \varepsilon^{\mu \nu \rho \sigma}F_{\mu\nu}F_{\rho\sigma}+ a_2 \, \varepsilon^{\mu \nu \rho \sigma}R^{\alpha\beta}_{\hspace{0.3cm} \mu \nu} R_{\alpha\beta \rho \sigma},
\end{equation}
which defines a boundary condition for the CWIs.
These results can also be extended to the non-abelian case. For a general chiral current $J^\mu_i$, we can write:
\begin{equation}
    \nabla_\mu \langle J_i^\mu \rangle = a_1\,D_{ijk} \, \varepsilon^{\mu \nu \rho \sigma} F^j_{\mu \nu} F^k_{\rho \sigma} + a_2 \,D_i \,\varepsilon^{\mu \nu \rho \sigma}R^{\alpha\beta}_{\hspace{0.3cm} \mu \nu} R_{\alpha\beta \rho \sigma},
\end{equation}
where we have introduced the anomaly tensors:
\begin{equation}
    D_{ijk} =\frac{1}{2} \, \text{Tr}[\{T_i,\,T_j\}\,T_k],\quad \qquad D_i=\text{Tr}[T_i],
\end{equation}
constructed with the non-abelian generators of the theory. In the case of the $D_i$'s, for example, in the Standard Model, where the symmetry is $SU(3)\times SU(2)\times U(1)_Y$, only the hypercharge $(U(1)_Y)$ contribution $\langle TTJ_Y\rangle$ is taken into account, since the $SU(2)$ and $SU(3)$ generators are traceless.
\\
Both the chiral $(F\tilde{F})$ and gravitational $(R\tilde{R})$ anomalies cancel once we sum over each generation of chiral fermions.
The cancellation of the gravitational anomaly in the Standard Model can be interpreted in two possible ways. On one hand, it shows the consistency of the coupling of the Standard Model to gravity, since the gauge currents are conserved in a gravitational background. On the other hand, the stress-energy tensor is just another operator of the Standard Model, and the conservation of the currents is required for the analysis of the mixing of such operators with the gauge currents at the perturbative level.\\
By applying two functional derivative with respect to the metric on eq$.$ \eqref{eq:anomaliachirale} and then Fourier transforming, one can determine the following anomalous constraint on the $TTJ_5$ correlator
\begin{equation}\label{eq:wigauge3}
	p_{3\mu_3}\braket{T^{\mu_1\nu_1}(p_1)T^{\mu_2\nu_2}(p_2)J_5^{\mu_3}(p_3)}= 4\, i \, a_2 \, (p_1 \cdot p_2) \left\{ \left[\varepsilon^{\nu_1 \nu_2 p_1 p_2}\left(g^{\mu_1 \mu_2}- \frac{p_1^{\mu_2} p_2^{\mu_1}}{p_1 \cdot p_2}\right) +\left( \mu_1 \leftrightarrow \nu_1 \right) \right] +\left( \mu_2 \leftrightarrow \nu_2 \right) \right\}.
\end{equation}
\normalsize
Such condition will be satisfied by the inclusion of an anomaly pole in the correlator.\\
We also need to require diffeomorphism and Weyl invariance which is encoded in the following equations
\begin{equation} \label{eq:WIdiffWeyl}
	\begin{aligned}
			&0=\nabla^{\mu}\left\langle T_{\mu \nu}\right\rangle-{F_B}_{ \n \m} \left\langle J_5^{\mu }\right\rangle+B_{\n}\nabla_{\mu}\left\langle J_5^{\mu }\right\rangle.\\
			&0=g_{\mu\nu}\langle T^{\mu\nu}\rangle
	\end{aligned}
\end{equation}
Note that, in this case, we can ignore the conformal anomaly contribution to the last equation since it does not affect our correlator.
The equations above in momentum space lead to the following constraints on the $TTJ_5$
\begin{equation} 
	\begin{aligned}
		&p_{i\mu_i}\,\braket{T^{\mu_1\nu_1}(p_1)T^{\mu_2\nu_2}(p_2)J_5^{\mu_3}(p_3)}=0,\qquad \quad && i=\{1,2\}.\\
		&\delta_{\mu_i\nu_i }\braket{T^{\mu_1\nu_1}(p_1)T^{\mu_2\nu_2}(p_2)J_5^{\mu_3}(p_3)}=0,\qquad \quad&& i=\{1,2\}\\
	\end{aligned}
\end{equation}

\subsection{The nonperturbative derivation}
In this section, first, we derive the most general expression for the $\langle TTJ_5\rangle$ that satisfies the (anomalous) conservation and trace Ward identities. We then proceed to fix the correlator by imposing invariance under the conformal group.
The analysis is performed by applying the L/T decomposition to the $\langle TTJ_5\rangle$.
We focus on a parity odd four-dimensional correlator, therefore its tensorial structure will involve the antisymmetric tensor $\varepsilon^{\mu\nu\rho\sigma}$.\\
We start by decomposing the energy-momentum tensor $T^{\m \n}$ and the current $J_5^\m$ in terms of their transverse-traceless part and longitudinal ones (also called "local")
\begin{align}
	T^{\mu_i\nu_i}(p_i)&= t^{\mu_i\nu_i}(p_i)+t_{loc}^{\mu_i\nu_i}(p_i),\label{decT}\\
	J_5^{\mu_i}(p_i)&= j_5^{\mu_i}(p_i)+j_{5 \, loc}^{\mu_i}(p_i),\label{decJ}
\end{align}
where
\begin{align}
	\label{loct}
	&t^{\mu_i\nu_i}(p_i)=\Pi^{\mu_i\nu_i}_{\alpha_i\beta_i}(p_i)\,T^{\alpha_i \beta_i}(p_i), &&t_{loc}^{\mu_i\nu_i}(p_i)=\Sigma^{\mu_i\nu_i}_{\alpha_i\beta_i}(p)\,T^{\alpha_i \beta_i}(p_i),\nn\\
	&j_5^{\mu_i}(p_i)=\pi^{\mu_i}_{\alpha_i}(p_i)\,J_5^{\alpha_i }(p_i), \hspace{1ex}&&j_{5\, loc}^{\mu_i}(p_i)=\frac{p_i^{\mu_i}\,p_{i\,\alpha_i}}{p_i^2}\,J_5^{\alpha_i}(p_i),
\end{align}
having introduced the transverse-traceless ($\Pi$), transverse $(\pi)$ and longitudinal ($\Sigma$) projectors, given respectively by 
\begin{align}
	\label{prozero}
	&\pi^{\mu}_{\alpha}  = \delta^{\mu}_{\alpha} - \frac{p^{\mu} p_{\alpha}}{p^2}, \\&
	%%%%%%%%%%%%%%%%%%%%%%%%%%%%%%%
	\Pi^{\mu \nu}_{\alpha \beta}  = \frac{1}{2} \left( \pi^{\mu}_{\alpha} \pi^{\nu}_{\beta} + \pi^{\mu}_{\beta} \pi^{\nu}_{\alpha} \right) - \frac{1}{d - 1} \pi^{\mu \nu}\pi_{\alpha \beta}\label{TTproj}, \\&
	%%%%%%%%%%%%%%%%%%%%%%%%%%%%%%%
	\Sigma^{\mu_i\nu_i}_{\alpha_i\beta_i}=\frac{p_{i\,\beta_i}}{p_i^2}\Big[2\delta^{(\nu_i}_{\alpha_i}p_i^{\mu_i)}-\frac{p_{i\alpha_i}}{(d-1)}\left(\delta^{\mu_i\nu_i}+(d-2)\frac{p_i^{\mu_i}p_i^{\nu_i}}{p_i^2}\right)\Big]+\frac{\pi^{\mu_i\nu_i}(p_i)}{(d-1)}\delta_{\alpha_i\beta_i}\label{Lproj}.
	%%%%%%%%%%%%%%%%%%%%%%%%%%%%%%%
\end{align}
Such decomposition allows to split our correlation function into the following terms
\begin{equation} \label{eq:splitlongttpart}
	\begin{aligned}
		\left\langle T^{\mu_{1} \n_{1}} T^{\mu_{2} \n_2} J_5^{\mu_{3}}\right\rangle=&\left\langle t^{\mu_{1} \n_{1}} t^{\mu_{2}\n_2} j_5^{\mu_{3}}\right\rangle+\left\langle T^{\mu_{1} \n_{1}} T^{\mu_{2}\n_2} j_{5\, l o c}^{\mu_{3}}\right\rangle+\left\langle T^{\mu_{1} \n_{1}} t_{l o c}^{\mu_{2}\n_2} J_5^{\mu_{3}}\right\rangle+\left\langle t_{l o c}^{\mu_{1} \n_{1}} T^{\mu_{2}\n_2} J_5^{\mu_{3}}\right\rangle \\
		&-\left\langle T^{\mu_{1} \n_{1}} t_{l o c}^{\mu_{2}\n_2} j_{5\, l o c}^{\mu_{3}}\right\rangle-\left\langle t_{l o c}^{\mu_{1} \n_{1}} t_{l o c}^{\mu_{2}\n_2} J_5^{\mu_{3}}\right\rangle-\left\langle t_{l o c}^{\mu_{1} \n_{1}} T^{\mu_{2}\n_2} j_{5\, l o c}^{\mu_{3}}\right\rangle+\left\langle t_{l o c}^{\mu_{1} \n_{1}} t_{l o c}^{\mu_{2}\n_2} j_{5\, l o c}^{\mu_{3}}\right\rangle .
	\end{aligned}
\end{equation}
Using the conservation and trace WIs derived in the previous section, it is then possible to completely fix all the longitudinal parts, i.e. the terms containing at least one $t^{\m \n}_{{loc}}$ or $j^\m_{5\, {loc}}$. We start by considering the non-anomalous equations
\begin{equation} \label{ppp}
	\begin{aligned}
		&\delta_{\mu_i\nu_i }\braket{T^{\mu_1\nu_1}(p_1)T^{\mu_2\nu_2}(p_2)J_5^{\mu_3}(p_3)}=0,\qquad \quad&& i=\{1,2\}\\
		&p_{i\mu_i}\,\braket{T^{\mu_1\nu_1}(p_1)T^{\mu_2\nu_2}(p_2)J_5^{\mu_3}(p_3)}=0,\qquad \quad && i=\{1,2\}.
	\end{aligned}
\end{equation}
Thanks to these WIs, we can eliminate most of terms on the right-hand side of equation \eqref{eq:splitlongttpart}, ending up only with two terms
\begin{equation}
	\left\langle T^{\mu_{1} \n_{1}} T^{\mu_{2} \n_2} J_5^{\mu_{3}}\right\rangle=\left\langle t^{\mu_{1} \n_{1}} t^{\mu_{2}\n_2} j_5^{\mu_{3}}\right\rangle+\left\langle t^{\mu_{1} \n_{1}} t^{\mu_{2}\n_2} j_{5\, l o c}^{\mu_{3}}\right\rangle.
\end{equation}
The remaining local term is then fixed by the anomalous WI of $J_5$.
First, we construct the most general expression in terms of tensorial structures and form factors
\beq \label{eq:locpartdecompgen}
\left\langle t^{\mu_{1} \n_{1}} t^{\mu_{2}\n_2} j_{5\, l o c}^{\mu_{3}}\right\rangle=
p_3^{\m_3}\, \Pi^{\m_1 \n_1}_{\a_1  \b_1}(p_1) \, \Pi^{\m_2 \n_2}_{\a_2 \b_2}(p_2) \,  \varepsilon^{\a_1 \a_2 p_1 p_2}\left( F_1 \, g^{\b_1 \b_2} + F_2 \, p_1^{\b_2} p_2^{\b_1} \right)
\eeq
where, due to the Bose symmetry, both $F_1$ and $F_2$ are symmetric under the exchange $\left(p_1\leftrightarrow p_2\right)$. Then, recalling the definition of $j_{5\, l o c}$ and the anomalous WI 
\small
\begin{equation}\label{eq:idwanomlp3}
	p_{3\mu_3}\braket{T^{\mu_1\nu_1}(p_1)T^{\mu_2\nu_2}(p_2)J_5^{\mu_3}(p_3)}= 4\, i \, a_2 \, (p_1 \cdot p_2) \left\{ \left[\varepsilon^{\nu_1 \nu_2 p_1 p_2}\left(g^{\mu_1 \mu_2}- \frac{p_1^{\mu_2} p_2^{\mu_1}}{p_1 \cdot p_2}\right) +\left( \mu_1 \leftrightarrow \nu_1 \right) \right] +\left( \mu_2 \leftrightarrow \nu_2 \right) \right\},
\end{equation}
\normalsize
we can write
\begin{equation} \label{eq:anompolettj}
	\left\langle t^{\mu_{1} \n_{1}} t^{\mu_{2}\n_2} j_{5\, l o c}^{\mu_{3}}\right\rangle= 4 ia_2 \frac{p_3^{\mu_3}}{p_3^2} \, (p_1 \cdot p_2) \left\{ \left[\varepsilon^{\nu_1 \nu_2 p_1 p_2}\left(g^{\mu_1 \mu_2}- \frac{p_1^{\mu_2} p_2^{\mu_1}}{p_1 \cdot p_2}\right) +\left( \mu_1 \leftrightarrow \nu_1 \right) \right] +\left( \mu_2 \leftrightarrow \nu_2 \right) \right\}.
\end{equation}
One can show that this formula coincides with eq$.$ \eqref{eq:locpartdecompgen} after contracting the projectors' indices and fixing the form factors in the following way
\begin{equation}
	\begin{aligned}
		&F_1=\frac{16 i a_2 (p_1\cdot p_2)}{p_3^2}, \qquad\qquad\qquad
		F_2=-\frac{16 i a_2}{p_3^2}.
	\end{aligned}
\end{equation}
Therefore, all the local terms of the $\langle TTJ_5\rangle$ are fixed. The only remaining term to be studied in order to reconstruct the entire correlator is the transverse-traceless part $\left\langle t^{\mu_{1} \n_{1}} t^{\mu_{2}\n_2} j_5^{\mu_{3}}\right\rangle$ that can be expressed as
\begin{equation} \label{eq:defX}
	\left\langle t^{\mu_1 \n_1}\left(p_1\right) t^{\mu_2 \n_2}\left(p_2\right) j_5^{\mu_3}\left(p_3\right)\right\rangle=\Pi_{\alpha_1 \beta_1}^{\mu_1 \nu_1}\left(p_1\right) \Pi_{\alpha_2 \b_2}^{\mu_2 \n_2}\left(p_2\right) \pi_{\alpha_3}^{\mu_3}\left(p_3\right) X^{\alpha_1 \beta_1 \alpha_2 \b_2 \alpha_3}
\end{equation}
where $X^{\alpha_1 \beta_1 \alpha_2 \b_2 \alpha_3}$ is a general rank five tensor built by products of metric tensors, momenta and the Levi-Civita symbol
with the appropriate choice of indices. As a consequence of the projectors in \eqref{eq:defX}, $X^{\alpha_1 \beta_1 \alpha_2 \b_2 \alpha_3}$ can
not be constructed by using $g_{\a_i \b_i}$, nor by $p_{i \, \a_i}$ with $i =\{1,2,3\}$. 
We also must keep in mind that, due to symmetries of the correlator, form factors associated with structures linked by a $(1\leftrightarrow 2)$ transformation (the gravitons exchange) are dependent. Additionally, the number of form factors can be further reduced by considering Schouten identities. In the end, the transverse-traceless part can be written in the minimal form
\begin{equation}\label{eq:decttnm}
	\begin{gathered}
		\left\langle t^{\mu_1 \nu_1}\left(p_1\right) t^{\mu_2 \nu_2}\left(p_2\right) j_5^{\mu_3}\left(p_3\right)\right\rangle=\Pi_{\alpha_1 \beta_1}^{\mu_1 \nu_1}\left(p_1\right) \Pi_{\alpha_2 \beta_2}^{\mu_2 \nu_2}\left(p_2\right) \pi_{\alpha_3}^{\mu_3}\left(p_3\right)\bigg[ \\
		A_1 \varepsilon^{p_1 \alpha_1 \alpha_2 \alpha_3} p_2^{\beta_1} p_3^{\beta_2}-A_1\left(p_1 \leftrightarrow p_2\right) \varepsilon^{p_2 \alpha_1 \alpha_2 \alpha_3} p_2^{\beta_1} p_3^{\beta_2} \\
		+A_2 \varepsilon^{p_1 \alpha_1 \alpha_2 \alpha_3} \delta^{\beta_1 \beta_2}-A_2\left(p_1 \leftrightarrow p_2\right) \varepsilon^{p_2 \alpha_1 \alpha_2 \alpha_3} \delta^{\beta_1 \beta_2} \\
		\left.+A_3 \varepsilon^{p_1 p_2 \alpha_1 \alpha_2} p_2^{\beta_1} p_3^{\beta_2} p_1^{\alpha_3}+A_4 \varepsilon^{p_1 p_2 \alpha_1 \alpha_2} \delta^{\beta_1 \beta_2} p_1^{\alpha_3}\right.\bigg]
	\end{gathered}
\end{equation}
where $A_3$ and $A_4$ are antisymmetric under the exchange $(p_1\leftrightarrow p_2)$. 
The four form factors appearing in such expressions can be fixed by imposing the invariance of the correlator under the conformal group.
The dilatations WIs are
\begin{align}
\label{eqD}
	\left(\sum_{i=1}^3\Delta_i-2d-\sum_{i=1}^2\,p_i^\mu\frac{\partial}{\partial p_i^\mu}\right)\braket{T^{\mu_1\nu_1}(p_1)T^{\mu_2\nu_2}(p_2)J_5^{\mu_3} (p_3)}=0.
\end{align}
while the special conformal WIs take the form
\begin{equation}
	\begin{aligned}
		0=&\, \mathcal{K}^\kappa\left\langle T^{\mu_1 \nu_1}\left(p_1\right) T^{\mu_2 \nu_2}\left(p_2\right) J_5^{\mu_3 }\left(p_3\right)\right\rangle \\
		&=\sum_{j=1}^2\left(2\left(\Delta_j-d\right) \frac{\partial}{\partial p_{j \kappa}}-2 p_j^\alpha \frac{\partial}{\partial p_j^\alpha} \frac{\partial}{\partial p_{j \kappa}}+\left(p_j\right)^\kappa \frac{\partial}{\partial p_j^\alpha} \frac{\partial}{\partial p_{j \alpha}}\right)\left\langle T^{\mu_1 \nu_1}\left(p_1\right) T^{\mu_2 \nu_2}\left(p_2\right) J_5^{\mu_3}\left(p_3\right)\right\rangle \\
		&\qquad+4\left(\delta^{\kappa\left(\mu_1\right.} \frac{\partial}{\partial p_1^{\alpha_1}}-\delta_{\alpha_1}^\kappa \delta_\lambda^{\left(\mu_1\right.} \frac{\partial}{\partial p_{1 \lambda}}\right)\left\langle T^{\left.\nu_1\right) \alpha_1}\left(p_1\right) T^{\mu_2 \nu_2}\left(p_2\right) J_5^{\mu_3 }\left(p_3\right)\right\rangle \\
		&\qquad+4\left(\delta^{\kappa\left(\mu_2\right.} \frac{\partial}{\partial p_2^{\alpha_2}}-\delta_{\alpha_2}^\kappa \delta_\lambda^{\left(\mu_2\right.} \frac{\partial}{\partial p_{2 \lambda}}\right)\left\langle T^{\left.\nu_2\right) \alpha_2}\left(p_2\right) T^{\mu_1 \nu_1}\left(p_1\right) J_5^{\mu_3 }\left(p_3\right)\right\rangle .
	\end{aligned}
\end{equation}
These equations have been solved in \cite{Coriano:2023gxa}. The result is determined in terms of special $3K$ integrals, which are parametric integrals of three Bessel functions.
Explicitly we have
\begin{equation}
	\begin{aligned}
		& A_1=-4 i a_2 p_2^2 I_{5\{2,1,1\}} \\
		& A_2=-8 i a_2 p_2^2\Big(p_3^2 I_{4\{2,1,0\}}-1\Big) \\
		& A_3=0 \\
		& A_4=0
	\end{aligned}
\end{equation}
Once again, one can observe how the solution of the (anomalous) chiral and conformal WIs is sufficient to solve for the entire correlator. This shows that the inclusion of the anomaly pole is crucial for the entire vertex. 
\subsection{The perturbative $TTJ_5$}\label{metric}
\begin{figure}[t]
	\includegraphics[scale=0.5]{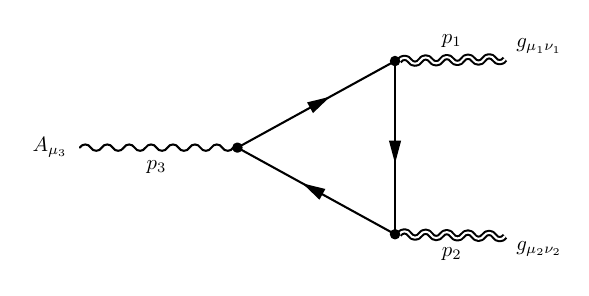}
	\includegraphics[scale=0.5]{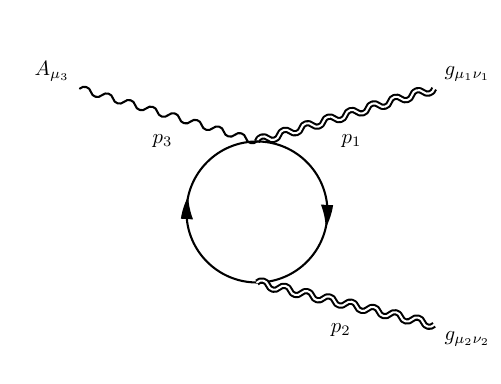}
	\includegraphics[scale=0.5]{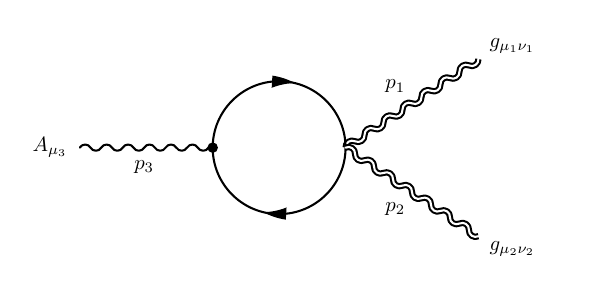}
	\includegraphics[scale=0.5]{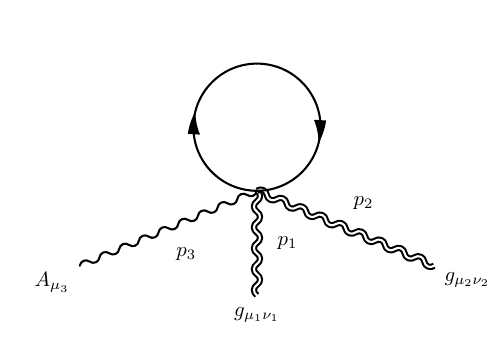}
	\caption{Feynman Diagrams of the three different topologies appearing in the perturbative expansion of the $TTJ_5$, responsible for the gravitational anomaly.} 
	\label{fig:feynmdiagr}
\end{figure}

The $\langle TTJ_5\rangle$ correlator perturbatively at one-loop, can be computed in a suitable regularization scheme. We consider the following action with a fermionic field in a gravitational and axial gauge field background
\begin{equation}
    S_0=\int d^d x \, e \left[\frac{i}{2} \bar{\psi} e_a^\mu \gamma^a\left(\partial_\mu \psi\right)-\frac{i}{2}\left(\partial_\mu \bar{\psi}\right) e_a^\mu \gamma^a \psi-g A_\mu  \bar{\psi}  e_a^\mu  \gamma^a\gamma_5\psi  +\frac{i}{4} \omega_{\mu a b} e_c^\mu \bar{\psi} \gamma^{a b c} \psi\right]
\end{equation}
with
\begin{equation}
    \gamma^{abc}=\{\Sigma^{ab},\gamma^c \}.
\end{equation}
Taking a first variation of the action with respect to the metric, one can construct the energy-momentum tensor as
\begin{equation}
    T^{\mu \nu}=-\frac{i}{2}\left[\bar{\psi} \gamma^{(\mu} \nabla^{\nu)} \psi-\nabla^{(\mu} \bar{\psi} \gamma^{\nu)} \psi-g^{\mu \nu}\left(\bar{\psi} \gamma^\lambda \nabla_\lambda \psi-\nabla_\lambda \bar{\psi} \gamma^\lambda \psi\right)\right]-g \bar{\psi}\left(g^{\mu \nu} \gamma^\lambda A_\lambda-\gamma^{(\mu} A^{\nu)}\right) \gamma_5 \, \psi.
\end{equation}
We can then proceed to evaluate the $TTJ_5$ correlator pertubatively by considering the diagrams depicted in Fig$.$ \ref{fig:feynmdiagr}. One can show that the perturbative result, expressed in terms of Feynman master integrals, is completely in agreement with the nonperturbative outcome described in the previous section (see \cite{Coriano:2023gxa} for the details).
\\
It is clear that at this stage, it is important to devise possible ways to characterize, both phenomenologically and experimentally, the methods that need to be implemented in order to gather information on the axion-like interactions that manifest in the conformal limit.
In the next section, we are going to discuss a possible method of detecting this state using polarimetry. The method is well-known in the context of axion physics, suggested by Sikivie and Harari \cite{Harari:1992ea}.
In the subsequent section, we offer an independent derivation of the equations that relate the rotation of the angle of polarization in terms of the axionic background in the eikonal approximation. We will be using axion electrodynamics in a local formulation.
Since the action is a consequence of the chiral WI, which is local, there is no information, in this analysis, on the entire structure of the interaction that, as we have shown in the previous sections, is characterized by a sum rule. We will come back to this point before our conclusions.

\section{Axion electrodynamics: the Faraday effect} 
\label{electro1}
In this section we review the derivation of the local anomaly action with the inclusion of a pseudoscalar, both in the case of a constant 
$\theta$ term and of a variable $\theta$, with $\theta\equiv \varphi/M$. Here $M$ is the scale defining the coupling of the axion field to the anomaly. We introduce also a dimensionfull coupling $\tilde{g}\equiv g/M$. 
We first discuss a well-known effect in axion physics, that is attributed to axion electrodynamics, i.e. the rotation of the plane of polarization of a light beam in the presence of a condensed axion in the background. The effect was pointed out in \cite{Harari:1992ea} and finds direct application in TIs as well. The derivation that we present is exact up to the last step, where we will resort to the eikonal approximation in order to solve the two coupled propagation equations for the effective $D$ and $H$ fields, that characterize the effective dynamics. In a subsequent \secref{leading} we point out the existence of some special features of the photon propagator in axion electrodynamics, which allow to determine over the entire light-cone surface the bi-tensor function that controls the residue of its leading singularity. We are going to show the presence of periodic oscillations in such function, by solving the recursion relations of its Hadamard expansion. The oscillations are parameterized by an angle which is proportional to the gradient of the background axion field. 
\\
We work in a rationalized system of units with ${4\pi, c\to 1}$. The Lagrangian for the coupling of an axion-like field to the $U(1)$ EM gauge field is given by
\beq 
\label{cc}
\mathcal{L}=-\frac{1}{4}F_{\mu\nu}F^{\mu\nu}+\frac{1}{2} \partial_\mu
\varphi \partial^\mu \varphi + \frac{1}{4}\widetilde{g}\,\varphi\,
F_{\mu\nu}\widetilde{F}^{\mu\nu},
\eeq
with the inclusion of a kinetic term for $\varphi$. \\
From the above Lagrangian the equations of motion for the axion field $\varphi$ are
\beq\label{motum1axion} \Box \varphi \, - \,
\frac{\widetilde{g}}{4}\,F^{\mu\nu}\widetilde{F}_{\mu\nu}=\,0.
\eeq

For the electromagnetic field, the covariant equation
\beq\label{EOM of EM}
\partial_\mu F^{\mu\nu}=J^\nu,
\eeq 
can be cast into the form  
\beq\label{coup} -\partial_\mu
F^{\mu\nu}+\widetilde{g}\,\partial_{\mu}(\widetilde{F}^{\mu\nu}\varphi)=\,0.
\eeq
The canonical gauge current is extracted from the variation
\beq
J^\nu=\frac{\delta \mathcal{L}}{\delta \partial_\mu A_\nu}\delta A_\nu \qquad \delta A_{\nu}=\partial_{\nu}\alpha(x), 
\eeq
modulo a vanishing four-divergence $(\partial_{\mu}\partial_{\nu}F^{\mu\nu}=0)$
and it is given by
\beq
 J^\nu=\widetilde{g}\,\partial_{\mu}(\widetilde{F}^{\mu\nu}\varphi) \eeq 
 which is conserved.

The two gauge invariants that we can build out of the field strength are 

\begin{align}
F^{\mu\nu}F_{\mu\nu}&=-2(\E^2-\B^2),\\
\label{-4EdotB}\widetilde{F}^{\mu\nu}F_{\mu\nu}&=-4 \E \cdot \B,
\end{align}
with 
\beq
F^{0i}=-\E^i \qquad F^{ij}=-\epsilon^{ijk}\B^k, \tilde{F}^{io}=\B^i\qquad \tilde{F}^{ij}=\epsilon^{ijk}\E^k.
\eeq
Rewritten in terms of the gauge invariant fields $\B$ and $\E$, the action 
\beq
\label{act}
\mathcal{L}=\int d^4 x \left( \frac{1}{2}\partial_\mu
\varphi \partial^\mu \varphi + \frac{1}{2}(\E^2-\B^2) - \tilde{g}\varphi \B\cdot \E\right)
\eeq 
is invariant under the duality transformation
\beq
\label{dd}
(\B,\E)\to(-\E,\B).  
\eeq
As we are going to show below, the equations describing the optical activity of the medium in the eikonal limit preserve this symmetry. It is also time reversal ($T$) invariant if the axion filed is odd under $T$. 

The Bianchi identity 
$\epsilon_{\mu\nu\alpha\beta}F^{\alpha \beta}=0$
gives the two homogenous equations 
\beq
\label{Beq}
\nabla\cdot \B=0 , 
\eeq
\beq
\label{Eeq}
\qquad \nabla\times \E=-\frac{\partial \B}{\partial t}.
\eeq
To obtain the last two equations, we separate the spatial from the temporal components of the 4-current $J^\mu$, obtaining 
\begin{align}\label{J^0}
J^0 & =\widetilde{g}\,\varphi\partial_i
\widetilde{F}^{i0}+\widetilde{g}\,\widetilde{F}^{i0}\partial_i
\varphi,\\ \label{J0} 
\end{align}
\begin{align}
J^j & =\widetilde{g}\,\varphi\partial_i
\widetilde{F}^{ij}+\widetilde{g}\,\widetilde{F}^{ij}\partial_i
\varphi,
\label{Jj}
\end{align}
explicitly given by
 \beq
 \label{JJ0}
J^0=\widetilde{g}\,\varphi \, \nabla \cdot
\B+\widetilde{g}\,\nabla\varphi \cdot \B,
\eeq 
and
\beqa
\label{JJj}
  {\bf J}=&&\widetilde{g} \,
\varphi\left(-\frac{\partial \, \B}{\partial t}-\nabla\times \E
\right)-\widetilde{g}\left(\B \frac{\partial \varphi}{\partial
t}-\E \times \nabla\varphi \right)\nonumber \\
=&& -\widetilde{g}\left(\B \frac{\partial \varphi}{\partial
t}-\E \times \nabla\varphi \right), \eeqa 
having used the Bianchi identiy \eqref{Eeq}.
The time component of the lhs of \eqref{EOM of EM}
\beq \partial_\mu F^{\mu
0}=\nabla \cdot \E, \eeq 
combined with \eqref{JJ0}
gives, after using the Bianchi identiy \eqref{Beq}, the modified Gauss law takes the form
 \beq\label{fieldeq4} \nabla \cdot \E=\widetilde{g} \,
\nabla\varphi\cdot \B.\eeq \\
The space part of \eqref{EOM of EM}
\beq
\partial_\mu F^{\mu j}=\partial_0 F^{0
j}+\partial_i F^{i j}=\partial_0 F^{0 j}+\epsilon^{i j
k}\partial_j B^k=\nabla \times \B-\frac{\partial \E}{\partial t}, \eeq 
combined with \eqref{Jj} finally gives the modified displacement law
\beq \label{fieldeq5} \nabla
\times \B-\frac{\partial \E}{\partial t}=-\widetilde{g} \B
\frac{\partial \varphi}{\partial t}+\widetilde{g} \E \times
\nabla\varphi. \eeq \vspace{1cm}
In summary, the four modified equations are the two Bianchi identities \eqref{Beq} and \eqref{Eeq}, together with 
\begin{align}
\Box\varphi\,&=\,-\widetilde{g}\,\E\cdot \B, \\
\nabla \cdot \E &=\widetilde{g} \, \nabla\varphi\cdot \B, \label{EE}\\
\nabla \times \B-\frac{\partial \E}{\partial t}&=-\widetilde{g} \B
\frac{\partial \varphi}{\partial t}+\widetilde{g} \E \times
\nabla\varphi,
\end{align}
corresponding, respectively, to the axion equation of motion, the modified Gauss law and the modified displacement law for the 
$\B$ and $\E$ fields. We can convert to the notation of \cite{Wilczek1987} by the replacement $\tilde{g}\varphi=-\kappa a$, where $a$ is the dimensionless 
axionic angle normalized as $M a =\varphi$, and $M$, as already mentioned, is the typical scale of the axionic interaction, with the three  coupled equations in the form

\begin{align}
\nabla\cdot \E =&-\kappa \nabla a \cdot \B \\
\nabla\times \B = &\frac{\partial \E}{\partial t}+\kappa\left(\dot{a} \B +\nabla a\times \E\right).
\end{align}

\subsection{Rotation of the polarization plane in the eikonal approximation}
We are interested in studying the effect of a slowly varying axionic background. We work in the geometrical optics approximation, where we neglect all the terms containing more than two derivatives on the axion field or gradient squared terms, of the form $(\nabla_i \phi)^2$ and $\nabla_i\nabla_j \phi$.
\subsection{Eikonal D}
We take the curl of the Ampere law \eqref{Eeq}
\beq
\nabla\times\,(\nabla\times\E)+\frac{\partial}{\partial
t}(\nabla\times \B)=0,\eeq in which we use the displacement law
\eqref{fieldeq5} \beq \nabla\times(\nabla\times
\E)+\frac{\partial}{\partial t}\left(\frac{\partial E}{\partial
t}+\widetilde{g} \,
\E\times\nabla\varphi-\widetilde{g}\B\frac{\partial
\varphi}{\partial t}\right).\eeq 
At this stage we use the identity
$\nabla\times(\nabla\times\E)=\nabla(\nabla\cdot\E)-(\nabla^2
\E)$, and in the eikonal approximation we obtain the relation
\beq
\nabla(\nabla\cdot\E)-(\nabla^2 \E)+\frac{\partial^2 \E}{\partial
t^2}+\widetilde{g}\frac{\partial \E}{\partial
t}\times\nabla\varphi-\widetilde{g}\frac{\partial \B}{\partial
t}\frac{\partial \varphi}{\partial t}=0, \eeq 
that is simplified using \eqref{fieldeq4}, to the form
\beq
\Box\E=-\nabla(\widetilde{g}\nabla\varphi\cdot\B)+\widetilde{g}\frac{\partial\B}{\partial
t}\frac{\partial\varphi}{\partial
t}-\widetilde{g}\frac{\partial\E}{\partial
t}\times\nabla\varphi.\eeq  A further simplification, using \eqref{fieldeq5} gives

\beqa\label{1claudiodimstep1}
\Box\E&=&-\nabla(\widetilde{g}\nabla\varphi\cdot\B)+\widetilde{g}\frac{\partial\B}{\partial
t}\frac{\partial\varphi}{\partial
t}-\widetilde{g}\left(\nabla\times\B-\left[\widetilde{g}\E\times\nabla\varphi\right] \right.
\nonumber \\
&& \left.
+\left[\widetilde{g}\B\frac{\partial\varphi}{\partial
t}\right]\right)\times\nabla\varphi. \eeqa
Terms in squared brackets $\left[\,\,\,\right]$ are neglected in the eikonal approxmation. 
Using the identity \beq
\nabla(\boldsymbol{X}\cdot\boldsymbol{Y})=(\boldsymbol{X}\cdot\nabla)\boldsymbol{Y}+
(\boldsymbol{Y}\cdot\nabla)\boldsymbol{X}+\boldsymbol{X}\times(\nabla\times\boldsymbol{Y})+
\boldsymbol{Y}\times(\nabla\times\boldsymbol{X}),\eeq setting $\boldsymbol{X}=\nabla\varphi$ e
$\boldsymbol{Y}=\B$, we obtain \beq
\nabla(\nabla\varphi\cdot\B)=(\nabla\varphi\cdot\nabla)\B+\nabla\varphi\times(\nabla\times\B).\eeq
that inserted into \eqref{1claudiodimstep1}, after some simplifications, yields the relation
 
\beq 
\Box
\E=-\widetilde{g}(\nabla\varphi\cdot\nabla)\B+\widetilde{g}\frac{\partial\B}{\partial
t}\frac{\partial\varphi}{\partial t}.\eeq
This can be covariantized in the form  
 \beq 
\label{eik0} 
 \Box
\E=\widetilde{g}\partial_\mu \varphi\, \partial^\mu \B.\eeq
It is possible to square this relation in order to remove the mixed derivative terms in $\phi$ and $B$, bringing it into the form 
\begin{align}
\label{EeD}
\Box (\E-\frac{1}{2}\widetilde{g}\varphi
\B)=&-\frac{1}{2}\widetilde{g}\varphi \, \Box \B .
\end{align}
Indeed, expanding the equation above 
\beq -\Box\E+\frac{1}{2}\widetilde{g}\partial_\mu(\
\partial^\mu \varphi\,\B+\varphi\partial^\mu\B)-\frac{1}{2}\widetilde{g}\varphi\Box\B=0,
\label{eik1}
\eeq 
and using the eikonal approximation, \eqref{eik1} reduces to  \eqref{eik0}. It is convenient to introduce the induced electric field 
\beq
\bold{D}=\E-\frac{1}{2}\widetilde{g}\varphi
\B
\eeq
and solve perturbatively the equation by an appropriate expansion. 
Therefore \eqref{EeD}, that we call eikonal-$D$, is the first 
equation useful for investigating the rotation of the angle of polarization of an incoming, linearly polarized plane wave on the material, 
in the presence of a slowly varying background axion field.  \\
\subsection{Eikonal H}
Before turning to the perturbative solution, we derive a similar equation in which $\E$ and $\B$ are interchanged.  
We take the $curl$ of \eqref{fieldeq5} \beq
\nabla\times\nabla\times\B-\frac{\partial}{\partial t} \nabla
\times \E=\widetilde{g}\nabla\times(\E\times\nabla\varphi)-
\widetilde{g}\nabla\times\left(\B\frac{\partial \varphi}{\partial t}\right),
\eeq 
that we can rewrite in the form
 \beq\label{step8}
\nabla(\nabla\cdot\B)-\nabla^2\B-\frac{\partial}{\partial
t} \nabla \times
\E=\widetilde{g}\nabla\times(\E\times\nabla\varphi)-
\widetilde{g}\nabla\times\left[\B\frac{\partial \varphi}{\partial t}\right].
\eeq 
The squared brackets in the last term above, indicate the we can apply the eikonal approximation and pull out the axion field from the action of the $curl$ and use the Bianchi identity \eqref{Eeq} \eqref{Beq} to obtain

\beq\label{step8}
-\nabla^2\B+\frac{\partial^2
\B}{\partial
t^2}=\widetilde{g}\nabla\times(\E\times\nabla\varphi)-
\widetilde{g}\frac{\partial \varphi}{\partial t}\nabla\times\B
.\eeq 
Using the identity
 \beq
\nabla\times(\boldsymbol{X}\times\boldsymbol{Y})=\boldsymbol{X}(\nabla\cdot\boldsymbol{Y})-
\boldsymbol{Y}(\nabla\cdot\boldsymbol{X})+(\boldsymbol{Y}\cdot\nabla)\boldsymbol{X}-
(\boldsymbol{X}\cdot\nabla)\boldsymbol{Y}\eeq 
we rewrite \beq
\nabla\times(\E\times\nabla\varphi)=(\nabla\varphi\cdot\nabla)\E-(\nabla\varphi)(\nabla\cdot\E)
,\eeq that allows to re-express \eqref{step8} in the form
\beq\label{step10} \Box
\B=\widetilde{g}\Big((\nabla\varphi\cdot\nabla)\E-\nabla\varphi(\nabla\cdot\E)\Big)-
\widetilde{g}(\nabla\times\B)\frac{\partial \varphi}{\partial t}.
\eeq
Notice that in the equation above, we can drop the term containing $\nabla\cdot \E$ if we 
 use\eqref{EE}, the modified Gauss law, performing a first eikonal approximation, deriving the expression

\beq \Box
\B=\widetilde{g}(\nabla\varphi\cdot\nabla)\E+\widetilde{g}\left(\frac{\partial
\E}{\partial t}+\E\times\nabla\varphi-\left[\B\frac{\partial
\varphi}{\partial t}\right]\right)\frac{\partial \varphi}{\partial t},
\eeq 
where we have re-expressed $\nabla\times \B$ in \eqref{step10} using the modified displacement law. The last term in the squared brackets can also be dropped in the eikonal limit, giving
\beq \Box
\B=\widetilde{g}(\nabla\varphi\cdot\nabla)\E-\widetilde{g}\frac{\partial
\E}{\partial t}\frac{\partial \varphi}{\partial t}. \eeq
This equation can be rewritten in the covariant form as 
\beq \Box
\B+\widetilde{g}\partial_{\mu}\E\,\partial^{\mu}\varphi=0. \eeq
As in the derivation of the equation for $D$, also in this case, the derivative terms bilinear in $\varphi$ and $\B$ are the leftover from the 
eikonal expansion of 
\begin{align}
\label{EeH}
\Box (\B+\frac{1}{2}\widetilde{g}\varphi
\E)=&\frac{1}{2}\widetilde{g}\varphi \, \Box \E.
\end{align}
Also in this case we define an effective magnetic field of the form 
\beq
\bold{H}=\B+\frac{1}{2}\widetilde{g}\varphi
\E.
\eeq
Notice that the system of the two eikonal equations \eqref{EeD} and \eqref{EeH}  is invariant under the duality transformation  \eqref{dd}. We summarize the equations in the forms 
\beq
\Box 
{\bold H}=-\frac{1}{2}\widetilde{g}\varphi \, \Box \B
\eeq
\beq
\Box {\bold D}=\frac{1}{2}\widetilde{g}\varphi \, \Box \E.
\eeq
When solving the equations perturbatively, at the lowest order with $\varphi$ held constant, we find that both $\Box \mathbf{E}$ and $\Box \mathbf{B}$ vanish, indicating that $\mathbf{E}$ and $\mathbf{B}$ satisfy the vacuum wave equation. Moving to the first order of approximation, we encounter new combinations: $\mathbf{E} + \widetilde{g}\varphi \frac{\mathbf{B}}{2}$ and $\mathbf{B} - \widetilde{g}\varphi \frac{\mathbf{E}}{2}$. These are denoted as $\mathbf{D}$ and $\mathbf{H}$, respectively. Notably, these combinations fulfill the homogeneous D'Alembert equation, indicating their propagation without distortion in free space.\\
This analytical progression implies that an electromagnetic wave, initially linearly polarized at the lowest order, will exhibit behavior at the first order where both $\mathbf{D}$ and $\mathbf{H}$ (and thus $\mathbf{B}$ and $\mathbf{E}$) oscillate along a fixed direction. This phenomenon leads to an intriguing consequence: with each incremental change in $\varphi$ by an amount $\Delta \varphi$ along the path of the electromagnetic wave, the vectors $\mathbf{E}$ and $\mathbf{B}$ undergo a rotation. The angle of this rotation, $\Delta \phi$, is directly proportional to the change in $\varphi$, as described by the equation:
\begin{equation}
\Delta \phi = \frac{1}{2}\widetilde{g}\, \Delta\varphi.
\end{equation}
This relationship highlights the coupling between the electromagnetic field and the scalar field $\varphi$, showcasing how variations in $\varphi$ across space lead to observable changes in the polarization angle of the wave, a phenomenon that could have implications for our understanding of wave propagation and field interactions in various physical contexts.

\subsection{Light-cone structure of the photon propagator and oscillations for a timelike axionic background}
\label{leading}
The equations of motion for the gauge fields are quite interesting in their behaviour around the light-cone 
\begin{equation}
\label{five}
\partial_\mu F^{\mu \nu} = b_0 \partial_\mu \varphi \epsilon^{\mu \nu \alpha \beta} \partial_\alpha A_\beta
\end{equation}
where $b_0\equiv 4  \tilde{g}$.
In the Lorentz gauge they take the form
\begin{equation}
\label{seven}
\Box A_\nu - k_{\nu \alpha \beta} \partial^\alpha A^\beta = 0,
\end{equation}
where
\begin{equation}
\label{eight}
k_{\nu \alpha \beta} = b_0 \partial_\mu \varphi \epsilon^{\mu \nu \alpha \beta}.
\end{equation}
Since the Gaussian operator is diagonal in the highest derivatives, the Hadamard expansion for the propagator of the gauge fields in the axionic background can be directly set up by introducing the Green's function $\hat{\Delta}$ for Eq. \eqref{seven}
\begin{equation}
\label{nine}
\Box \hat{\Delta}_{\nu \rho} - k_{\nu \alpha \beta} \partial^\alpha \hat{\Delta}^\beta_\rho = \delta_{\nu \rho} \delta^{4}(z),
\end{equation}
where $z = x - y$.
We will investigate the structure of the Feynman propagator around the light-cone showing the presence of some special features of the photon propagator. \\
The standard ansatz for its behaviour around the light cone is given by the Hadamard expansion
\begin{eqnarray}
\label{ten}
\hat{\Delta}_{\nu \rho}(x, y) &=& G^{(0)}_{\nu \rho}(x, y) D_F  \nonumber \\
&& - \frac{i}{16 \pi^2} \ln \left( \frac{z^2}{\mu^2} - i \epsilon \right) \sum_{n = 0}^{\infty} \left( \frac{z^2}{4} \right)^n \frac{1}{n!} G_{\nu \rho}^{(n + 1)}(x, y)
\end{eqnarray}
Here, $D_F$ is the Feynman free propagator in coordinate space:
\begin{equation}
\label{eleven}
D_F = \frac{1}{4 \pi^2 i (z^2 - i \epsilon)}
\end{equation}
and $\mu$ is a mass parameter introduced to keep the argument of the logarithm dimensionless.\\
We are going to investigate the behaviour of the coefficient of the leading pole singularity on the light cone, 
which is controlled by the anomaly. 
The recursion relations for the coefficients of the expansion can be readily obtained. Setting the singularities of the Green function to zero independently, when applied to the expansion \ref{ten}, we obtain 
a leading singularity equation for $z^{-4}$
\begin{equation}
\label{thirteen}
2 z^{\mu} \partial_\mu G^{(0)}_{\nu \rho}(x, y) - z^{\alpha} k_{\nu\alpha\beta} G^{(0)}_{\beta\rho}(x, y) = 0
\end{equation}
a $D_F$ (i.e., $z^{-2}$) singularity equation
\begin{eqnarray}
\label{fourteen}
\Box G^{(0)}_{\nu\rho} + G^{(1)}_{\nu\rho} + z^{\mu}\partial_\mu G^{(1)}_{\nu\rho} -
k_{\nu\alpha\beta}\partial^{\alpha}G^{(0)}_{\beta\rho}  & & \nonumber \\
- {z^\alpha \over 2} k_{\nu\alpha\beta} {G^{(1)}}^{\beta}_{\rho} &=& 0
\end{eqnarray}
and a log-equation (for $z^{2n}\ln({z^2\over {\mu}^2})$):
\begin{eqnarray}
\label{fifteen}
z^{\mu} \partial_\mu G^{n+2}_{\nu\rho} + (n+2)G^{(n+1)}_{\nu\rho} +
\Box G^{(n+1)}_{\nu\rho} &  & \nonumber \\
- \frac{z^{\alpha}}{2} k_{\nu\alpha\beta}
G^{(n+2)\beta}_{\rho} -
k_{\nu\alpha\beta} \partial^{\alpha} G^{(n+1)}_{\beta\rho} &=& 0
\end{eqnarray}
Defining $G^{(-1)} = 0$ for $n = -1, 0, 1, ...$, all the equations can be summarized in the form
\begin{eqnarray}
\label{sixteen}
z^{\mu} \partial_\mu G^{(n+1)}_{\nu\rho} + (n+1)G^{(n+1)}_{\nu\rho} + \Box
G^{(n)}_{\nu\rho}  &  & \nonumber \\
- \frac{z^{\alpha}}{2} k_{\nu\alpha\beta}
G^{(n+1)}_{\beta\rho} - k_{\nu\alpha\beta}\partial^{\alpha}G^{(n) \beta}_{\rho}
&=& 0.
\end{eqnarray}
Here, we observe that the axion field acts as a background field, affecting the propagation of the leading singularity in a non-trivial manner (through $k_{\nu\alpha\beta}$).\\
For a constant $\varphi$-angle, the equation for the propagator of the leading singularity becomes
\begin{equation}
\label{seventeen}
z^\mu \partial_\mu G^{(0)}_{\nu\rho} = 0
\end{equation}
holds true on the light cone surface. The initial condition is:
\begin{equation}
\label{eighteen}
G^{(0)}_{\nu\rho} (x, x) = \delta_{\nu\rho}
\end{equation}
The solution is simply the Kronecker delta, making the strength diagonal across the entire characteristic surface. In general, the propagator for separate points is given by 
\begin{equation}
\label{twenty}
\Box^{-1} = \frac{G^{(0)}_{\nu\rho}}{(z^2 - i \epsilon) 4\pi^2 i} = \frac{\delta_{\nu\rho}}{(z^2 - i \epsilon) 4\pi^2 i}.\end{equation}

\subsection{Leading singularity dynamics}

We now focus on Eq. (16), which describes the leading behavior of the electromagnetic field dynamics in the presence of a local vacuum angle.\\
We will demonstrate that for a linearly increasing axion field in the timelike case, some components of the leading singularity tensor ${G^{(0)}}_{\nu\rho}$ remain constant on the light cone surface.\\
The reasoning is straightforward and applies to both abelian and non-abelian cases. Contracting both sides of \eqref{sixteen} with $\partial_{\nu}\theta$ and utilizing the antisymmetry of $k$ (defined in \eqref{eight}), we obtain:
\begin{equation}
\label{twenty-one}
\partial_\nu \theta z^\mu \partial_\mu {G^{(0)}}_{\nu\rho}(x(s), 0) = 0.
\end{equation}
We introduce the parameterization
\begin{equation}
\label{twenty-two}
x^\mu(s) = s x^\mu, \quad 0 < s < 1; \quad y = 0
\end{equation}
for a straight line on the light-cone surface, we can rewrite \eqref{twenty-one} as:
\begin{equation}
\label{twenty-three}
\partial_\nu \theta(x) \frac{d}{ds} {G^{(0)}}_{\nu\rho}(x(s), 0) = 0
\end{equation}
Repeating this process again yields:
\begin{equation}
\label{twenty-four}
x^\nu \frac{d}{ds} G^{(0)}_{\nu\rho}(x(s), 0) = 0
\end{equation}
This implies that certain components of the leading singularity tensor ${G^{(0)}}_{\nu\rho}$ are constant along straight lines on the light-cone surface, for a timelike gradient of the axion field.

\subsection{Timelike axion field and the leading singularity}

It can be shown that if the variation of the $\theta$ field is linear in a timelike direction, a frame exists where it only has a time dependence:
\begin{equation}
\partial_\mu \theta = (a_0, 0, 0, 0)
\end{equation}
In particular, for the timelike components, we obtain:
\begin{equation}
\frac{d}{ds} G^{(0)}_{0\rho} = 0
\end{equation}
Therefore, from the initial condition (Eq. 21), we get
\begin{equation}
G^{(0)}_{0i} = 0 \quad \quad G^{(0)}_{00} = 1
\end{equation}
The remaining components of the leading singularity tensor can also be determined.
Equation \eqref{sixteen} for the leading singularity can be expressed as:
\begin{equation}
\frac{d}{ds} G^{(0)}_{ij}(x(s), 0) - \frac{ca_0}{2} x^k \epsilon^{0ikl} G^{(0)}_{lj}(x(s), 0) = 0 \\
\quad \quad i, j, k, l = 1, 2, 3
\end{equation}
For a given timelike vector $x^\mu$, this equation has two first integrals. Define:
\begin{equation}
G^{(0)}_{ij}(x(s), 0) = \hat{v}_j; \quad \quad \hat{\omega}^k = \frac{a_0}{2} x^k
\end{equation}
where $\hat{\omega}$ is constant for a fixed $x^\mu$. Rewrite the equation as:
\begin{equation}
\frac{d}{ds} \hat{v}_j = \hat{\omega} \wedge \hat{v}_j
\end{equation}
Here, the two first integrals of motion along the s-line are:
\begin{equation}
\hat{\omega} \cdot \hat{v}_j = \frac{a_0}{2} x^k G^{(0)}_{kj} (x(s), 0) = d_j
\end{equation}
\begin{equation}
\hat{v}^2_j = \sum_i G^{(0)}_{ij}(x(s), 0) G^{(0)}_{ij} (x(s), 0) = l_j
\end{equation}
This approach allows for determining the remaining components of the leading singularity tensor $G^{(0)}_{ij}$ based on the initial conditions and the specific form of the timelike axion field.\\
We identify the following expressions for the Feynman propagators of electromagnetic fields in the linearly growing, timelike $\theta$ vacuum, with $r=|\vec{x}|$
\begin{equation}
<A_i(x) A_j(0)>_{\theta} =  [(\delta_{ij} - \frac{x^{ij}}{r^2}) \cos \left( \frac{a_0 r}{2} \right) + \epsilon^{ikj} \frac{x^k}{r} \sin \left( \frac{a_0 r}{2} \right) + \frac{x^i x^j}{r^2}] \frac{1}{4 \pi^2 i (x^2 - i 0)} + \text{log. terms}
\end{equation}
\begin{equation}
<A_0(x) A_i(0)>_{\theta} = \frac{\delta_{0i}}{4 \pi^2 i (x^2 - i 0)} + G^{(1)}_{0i}(x, 0) \ln (x^2 - i 0),
\end{equation}
with a typical oscillating behaviour on the leading singularity induced by the anomaly around the light-cone surface. The oscillations are present if the gradient of the same background is nonzero. In general, the structure of the bitensor characterising the residue of the leading singularity of a given propagator cannot be determined globally, but only at coincidence points, as clear from the recursion relations derived from the Hadamard expansion. The propagation, in this case, is quite interesting, since $G^{(0)}_{\mu\nu}(x,y)$ can be determined globally. We expect that the specific features of this propagation may provide guidance for the  identification of the anomaly behaviour in real experimental settings, given the singularity of the phenomenon. 

\section{Perspectives and conclusions}
In conclusion, our investigation has shed light on the intricate dynamics of chiral anomaly-driven interactions within the framework of conformal field theory (CFT) in four-dimensional spacetime. \\
Throughout our discussion, we have elucidated the fundamental aspects of these interactions, particularly focusing on chiral anomalies in both vector currents and gravitons. We have highlighted their emergence in topological materials, where gravitational chiral anomalies arise from thermal gradients, as elucidated by the  Luttinger relation.\\
Moreover, within the framework of CFT, we have uncovered the mediation of these interactions through quasiparticle excitations in the form of anomaly poles, discerned through a nonlocal effective action derived via perturbation theory. \\
Our examination of conformal Ward identities (CWIs) in momentum space has provided insights into the longitudinal and transverse sectors of chiral anomaly interactions, demonstrating their intimate coupling and the role of intermediary particles such as chiral fermions or bilinear Chern-Simons currents.
We have discussed the transformation of anomaly poles into cuts in the presence of fermion mass corrections, while emphasizing the retention of mass-independent sum rules in the axial-vector channel in all the chiral cases. Lastly, our exploration has highlighted the importance of investigating these sum rules alongside experimental observations, such as the rotation of the plane of polarization of incident light, in the quest to detect axion-like/quasiparticle entities in topological materials.\\
In retrospect, our journey through the intricacies of chiral anomaly-driven interactions underscores their profound implications across various disciplines, serving as a testament to the richness of theoretical frameworks like CFT in elucidating fundamental phenomena in nature.

\centerline{\bf Acknowledgements}
We thank Stefania D'Agostino and Alessandro Bramanti for discussions.
This work is partially supported by INFN within the Iniziativa Specifica QG-sky.  
The work of C. C. and S.L. is funded by the European Union, Next Generation EU, PNRR project "National Centre for HPC, Big Data and Quantum Computing", project code CN00000013. This work is partially supported by the the grant PRIN 2022BP52A MUR "The Holographic Universe for all Lambdas" Lecce-Naples.

\appendix

\section{Special conformal Ward identities in the operatorial approach}
In this appendi, for the sake of completeness, we present an operator-based derivation of conformal Ward Identities (CWIs) for three-point function correlators that include the stress-energy tensors.\\
To start, we consider an infinitesimal transformation:
\begin{equation}
x^\mu \to x'^\mu = x^\mu + v^\mu(x),
\end{equation}
which is categorized as an isometry if it preserves the form of the metric tensor \(g_{\mu \nu}(x)\). The transformed metric \(g'_{\mu\nu}(x')\) in the new coordinate system \(x'\) maintains the form:
\begin{equation}
g'_{\mu\nu}(x') = g_{\mu\nu}(x').
\end{equation}
Implementing this condition into the standard covariant transformation rule for \(g_{\mu\nu}(x)\) results in:
\begin{equation}
g'_{\mu\nu}(x') = \frac{\partial x^\rho}{\partial x'^\mu} \frac{\partial x^\sigma}{\partial x'^\nu} g_{\rho\sigma}(x) = g_{\mu\nu}(x'),
\end{equation}
leading to the derivation of the Killing equation:
\begin{equation}
v^\alpha \partial_\alpha g_{\mu\nu} + g_{\mu\sigma} \partial_\nu v^\sigma + g_{\sigma \nu} \partial_\mu v^\sigma = 0.
\end{equation}
For conformal transformations, the metric condition is modified to:
\begin{equation}
g'_{\mu\nu}(x') = \Omega^{-2} g_{\mu\nu}(x'),
\end{equation}
yielding the conformal Killing equation:
\begin{equation}
v^\alpha\partial_\alpha g_{\mu\nu} + g_{\mu\sigma} \partial_\nu v^\sigma + g_{\sigma \nu} \partial_\mu v^\sigma = 2 \sigma g_{\mu\nu}.
\end{equation}
In the limit of flat spacetime, this simplifies to:
\begin{equation}
\partial_\mu v_\nu + \partial_\nu v_\mu = 2 \sigma \eta_{\mu\nu}, \quad \sigma = \frac{1}{d} \partial \cdot v.
\end{equation}
Transitioning to the Euclidean case and omitting the index positions, we express any conformal transformation via a local rotation matrix:
\begin{equation}
R^\mu_\alpha = \Omega \frac{\partial x'^\mu}{\partial x^\alpha},
\end{equation}
which allows us to expand \(R\) around the identity matrix:
\begin{equation}
R = \mathbf{1} + [\mathbf{\epsilon}] + \ldots,
\end{equation}
where \([\epsilon]\) is an antisymmetric matrix, reformulated in terms of antisymmetric parameters \(\tau_{\rho\sigma}\) and the \(SO(d)\) rotation group generators \(\Sigma_{\rho\sigma}\) as:
\begin{align}
[\epsilon]_{\mu\alpha} &= \frac{1}{2} \tau_{\rho\sigma} (\Sigma_{\rho\sigma})_{\mu\alpha}, \\
(\Sigma_{\rho\sigma})_{\mu\alpha} &= \delta_{\rho\mu}\delta_{\sigma\alpha} - \delta_{\rho\alpha}\delta_{\sigma\mu}.
\end{align}
From this, we derive a relation between the conformal transformation parameters \(v\) and the rotation matrix parameters \(\tau_{\mu\alpha}\):
\begin{equation}
R_{\mu\alpha} = \delta_{\mu\alpha} + \tau_{\mu\alpha} = \delta_{\mu\alpha} + \frac{1}{2}\partial_{[\alpha}v_{\mu]}.
\end{equation}
Considering the scaling dimensions \(\Delta_A\) of a vector field \(A_\mu(x)'\), its transformation under a conformal change is captured by:
\begin{align}
A'^\mu(x') &= \Omega^{\Delta_A} R_{\mu \alpha} A^\alpha(x), \\
&= (1-\sigma+\ldots)^{\Delta_A}(\delta_{\mu\alpha}+\frac{1}{2}\partial_{[\alpha}v_{\mu]} +\ldots) A^\alpha(x),
\end{align}
leading to an expression for the variation of \(A^\mu(x)\):
\begin{equation}
\delta A^\mu(x) = -(v\cdot \partial +\Delta_A \sigma)A^\mu(x) + \frac{1}{2} \partial_{[\alpha}v_{\mu]}A^\alpha(x),
\end{equation}
defined as the Lie derivative \(L_v A^\mu(x) = -\delta A^\mu(x)\), aside from a sign difference.
\\
Exploring a general rank-2 tensor field \(\phi^{IK}\) with scaling dimension \(\Delta_\phi\) and transforming under the representation \(D^I_J(R)\) of the rotation group \(SO(d)\), we adapt the previous form to:
\begin{equation}
\phi'^{IK}(x') = \Omega^{\Delta_{\phi}} D^I_{I'}(R) D^K_{K'}(R) \phi^{I'K'}(x).
\end{equation}
For the stress-energy tensor, utilizing its scaling dimension \(\Delta_T\) and considering a special conformal transformation (SCT) parameterized by \(b_\mu\), we derive its impact on \(T^{\mu\nu}(x)\), culminating in the formulation of the SCT operator \(\mathcal{K}^\kappa\) acting on \(T\) in a finite form.\\
For a special conformal transformation (SCT), characterized by
\begin{equation}
v_\mu(x) = b_\mu x^2 - 2 x_\mu b \cdot x,
\end{equation}
the transformation of the stress-energy tensor becomes
\be
\delta T^{\mu\nu}(x)=-(b^\alpha x^2 -2 x^\alpha b\cdot x )\, \partial_\alpha  T^{\mu\nu}(x)   - \Delta_T \sigma T^{\mu\nu}(x)+
2(b_\mu x_\alpha- b_\alpha x_\mu)T^{\alpha\nu} + 2 (b_\nu x_\alpha -b_\alpha x_\nu)\, T^{\mu\alpha}(x).
\ee
It is sufficient to differentiate this expression respect to $b_\kappa$ in order to derive the form of the 
SCT  $K^\kappa$ on $T$ in its finite form 
\bea
\mathcal{K}^\kappa T^{\mu\nu}(x)&\equiv &\delta_\kappa T^{\mu\nu}(x) =\frac{\partial}{\partial b^\kappa} (\delta T^{\mu\nu})\nonumber \\
&=& -(x^2 \partial_\kappa - 2 x_\kappa x\cdot \partial) T^{\mu\nu}(x) + 2\Delta_T x_\kappa T^{\mu\nu}(x) +
2(\delta_{\mu\kappa}x_\alpha -\delta_{\alpha \kappa}x_\mu) T^{\alpha\nu}(x) \nonumber \\ 
&& + 2 (\delta_{\kappa\nu} x_{\alpha} -\delta_{\alpha \kappa} x_\nu )T^{\mu\alpha}. 
\label{ith}
\eea
The approach can be generalized to correlators built out of several operators. 
In the case of a $TJJ$ correlator, 
\be
\Gamma^{\mu\nu\alpha\beta}(x_1,x_2,x_3)=\langle T^{\mu\nu}(x_1) J^\alpha(x_2)J^\beta(x_3)\rangle
\ee
with a vector current of dimension $\Delta_J$, the CWI's  take the explicit form 
\bea
\mathcal{K}^\kappa \Gamma^{\mu\nu\a\b}(x_1,x_2,x_3) 
&=& \sum_{i=1}^{3} {K_i}^{ \kappa}_{scalar}(x_i) \Gamma^{\mu\nu\a\b}(x_1,x_2,x_3) \nn \\
&& + 2 \left(  \delta^{\mu\kappa} x_{1\rho} - \delta_{\rho}^{\kappa }x_1^\mu  \right)\Gamma^{\rho \nu\alpha\beta}
 + 2 \left(  \delta^{\nu\kappa} x_{1\rho} - \delta_{\rho}^{\kappa }x_1^\nu  \right)\Gamma^{\mu\rho \alpha\beta}\nonumber
\\ &&  2 \left(  \delta^{\a\kappa} x_{2\rho} - \delta_{\rho}^{\kappa }x_2^\a  \right)\Gamma^{\mu\nu \rho\beta}
 +  2 \left(  \delta^{\beta\kappa} x_{3\rho} - \delta_{\rho}^{\kappa }x_3^\beta  \right)\Gamma^{\mu\nu \alpha\rho}=0,
 \eea
where 
\be
\label{ki}
{\mathcal{K}_i}^{\kappa}_{scalar}=-x_i^2 \frac{\partial }{\partial x_\kappa} + 2 x_i^\kappa x_i^\tau\frac{\partial}{\partial x_i^\tau} + 2 \Delta_i x_i^\kappa
\ee
is the scalar part of the special conformal operator acting on the $i_{\textrm{th}}$ coordinate and $\Delta_i\equiv(\Delta_T,\Delta_J,\Delta_J)$ are the scaling dimensions of the operators in the correlation function.

 \section{Scalar 3-point functions: examples of solutions}

To elucidate the structure of the solutions of the CWIs, as an example, we examine the scalar correlator $\Phi(p_1,p_2,p_3)$, defined by two homogeneous conformal equations:
\begin{equation}
K_{31}\Phi=0 \quad \text{and} \quad K_{21}\Phi=0,
\end{equation}
combined with the scaling equation:
\begin{equation}
\label{scale}
\sum_{i=1}^3 p_i\frac{\partial}{\partial p_i} \Phi=(\Delta-2 d) \Phi.
\end{equation}
Following the approach outlined in \cite{Coriano:2013jba}, we adopt an ansatz for the solution in the form:
\begin{equation}
\label{ans}
\Phi(p_1,p_2,p_3)=p_1^{\Delta - 2 d} x^{a}y^{b} F(x,y),
\end{equation}
where $x=\frac{p_2^2}{p_1^2}$ and $y=\frac{p_3^2}{p_1^2}$. Here, $p_1$ is chosen as the "pivot" in the expansion, although any of the three momentum invariants could be chosen equivalently. The homogeneity of $\Phi$ of degree $\Delta-2d$ under a scale transformation, as dictated by \eqref{scale}, is accommodated by the factor $p_1^{\Delta - 2 d}$ in \eqref{ans}. Employing scale-invariant variables $x$ and $y$ leads to the hypergeometric form of the solution.

Upon analysis, we find the equations:
\begin{align}
K_{21}\Phi &= 4 p_1^{\Delta -2d -2} x^a y^b \notag \\
&\quad\times\left(  x(1-x)\frac{\partial }{\partial x \partial x}  + (A x + \gamma)\frac{\partial }{\partial x} -
2 x y \frac{\partial^2 }{\partial x \partial y}- y^2\frac{\partial^2 }{\partial y \partial y} + 
D y\frac{\partial }{\partial y} + (E +\frac{G}{x})\right) F(x,y) = 0,
\label{red}
\end{align}
where
\begin{align}
&A = D = \Delta_2 +\Delta_3 - 1 -2 a -2 b -\frac{3 d}{2}, \notag \\
&\gamma(a) = 2 a +\frac{d}{2} -\Delta_2 + 1, \notag \\
&G = \frac{a}{2}(d +2 a - 2 \Delta_2), \notag \\
&E = -\frac{1}{4}(2 a + 2 b +2 d -\Delta_1 -\Delta_2 -\Delta_3)(2 a +2 b + d -\Delta_3 -\Delta_2 +\Delta_1).
\end{align}
Similar constraints are obtained from the equation $K_{31}\Phi=0$, with the exchanges $(a,b,x,y)\to(b,a,y,x)$:
\begin{align}
K_{31}\Phi &= 4 p_1^{\Delta -2 d -2} x^a y^b \notag \\
&\quad\times\left(  y(1-y)\frac{\partial }{\partial y \partial y}  + (A' y + \gamma')\frac{\partial }{\partial y} -
2 x y \frac{\partial^2 }{\partial x \partial y}- x^2\frac{\partial^2 }{\partial x \partial x} + 
D' x\frac{\partial }{\partial x} + (E' +\frac{G'}{y})\right) F(x,y) = 0,
\label{red}
\end{align}
where
\begin{align}
&A'=D'= A, \quad \gamma'(b) = 2 b +\frac{d}{2} -\Delta_3 + 1, \notag \\
&G' = \frac{b}{2}(d +2 b - 2 \Delta_3), \quad E' = E.
\end{align}
To reduce the equations to hypergeometric form, in \eqref{red}, we set $G/x=0$ and $G'/y=0$, implying:
\begin{equation}
\label{cond1}
a=0\equiv a_0 \quad \text{or} \quad a=\Delta_2 -\frac{d}{2}\equiv a_1,
\end{equation}
and
\begin{equation}
\label{cond2}
b=0\equiv b_0 \quad \text{or} \quad b=\Delta_3 -\frac{d}{2}\equiv b_1.
\end{equation}
The four independent solutions of the CWIs are characterized by the same four pairs of indices $(a_i,b_j)$, where $i,j=1,2$. Defining:
\begin{align}
&\alpha(a,b)= a + b + \frac{d}{2} -\frac{1}{2}(\Delta_2 +\Delta_3 -\Delta_1), \notag \\
&\beta(a,b)= a +  b + d -\frac{1}{2}(\Delta_1 +\Delta_2 +\Delta_3),
\label{alphas}
\end{align}
we have:
\begin{equation}
E = E' = -\alpha(a,b)\beta(a,b), \quad A = D = A' = D' = -\left(\alpha(a,b) +\beta(a,b) +1\right).
\end{equation}
Apologies for the abrupt ending. Continuing from where we left off:

The solutions take the form:
\begin{align}
\label{F4def}
&F_4(\alpha(a,b), \beta(a,b); \gamma(a), \gamma'(b); x, y) \notag \\
&\quad= \sum_{i = 0}^{\infty}\sum_{j = 0}^{\infty} \frac{(\alpha(a,b), {i+j}) \, 
	(\beta(a,b),{i+j})}{(\gamma(a),i) \, (\gamma'(b),j)} \frac{x^i}{i!} \frac{y^j}{j!}.
\end{align}
We denote $\alpha\ldots \gamma'$ as the first$\ldots$ fourth parameters of $F_4$. The four independent solutions are then all of the form $x^a y^b F_4$, where the hypergeometric functions will take some specific values for its parameters, with $a$ and $b$ fixed by \eqref{cond1} and \eqref{cond2}. Specifically, we have:
\begin{equation}
\Phi(p_1,p_2,p_3)=p_1^{\Delta-2 d} \sum_{a,b} c(a,b,\vec{\Delta})\,x^a y^b \,F_4(\alpha(a,b), \beta(a,b); \gamma(a), \gamma'(b); x, y), 
\label{compact}
\end{equation}
where the sum runs over the four values $a_i, b_i$ $(i=0,1)$ with arbitrary constants $c(a,b,\vec{\Delta})$, and $\vec{\Delta}=(\Delta_1,\Delta_2,\Delta_3)$. Notice that \eqref{compact} is a very compact way to write down the solution. However, once this type of solutions of a homogeneous hypergeometric system are inserted into an inhomogeneous system of equations, the sum over $a$ and $b$ needs to be made explicit. For this reason, it is convenient to define:
\begin{align}
&\alpha_0\equiv \alpha(a_0,b_0)=\frac{d}{2}-\frac{\Delta_2 + \Delta_3 -\Delta_1}{2},\, && \beta_0\equiv \beta(b_0)=d-\frac{\Delta_1 + \Delta_2 +\Delta_3}{2},  \notag \\
&\gamma_0 \equiv \gamma(a_0) =\frac{d}{2} +1 -\Delta_2,\, \nn \\
&\gamma'_0\equiv \gamma(b_0) =\frac{d}{2} +1 -\Delta_3.
\end{align}
These are the four basic (fixed) hypergeometric parameters, and all the remaining ones are defined by shifts with respect to these. The four independent solutions can be re-expressed in terms of the parameters above as:

\begin{align}
\label{F4def}
&S_1(\alpha_0, \beta_0; \gamma_0, \gamma'_0; x, y)\equiv F_4(\alpha_0, \beta_0; \gamma_0, \gamma'_0; x, y) = \sum_{i = 0}^{\infty}\sum_{j = 0}^{\infty} \frac{(\alpha_0,i+j) \, 
(\beta_0,i+j)}{(\gamma_0,i )\, (\gamma'_0,j)} \frac{x^i}{i!} \frac{y^j}{j!} 
\end{align}

and

\begin{align}
\label{solutions}
&S_2(\alpha_0, \beta_0; \gamma_0, \gamma'_0; x, y) = x^{1-\gamma_0} \, F_4(\alpha_0-\gamma_0+1, \beta_0-\gamma_0+1; 2-\gamma_0, \gamma'_0; x,y) \,, \nn \\
&S_3(\alpha_0, \beta_0; \gamma_0, \gamma'_0; x, y) = y^{1-\gamma'_0} \, F_4(\alpha_0-\gamma'_0+1,\beta_0-\gamma'_0+1;\gamma_0,2-\gamma'_0 ; x,y) \,, \nn \\
&S_4(\alpha_0, \beta_0; \gamma_0, \gamma'_0; x, y) = x^{1-\gamma_0} \, y^{1-\gamma'_0} \, 
F_4(\alpha_0-\gamma_0-\gamma'_0+2,\beta_0-\gamma_0-\gamma'_0+2;2-\gamma_0,2-\gamma'_0 ; x,y) \, . \nn
\end{align}

Notice that in the scalar case, one is allowed to impose the complete symmetry of the correlator under the exchange of the 3 external momenta and scaling dimensions, as discussed in \cite{Coriano:2013jba}. This reduces the four constants to just one.

%\bibliographystyle{jhep}
%\bibliography{TJJdilatonHprime}
\providecommand{\href}[2]{#2}\begingroup\raggedright\endgroup

\end{document}